\def\nk{n_{\rm b}}

\def\cu{\cos u}
\def\su{\sin u}

\def\rfr#1{Equation~(\ref{#1})}
\def\rfrs#1#2{Equations~(\ref{#1})~to~(\ref{#2})}

\def\derp#1#2{\rp{\partial{#1}}{\partial{#2}}}
\def\dert#1#2{\frac{{{\textrm{d}}}{#1}}{{{\textrm{d}}}{#2}}}

\def\virg#1{``#1"}

\def\eqi{\begin{equation}}
\def\eqf{\end{equation}}
\def\eqia{\begin{eqnarray}}
\def\eqfa{\end{eqnarray}}

\def\rp#1#2{{#1\over#2}}
\def\lb#1{\label{#1}}

\def\bds#1{\boldsymbol{#1}}

\def\cO{\cos\Omega}
\def\sO{\sin\Omega}

\def\cI{\cos I}
\def\sI{\sin I}

\def\ton#1{\left(#1\right)}
\def\qua#1{\left[#1\right]}
\def\grf#1{\left\{#1\right\}}
\def\ang#1{\left\langle #1\right\rangle}
\documentclass{aastex}

\usepackage{float}
\usepackage{amsmath,textgreek,w-greek,wasysym}
\usepackage{amscd,lineno}
\usepackage{amssymb}
\usepackage{graphicx,epsfig}
\usepackage{txfonts}
\usepackage{xr-hyper}
\RequirePackage{color}

\newcommand{\emaila}{lorenzo.iorio@libero.it}

\allowdisplaybreaks[1]

\begin{document}

\title{Preliminary constraints on the location of the recently hypothesized new planet of the Solar System from planetary orbital dynamics}

\shortauthors{L. Iorio}

\author{Lorenzo Iorio\altaffilmark{1} }
\affil{Ministero dell'Istruzione, dell'Universit\`{a} e della Ricerca
(M.I.U.R.)-Istruzione
\\ Permanent address for correspondence: Viale Unit\`{a} di Italia 68, 70125, Bari (BA),
Italy}

\email{\emaila}

\begin{abstract}
It has been recently proposed that the observed  grouping of either the perihelia and the orbital planes  of some observed distant Kuiper Belt Objects (KBOs) can be explained by the shepherding influence of a remote ($150~\textrm{au}\lesssim q_\textrm{X}\lesssim 350~\textrm{au}$), still unseen massive object PX having  planetary size ($5~m_\oplus\lesssim m_\textrm{X}\lesssim 20~m_\oplus$) and moving along an ecliptically inclined ($22~\textrm{deg}\lesssim I_\textrm{X}\lesssim 40~\textrm{deg}$), eccentric ($380~\textrm{au}\lesssim a_\textrm{X}\lesssim 980~\textrm{au}$) Heliocentric bound orbit located in space at $80~\textrm{deg}\lesssim \Omega_\textrm{X}\lesssim 120~\textrm{deg}$ and which is anti-aligned ($120~\textrm{deg}\lesssim \omega_\textrm{X}\lesssim 160~\textrm{deg}$) with those of the considered KBOs.
The trajectory of Saturn is nowadays known at essentially the same accuracy level of the inner planets due to the telemetry of the Cassini spacecraft. Thus, the expected perturbations $\dot\varpi,~\dot\Omega$ due to PX on the Kronian apsidal and draconitic orbital motions  are theoretically investigated to tentatively constrain the configuration space of PX itself. To this aim, we compare our predictions $\dot\varpi_\textrm{theo},~\dot\Omega_\textrm{theo}$ to the currently available experimental intervals of values $\Delta\dot\Omega_\textrm{obs},~\Delta\dot\varpi_\textrm{obs}$  determined by astronomers in the recent past without explicitly modeling and solving for PX itself. As such, our results, despite being plausible and in agreement to a large extent with other constraints released in the literature, should be regarded as proof-of-principle investigations aimed to encourage more accurate analyses in future. It turns out that the admissible region in its configuration space is moderately narrow as far as its position along its orbit, reckoned by the true anomaly $f_\textrm{X}$, is concerned, being concentrated around approximately $130~\textrm{deg}\lesssim f_\textrm{X}\lesssim 240~\textrm{deg}$. PX is certainly far from its perihelion ($f_\textrm{X}=0~\textrm{deg}$), in agreement with other recent studies. The future analysis of the data from the ongoing New Horizons mission might be helpful in further constraining the scenario considered here for PX. Its impact on the spaceraft's range over a multi-year span is investigated with a preliminary sensitivity analysis.
\end{abstract}


%

keywords{
Oort Cloud--Kuiper belt: general--celestial mechanics--ephemerides--gravitation
}
\section{Introduction}
Recently, indirect evidence for a distant body, dubbed\footnote{Here and in the following, \virg{X} is used with the meaning of \virg{unknown}, not as the Roman numeral for ten.} PX or X in the following for simplicity, having planetary size and lurking in the outskirts of the Solar System has been claimed once more \citep{BaBroAJ2016}. This time, its putative existence has been inferred from its possible shepherding gravitational action on some  of the observed distant Kuiper Belt Objects (KBOs) which turned out to exhibit a peculiar grouping pattern either in the values of their arguments of perihelia $\omega$ \citep{2014Natur.507..471T} and in the physical space itself \citep{BaBroAJ2016}. Its  physical and orbital features originally put forth by \citet{BaBroAJ2016} are resumed in Table~\ref{kepelemsX}.
Later, \citet{BroBaAJ2016} explored the parameter space of the hypothesized planet in more depth by allowing  for a certain range of variability in its possible physical and orbital features as a result of several numerical simulations of the dynamical evolution of different populations of KBOs. Examples of such an enlarged parameter space are displayed from Table~\ref{largo}~to~\ref{uffa}. Furthermore, \citet{2016AJ....152..126B} suggested that a distant pointlike pertuber with the characteristics envisaged in \citet{BroBaAJ2016} may explain the  observed obliquity of the Sun's equator as well as the specific pole position of the Solar spin axis from a nearly aligned initial state.

The publication of the seminal paper by \citet{BaBroAJ2016} gave vent to a lingering flood of works dealing with several different aspects concerning such a revamped version of the time-honoured hypothesis that the Sun has more planet(s) than those currently known; see, e.g.,
\citet{2016ApJ...822L...2C, 2016ApJ...824L..25F,2016ApJ...825...33K,2016ApJ...826...64B,2016AJ....152..126B,2016MNRAS.460L.109M,2016CeMDA.tmp...31S, 2016PASA...33...33P,2016MNRAS.460.1270D,2016arXiv160802873P,2016ApJ...827L..35N,2016ApJ...827L..24C,2016ApJ...827..125L,2016arXiv160704895W,
2016arXiv160701777K,2016ApJ...825L..13S,2016A&A...589A.134L,2016A&A...587L...8F,2016A&A...589A..63P,2016ApJ...822L..11G,2016MNRAS.458.2962R,
2016ApJ...824L..22M,2016arXiv160801421L,2016arXiv160705111G,2016Ap&SS.361..230S,
2016A&A...590L...2B,2016arXiv160506575L,2016ApJ...823L...3L,2016Ap&SS.361..212G,2016AJ....152...80H,2016AJ....152...94H,
2016MNRAS.463.2958V,2016MNRAS.460L..64D,2016MNRAS.459L..66D,2016MNRAS.462.1972D,2016MNRAS.460L.123D,
2017MNRAS.464.2290M,2016arXiv160908614M,2016arXiv160808772S,2016A&A...592A..86T,2016arXiv161004251S,2016arXiv161004992B,2017MNRAS.464.1709G}, and counting.

In this work, we propose to independently put to the test such an intriguing scenario by looking at the  perturbing effects that such a hypothesized distant planet would induce on the orbital motion of Saturn. Indeed, the ringed gaseous giant is the farthest known major body of the Solar System for which extended data records, whose level of accuracy is nowadays comparable with that of the inner planets thanks to the multi-year precise radiotracking to the Cassini spacecraft, are available.

In particular, we will attempt to preliminary narrow the configuration space of the putative remote planet of the Solar System by looking at the secular precessions of the Kronian longitudes of the perihelion $\varpi$ and of the node $\Omega$ induced by a distant pointlike perturber and at the most recent observational determinations of the rates of such orbital elements of Saturn. Such an approach is novel, being complementary to that recently followed by \citet{2016A&A...587L...8F}, who suitably re-processed the radiotracking ranging data to the Cassini spacecraft collected over the the decade 2004-2104 by explicitly modeling the gravitational pull exerted by the putative new planet, and produced new residuals of the Earth-Saturn range. Also \citet{2016AJ....152...94H} looked at the Cassini range measurements by simulating them under the action of the hypothesized distant perturber. Then, they contrasted their fictitious ranges with those of the INPOP ephemerides used in \citet{2016A&A...587L...8F} in a least-square fashion by producing a set of simulated post-fit range residuals.

As far as the denomination of such a still undiscovered major body of the our planetary system, several names have become more or less popular over the years in either the specialized literature and in popular accounts. The late entrant in the literature is Planet Nine, so that the subscript \virg{9} is often appended to the symbols denoting its physical and orbital parameters; it would have been likely Planet Ten if Pluto had not been demoted from  planet status in 2006. Given the remarkable distance envisaged by \citet{BaBroAJ2016,BroBaAJ2016} for it, we propose  the name
Telisto\footnote{From $\uptau\acute{\upeta}\uplambda\upiota\upsigma\uptau\upo\upvarsigma$: farthest, most remote. It should not be confused with $\textsc{T}\upepsilon\uplambda\upepsilon\upsigma\uptau\acute{\upomega}$, Telesto, a minor deity of the Greek mythology, whose name was given in 1983 to one of the irregular small satellites of Saturn (\url{http://www.cbat.eps.harvard.edu/iauc/03800/03872.html}). Ours is just a suggestion which we unpretentiously offer to the astronomical community. See \citet{1958ASPL....8....9G} for an interesting historical account on how Uranus and Neptune finally took their current names after their discoveries. For Pluto, see \citet{PLUTO}. It can be seen that not always the names originally proposed by their discoverers actually became of common usage.} which, among other things, circumvents any possible future classification issues and addresses specifically a distinctive feature of the hypothetical body under consideration.

The paper is organized as follows. In Section, \ref{supple} we introduce the corrections to the standard nodal and apsidal precessions of Saturn determined from observations with some of the most recent planetary ephemerides by independent teams of astronomers. Our analytical model for the long-term rates of change of the node and the pericenter induced by a distant, pointlike perturber is elucidated in Section~\ref{modello}. Our preliminary constraints from the Kronian perihelion and node on the perturber's position along its orbit  are inferred in Section~\ref{costra}.
Section~\ref{lite} contains a discussion of the method adopted in Section~\ref{costra} by comparing it to other approaches  followed in the literature.
Constraints on the mass of PX are posed in Section~\ref{massa} from the Kronian orbital precessions, independently of the KBOs-based assumptions by \citet{BaBroAJ2016,BroBaAJ2016}.
The possibility of using the ongoing telemetry of New Horizons, a further complementary feature of this work with respect to \citet{2016A&A...587L...8F, 2016AJ....152...94H}, is discussed in Section~\ref{Niù}. Section~\ref{fine} summarizes our findings.

Finally, as a  general consideration on the method adopted, we remark that, since, as discussed in more detail in Section~\ref{lite}, the orbital precessions of Saturn which we  use in the present study were determined without explicitly modeling the dynamical action of PX, the results of the current analysis should be considered as proof-of-principle investigations aimed to encourage more accurate future studies. On the other hand, despite their necessarily preliminary character, it  turns out that the bounds  obtained by us are plausible and compatible to a non-negligible extent with those inferred in existing studies in which genuine solve-for data reductions were performed.
\section{The supplementary periehlion and node precessions}\lb{supple}
Table~\ref{perihelia} lists the most recent determinations of the admissible intervals of values $\Delta\dot\varpi_{\rm obs},~\Delta\dot\Omega_{\rm obs}$ for   potential anomalous perihelion and node precessions $\Delta\dot\varpi,~\Delta\dot\Omega$ of Saturn obtained by two independent teams of astronomers \citep{2013MNRAS.432.3431P, 2011CeMDA.111..363F, 2016arXiv160100947F} by processing long observational records collected during the last century without modeling the gravitational pull of any additional major body with respect to the known ones.
Such observations-based ranges for the Kronian supplementary apsidal and nodal precessions $\Delta\dot\varpi,~\Delta\dot\Omega$ with respect to the standard ones due to known dynamical effects of either Newtonian and relativistic nature are statistically compatible with zero. Thus, they can be used to put tentative bounds on the  planetary orbital perturbations exerted by a still unmodelled pointlike object by comparing them to the corresponding theoretical predictions $\dot\varpi_{\rm theo},~\dot\Omega_{\rm theo}$ for the precessions caused by such a kind of perturber on a particle at heliocentric distance $r$. Equation~(9) and Equation~(13) in \citet{2012CeMDA.112..117I}, derived from a perturbing potential \citep{1991AJ....101.2274H} accurate to order $\mathcal{O}\ton{r^2/r^2_\textrm{X}}$, are just what we need since they yield  analytical expressions for the long-term precessions  due to a distant pointlike perturber X whose orbital period is assumed to be much longer than that of the perturbed planet. Such formulas are functions of either the orbital elements of the perturbed test particle and of the position unit vector ${\bds{\hat{l}}}^\textrm{X}$ of PX, and they have a general validity  since they hold for arbitrary orbital geometries of both the perturbed and the perturbing bodies.

\section{The analytical model for the perihelion and node precessions due to a distant, pointlike perturber}\lb{modello}
Here, we offer an outline of the procedure followed to obtain Equation~(9) and Equation~(13) in \citet{2012CeMDA.112..117I}.

The perturbing potential by \citet{1991AJ....101.2274H} is
\eqi U_\textrm{X} = \rp{Gm_\textrm{X}}{2r^3_\textrm{X}}\qua{r^2 - 3\ton{\bds r\bds\cdot{\bds{\hat{l}}}^\textrm{X} }^2}. \lb{hogg}\eqf
In order to analytically work out the long-term orbital perturbations induced by \rfr{hogg}, it has, first, to be evaluated onto the Keplerian trajectory of the disturbed test particle, characterized by
\begin{align}
r &=a (1 -e \cos E), \\ \nonumber \\
x & = r\ton{\cO \cu - \cI\sO\su}, \\ \nonumber \\
y &= r\ton{\sO \cu + \cI\cO \su}, \\ \nonumber\\
z & = r\sI\su,
\end{align}
 where $E,~a,~e,~I,~\Omega,~u$ are the eccentric anomaly, the semimajor axis, the eccentricity, the inclination to the reference $\grf{x,~y}$ plane, the longitude of the ascending node, the argument of latitude of the test particle, respectively. The latter one is defined as
$ u \doteq \omega + f$, in which $\omega$ is the argument of pericenter, and $f$ is the true anomaly connected with $E$ through
\begin{align}
\cos f & = \rp{\cos E-e}{1-e\cos E}, \\ \nonumber \\
\sin f & = \rp{\sqrt{1-e^2}\sin E}{1-e\cos E}.
\end{align}
Then, the average over an orbital period $P_\textrm{b}=2\uppi/\nk$ of the test particle has to be calculated by means of
\eqi \dert{t}{E} = \rp{1-e\cos E}{\nk};\eqf the orbital elements of PX entering $r_\textrm{X},~{\bds{\hat{l}}}$ have to be kept constant, as it is justified by the assumed much larger distance $r_\textrm{X}$ with respect to $r$ (see the discussion at the end of the present Section).
Finally, one obtains
\begin{equation}
\left\langle U_\textrm{X} \right\rangle  \doteq \left(\rp{\nk}{2\uppi}\right)\int_0^{P_{\rm b}} U_\textrm{X}\ dt =  \rp{Gm_\textrm{X}a^2}{r^3_\textrm{X}32}\mathcal{U}\left(I,~\Omega,~\omega;~{\bds{\hat{l}}}^\textrm{X}\right),\lb{us}
\end{equation}
with
\begin{equation}
\begin{array}{lll}
\mathcal{U} &\doteq & - \left(2 + 3 e^2\right)  \left( -8  + 9 \hat{l}_x^2 + 9 \hat{l}_y^2 +
        6 \hat{l}_z^2\right)   -120 e^2 \sin 2\omega \left(\hat{l}_x \cos\Omega + \hat{l}_y \sin\Omega\right)  \left[\hat{l}_z \sin I +\right. \\ \\
    &+&\left.\cos I \left(\hat{l}_y \cos\Omega - \hat{l}_x \sin\Omega\right) \right]  - 15 e^2 \cos 2\omega \left[3
 \left(\hat{l}_x^2 - \hat{l}_y^2\right)  \cos 2\Omega + 2 \left(
              \hat{l}_x^2 + \hat{l}_y^2 - 2
                    \hat{l}_z^2\right)  \sin^2 I -\right. \\ \\
                     &-&\left. 4 \hat{l}_z \sin 2I \left(\hat{l}_y
\cos\Omega - \hat{l}_x \sin\Omega\right)  + 6 \hat{l}_x \hat{l}_y \sin 2\Omega\right]  -
          6 \left(2 + 3 e^2\right)  \left[ \left(\hat{l}_x^2 - \hat{l}_y^2\right)   \cos 2\Omega \sin^2 I +\right. \\ \\
           &+&\left. 2 \hat{l}_z \sin 2I \left(\hat{l}_y
                    \cos\Omega - \hat{l}_x \sin\Omega\right)  +
                2 \hat{l}_x \hat{l}_y \sin^2 I \sin 2\Omega\right]  - 3\cos 2I \left\{ \left(2 + 3
                    e^2\right)  \left(\hat{l}_x^2 + \hat{l}_y^2 - 2 \hat{l}_z^2\right)  +\right. \\ \\
                     &+&\left. 5 e^2 \cos 2\omega \left[ \left(\hat{l}_x^2 - \hat{l}_y^2\right)   \cos 2\Omega + 2 \hat{l}_x \hat{l}_y \sin 2\Omega\right] \right\}.
\end{array}\lb{uavera}
\end{equation}
In \rfr{uavera}, $\hat{l}_x,~\hat{l}_y,~\hat{l}_z$ are to be intended as the components of ${\bds{\hat{l}}}$.

As far as the longitude of the ascending node $\Omega$ is concerned, the Lagrange planetary equation for its long-term variation \citep{2003ASSL..293.....B}
\eqi
\ang{\dert\Omega t} = -\rp{1}{\nk a^2\sqrt{1-e^2}\sin I}\derp{\ang{U_\textrm{pert}}}I\lb{Lagnodo},
\eqf
computed by means of \rfrs{us}{uavera},
allows to obtain Equation~(9) of \citet{2012CeMDA.112..117I} for the long-term precession of the node due to a distant, pointlike perturber as
\eqi
\dot\Omega_\textrm{theo} = -\rp{Gm_\textrm{X}}{4a^3_\textrm{X}\nk\sqrt{1-e^2}}\mathcal{N}\left(I,~\Omega,~\omega;~\bds{\hat{l}}\right),\lb{nodocazzo}
\eqf
with
\begin{equation}
\begin{array}{lll}
\mathcal{N} & \doteq & 3  \csc I \left[\hat{l}_z \cos I + \sin I \left( -\hat{l}_y  \cos\Omega +
\hat{l}_x \sin\Omega\right) \right]  \left\{ -2  \hat{l}_z \sin I +\right. \\ \\
 &+&\left.\cos I \left[ -2  +e^2\left( -3  + 5  \cos 2\omega\right)\right]  \left(\hat{l}_y \cos\Omega -
       \hat{l}_x \sin\Omega\right)  + \right. \\ \\
 &+&\left.      e^2 \left[\hat{l}_z \left( -3  + 5 \cos 2\omega\right)  \sin I - 5 \sin 2\omega \left(\hat{l}_x \cos\Omega + \hat{l}_y \sin\Omega\right) \right] \right\}.\lb{Enne}
\end{array}
\end{equation}
The Lagrange planetary equation for the long-term rate of change of the longitude of the pericenter $\varpi\doteq\Omega+\omega$ \citep{2003ASSL..293.....B}
\eqi
\ang{\dert{\varpi}{t}} = -\rp{1}{\nk a^2}\qua{ \ton{\rp{\sqrt{1-e^2}}{e}}\derp{\ang{U_\textrm{pert}}}{e} +\rp{\tan\ton{I/2}}{\sqrt{1-e^2}}\derp{\ang{U_\textrm{pert}}}{I}  }
\eqf
can be used to derive our $\dot\varpi_\textrm{theo}$ by replacing $\ang{U_\textrm{pert}}$ with \rfrs{us}{uavera}. Thus, we finally get Equation~(13) of \citet{2012CeMDA.112..117I} for the long-term apsidal precession induced by a distant pointlike disturbing body, which is
\eqi \dot\varpi_\textrm{theo} =  \frac{Gm_\textrm{X}}{128 r^3_\textrm{X} \nk\sqrt{1-e^2}}\left[\mathcal{G}\left(I,~\Omega,~\omega;~\bds{\hat{l}}\right)+\mathcal{H}\left(I,~\Omega,~\omega;~\bds{\hat{l}}\right)\right],\lb{pericazzo}\eqf
where
\begin{equation}
\begin{array}{lll}
-\rp{\mathcal{G}}{24\left(1-e^2\right)} & \doteq &
8 - 9 \hat{l}_x^2 - 9 \hat{l}_y^2 - 6 \hat{l}_z^2 -
          10 \left(\hat{l}_x^2 + \hat{l}_y^2 - 2 \hat{l}_z^2\right)  \cos 2\omega \sin^2 I +\\ \\
          &+&
          8 \hat{l}_z \cos\Omega \sin I \left[\hat{l}_y \cos I \left( -3  + 5 \cos 2\omega\right)  -
                5 \hat{l}_x \sin 2\omega\right]  +\\ \\
          &+&\cos 2\Omega \left[ -3  \left(\hat{l}_x^2 - l^2_y\right)   \left(5 \cos 2\omega + 2 \sin^2 I\right)  - 40 \hat{l}_x \hat{l}_y \cos I \sin 2\omega\right]  -\\ \\
          &-& 8 \hat{l}_z \sin I \left[\hat{l}_x \cos I \left( -3  + 5 \cos 2\omega\right)  +
                5 \hat{l}_y \sin 2\omega\right]  \sin\Omega +\\ \\
                &+&
          2 \left[ -3  \hat{l}_x \hat{l}_y \left(5 \cos 2\omega +
                      2 \sin^2 I\right)  +
                10 \left(\hat{l}_x^2 - \hat{l}_y^2\right)   \cos I \sin 2\omega\right]  \sin 2\Omega -\\ \\
                &-&
          \cos 2I \left\{3 \left(\hat{l}_x^2 + \hat{l}_y^2 - 2 \hat{l}_z^2\right)  +
                5 \cos 2\omega \left[ \left(\hat{l}_x^2 - \hat{l}_y^2\right)   \cos 2\Omega +
                      2 \hat{l}_x \hat{l}_y \sin 2\Omega\right] \right\}   \lb{Gi},
\end{array}
\end{equation}
and
\begin{equation}
\begin{array}{lll}
\mathcal{H} & \doteq & -96 \left[\hat{l}_z \cos I +
          \sin I \left( -\hat{l}_y  \cos\Omega +
                \hat{l}_x \sin\Omega\right) \right]  \left\{ -2  \hat{l}_z \sin I +\right. \\ \\
          &+&\left.\cos I \left[ -2  +e^2\left(- 3  +
                5  \cos 2\omega\right)\right]  \left(\hat{l}_y \cos\Omega -
                \hat{l}_x \sin\Omega\right)  +\right. \\ \\
          &+&\left. e^2 \left[\hat{l}_z \left( -3  + 5 \cos 2\omega\right)  \sin I - 5 \sin 2\omega \left(\hat{l}_x \cos\Omega +
                      \hat{l}_y \sin\Omega\right) \right] \right\} \tan\left(I/2\right)\lb{Hacca}.
\end{array}
\end{equation}

As far as the validity of the assumptions on which \rfr{us} is based, the standard relation\footnote{In it, $t_\textrm{p}$ denotes the time of the passage at pericenter.} between $t$ and $f$
\eqi \nk\ton{t - t_\textrm{p}}  = 2\arctan\ton{\sqrt{\rp{1-e}{1+e}}\tan\rp{f}{2}   }- \rp{e \sqrt{1 - e^2}\sin f  }{1 + e \cos f},\eqf allows to infer that, for the orbital configuration of Table~\ref{kepelemsX}, the true anomaly $f_\textrm{X}$ changes of about $\Delta f_\textrm{X}\approx 10~\textrm{deg}$ over $\Delta t = 100~\textrm{yr}$, which is just the length of the data records processed to obtain the supplementary orbital precessions of Table~\ref{perihelia}; even smaller angular ranges $\Delta f_\textrm{X}$ for $f_\textrm{X}$ can be obtained for either smaller values of $e_\textrm{X}$ or larger values of $a_\textrm{X}$. Since the lingering uncertainty in $f_\textrm{X}$ is much larger than $\Delta f_\textrm{X}$, according to both the present study (see next Sections) and other ones in the literature (see, e.g., \citep{2016A&A...587L...8F}), we can reasonably conclude that the approximation used to calculate the average orbital precessions of \rfr{nodocazzo} and \rfr{pericazzo} is  justified.
\section{Constraining the location of PX with the perihelion and the node of Saturn}\lb{costra}
\citet{BaBroAJ2016} did  not constrain  the true anomaly $f_\textrm{X}$ of PX;
furthermore, in linking its longitude of the ascending node $\Omega_\textrm{X}$ to the position of the orbital planes of the KBOs used in their analyses, they left room for a certain range of variability in the perturber's node. Later, \citet{BroBaAJ2016} explored in more detail the possible parameter space of PX allowing for a certain variability of its other orbital elements as well. Thus, in view also of the still small number of Trans-Neptunian Objects which \citet{BaBroAJ2016, BroBaAJ2016}  built upon their conclusions, we consider our theoretical expressions of \rfr{nodocazzo} and \rfr{pericazzo}, calculated with the values of some selected orbital elements of PX, as  functions of just $x=f_\textrm{X}$ and possibly other (partially) unconstrained orbital elements as  independent variables $x,~y,~\ldots$, and look for those domains in the $\grf{f_\textrm{X},~y,~\ldots}$ hypervolumes which are allowed by $\Delta\dot\varpi_{\rm obs},~\Delta\dot\Omega_{\rm obs}$ in Table~\ref{perihelia}. To this aim, we parameterize ${\bds{\hat{l}}}$ as
\begin{align}
{\hat{l}}_x \lb{lX} & =\cos\Omega_\textrm{X}\cos u_\textrm{X} -\cos I_\textrm{X}\sin\Omega_\textrm{X} \sin u_\textrm{X}, \\ \nonumber \\
{\hat{l}}_y & =\sin\Omega_\textrm{X}\cos u_\textrm{X} +\cos I_\textrm{X}\cos\Omega_\textrm{X} \sin u_\textrm{X}, \\ \nonumber \\
{\hat{l}}_z \lb{lZ}& =\sin I_\textrm{X}\sin u_\textrm{X},
\end{align}
in which  the heliocentric distance of PX is modelled as
\eqi
r_\textrm{X}=\rp{a_\textrm{X}(1-e_\textrm{X}^2)}{1+e_\textrm{X}\cos f_\textrm{X}}.
\eqf
Since the figures in Table~\ref{perihelia} were obtained by using the International Celestial Reference Frame (ICRF), we  convert Eqs \ref{lX}~to~\ref{lZ}, computed for the ecliptic values of Table~\ref{kepelemsX}, to the corresponding equatorial ones ${\bds{\hat{l}}}^{'}$.
\subsection{Working with the originally proposed configuration of PX}
\begin{figure*}
\centerline{
\vbox{
\begin{tabular}{cc}
\epsfysize= 7.5 cm\epsfbox{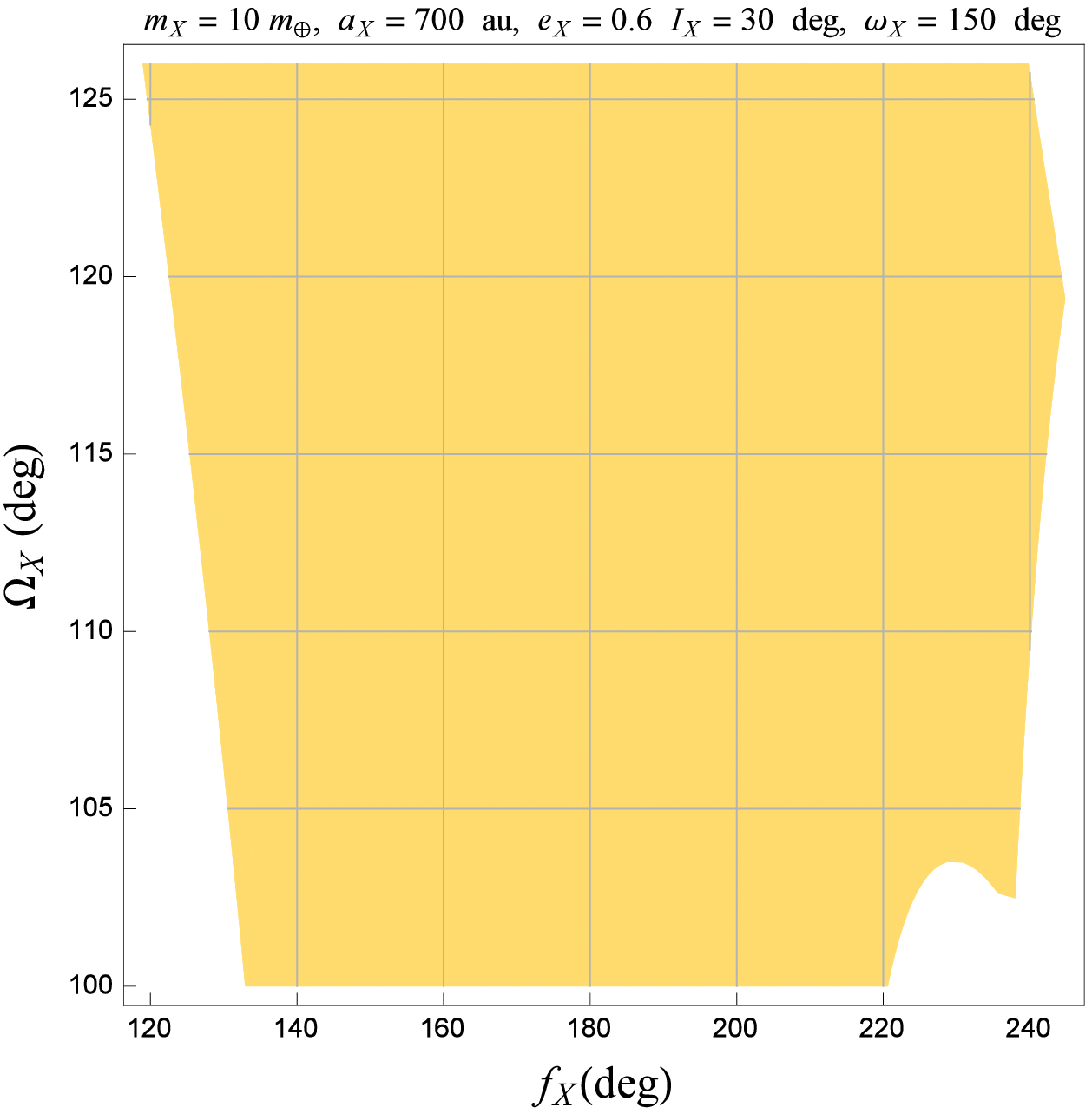} & \epsfysize= 7.5 cm\epsfbox{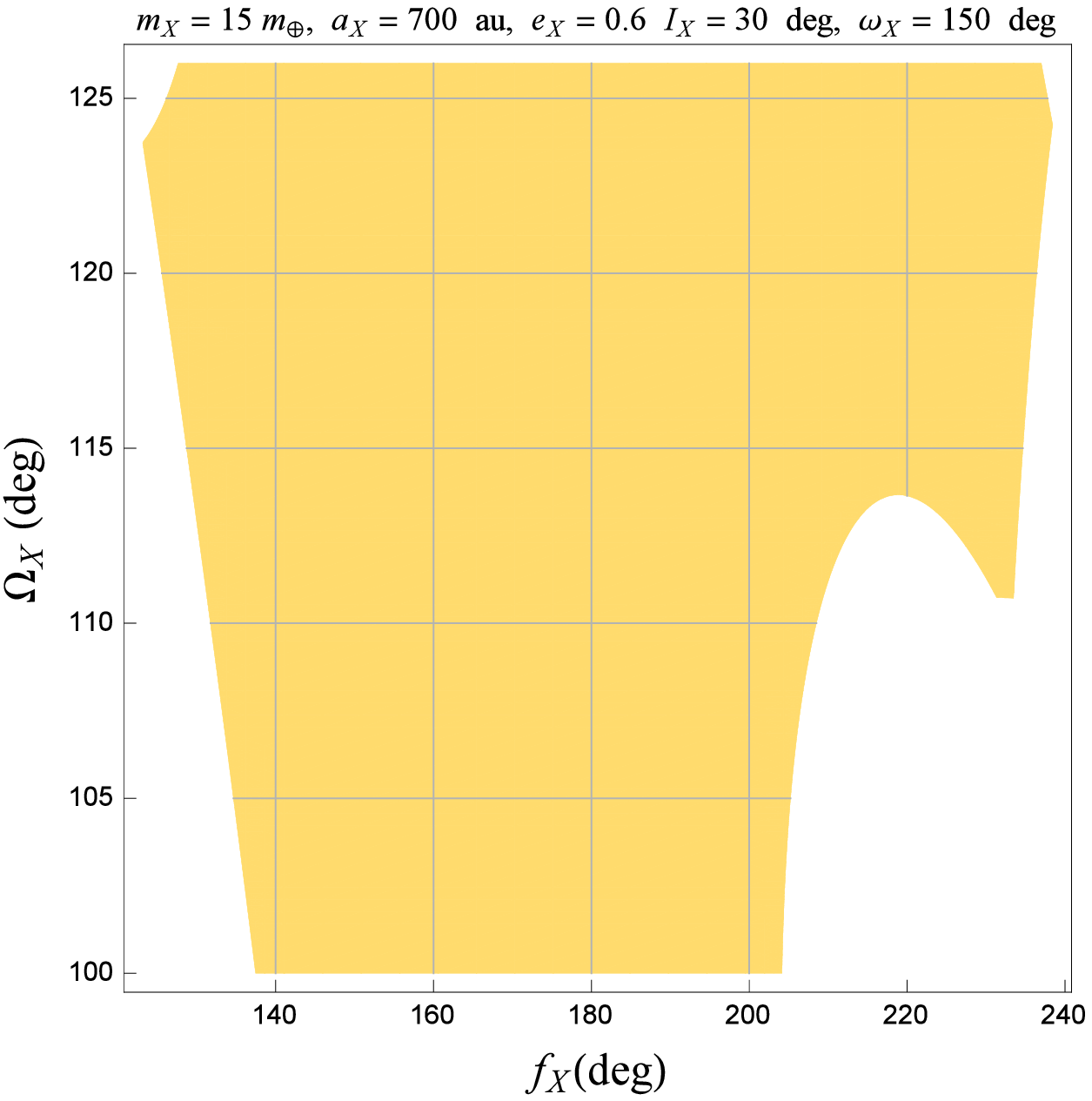} \\
\epsfysize= 7.5 cm\epsfbox{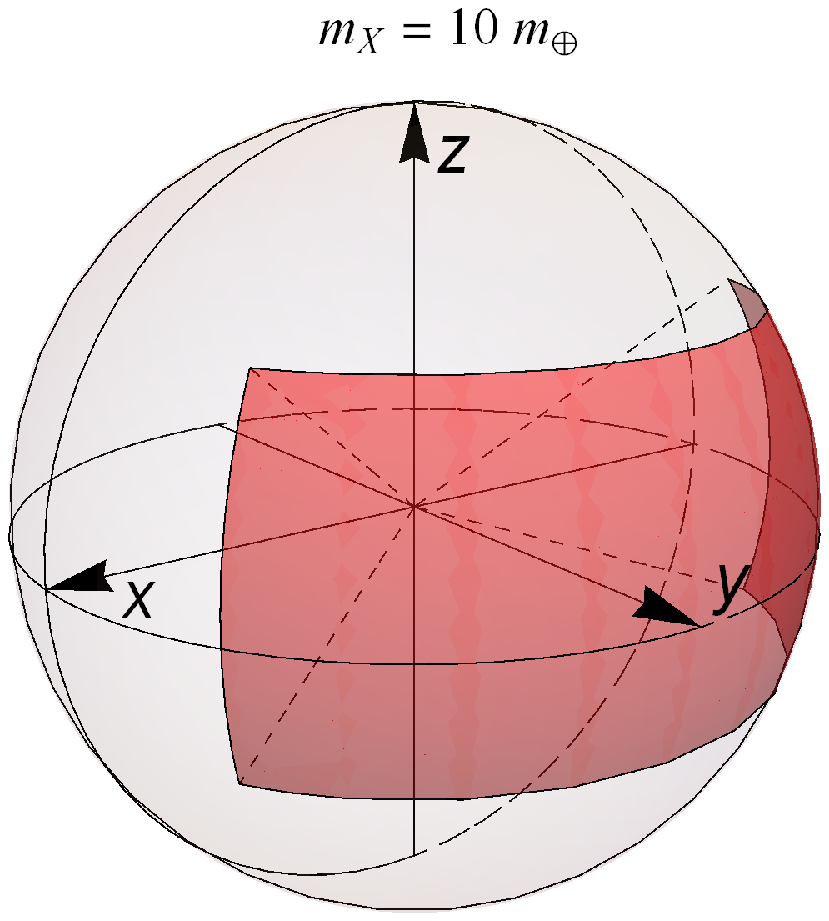} & \epsfysize= 7.5 cm\epsfbox{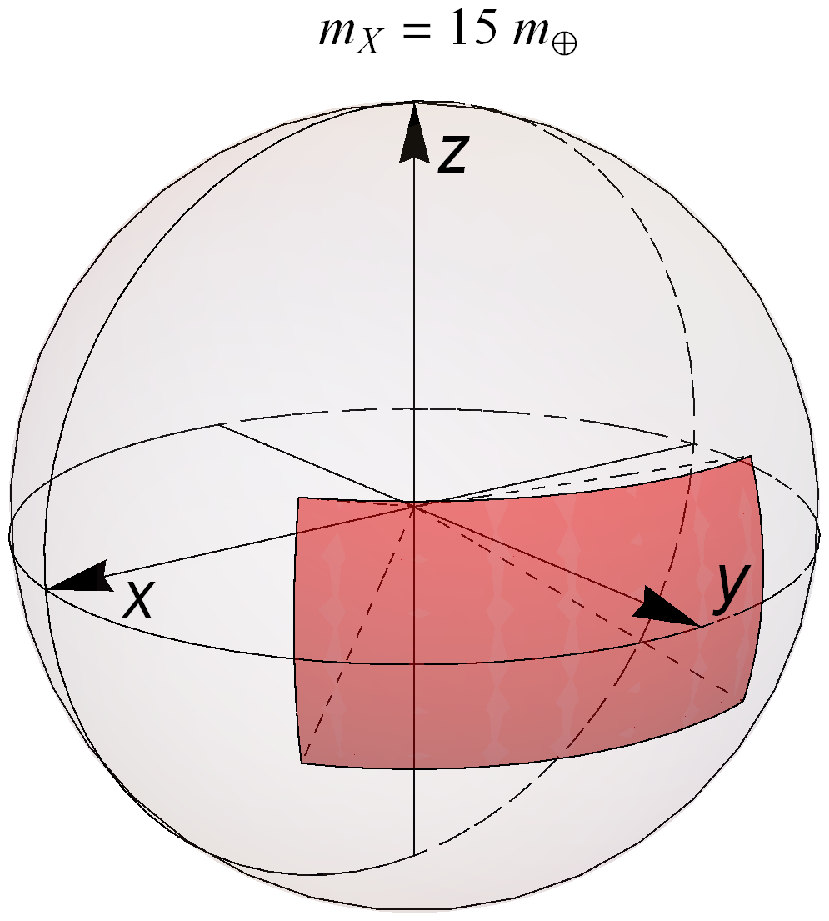} \\
\end{tabular}
}
}
\caption{Upper row: intersections $\mathcal{D}=\mathcal{A}_\varpi\cap \mathcal{A}_\Omega$ of the allowed regions in the $\grf{f_\textrm{X},~\Omega_\textrm{X}}$ plane determined by the  INPOP10a ephemerides \citep{2011CeMDA.111..363F} through a comparison of \rfrs{nodocazzo}{Enne} and \rfrs{pericazzo}{Hacca}, computed with \rfrs{lX}{lZ} and the nominal orbital values in Table~\ref{kepelemsX} (Left: $m_\textrm{X} = 10~m_\oplus$. Right: $m_\textrm{X} = 15~m_\oplus$), with $\Delta\dot\Omega_{\rm obs},\Delta\dot\varpi_{\rm obs}$ quoted in Table~\ref{perihelia}. As per the variability of $\Omega_\textrm{X}$, we restrict ourselves to the interval of values which seems to be favored in \citet{BaBroAJ2016}, i.e. $\Omega_\textrm{X}=113\pm 13~\textrm{deg}$. The uncertainty in $\Omega_\textrm{X}$ does not affect too much the allowed ranges for $f_\textrm{X}$, especially for $m_\textrm{X} = 10~m_\oplus$. In this case, it is basically  $130~\textrm{deg}\lesssim f_\textrm{X}\lesssim 240~\textrm{deg}$. For $m_\textrm{X} = 15~m_\oplus$, the permitted domain of $f_\textrm{X}$ is essentially equal to that of $m_\textrm{X} = 10~m_\oplus$ when $110~\textrm{deg}\lesssim\Omega_\textrm{X}\lesssim 126~\textrm{deg}$. Instead, for $\Omega_\textrm{X} = 100~\textrm{deg}$, it shrinks down to $140~\textrm{deg}\lesssim f_\textrm{X}\lesssim 200~\textrm{deg}$. Lower row: corresponding allowed regions onto the Celestial Sphere; the left panel depicts the area $31.1~\textrm{deg} \lesssim\alpha_\textrm{X}\lesssim 160.2~\textrm{deg},~-18.9~\textrm{deg}\lesssim\delta_\textrm{X}\lesssim 32.6~\textrm{deg}$. The right panel displays the region $40.1~\textrm{deg} \lesssim\alpha_\textrm{X}\lesssim 101.8~\textrm{deg},~-13.9~\textrm{deg} \leq\delta_\textrm{X}\leq 18.4~\textrm{deg}$ corresponding to the case $m_\textrm{X} = 15~m_\oplus,~100~\textrm{deg}\lesssim\Omega_\textrm{X}\lesssim 110~\textrm{deg}$. The $\grf{x,~y}$ plane is the Celestial Equator, while the $z$ axis points towards the Celestial North Pole.}\label{nodo_peri}
\end{figure*}

\lb{BaBro}
Let us describe in detail our procedure in the case of the orbital configuration of Table~\ref{kepelemsX} put forth for the first time in \citet{BaBroAJ2016}; in this case, the unconstrained or partially constrained orbital parameters, assumed as independent variables in our theoretical expressions, are $x=f_\textrm{X},~y=\Omega_\textrm{X}$, with $\Omega_\textrm{X} = 113\pm 13~\textrm{deg}$ \citep{BaBroAJ2016}. As a first step, we generate the surfaces representing the graphs of $\dot\Omega_\textrm{theo}\ton{f_\textrm{X},~\Omega_\textrm{X}},~\dot\varpi_\textrm{theo}\ton{f_\textrm{X},~\Omega_\textrm{X}}$, not shown here, by using \rfr{nodocazzo} and \rfr{pericazzo} calculated with the values of the orbital elements of Saturn and of PX as listed in Table~\ref{kepelemsX}  and for fixed values of the perturber's mass $m_\textrm{X}$,. Then, by posing $\dot\Omega_\textrm{theo}=\Delta\dot\Omega_\textrm{obs},~\dot\varpi_\textrm{theo}=\Delta\dot\varpi_\textrm{obs}$, we section such plots with horizontal planes corresponding to the maximum and the minimum values of  $\Delta\dot\Omega_\textrm{exp},~\Delta\dot\varpi_\textrm{exp}$ in Table~\ref{perihelia}. The resulting contours, collapsed all together in the $\grf{f_\textrm{X},~\Omega_\textrm{X}}$ plane, delimit two allowed regions $\mathcal{A}_\varpi,~\mathcal{A}_\Omega$ in it, not reproduced here separately  for the sake of simplicity. It turns out that each of them allows certain regions in the $\grf{f_\textrm{X},~\Omega_\textrm{X}}$ plane which are, instead, forbidden by the other precession. Finally, we take the intersection $\mathcal{D}=\mathcal{A}_\varpi\cap \mathcal{A}_\Omega$, which is not empty, in order to further restrict the parameter space of PX in the  $\grf{f_\textrm{X},~\Omega_\textrm{X}}$ plane. The results are the shaded regions, representing $\mathcal{D}$ for $m_\textrm{X}=10-15~m_\oplus$,  depicted in the upper row of Figure~\ref{nodo_peri}.
The resulting allowed range for the true anomaly of PX is essentially equal to $130~\textrm{deg}\lesssim f_\textrm{X}\lesssim 240~\textrm{deg}$ for both the values of $m_\textrm{X}$ considered and for almost the entire domain of variation of $\Omega_\textrm{X}$ quoted in Table~\ref{kepelemsX}, apart from the case $m_\textrm{X}=15~m_\oplus,~100~\textrm{deg}\lesssim\Omega_\textrm{X}\lesssim 110~\textrm{deg}$ when it reduces to $140~\textrm{deg}\lesssim f_\textrm{X}\lesssim 200~\textrm{deg}$. For a bird's eye view of such constraints, see Figure~\ref{ellis}. As far as the position in the sky in terms of the right ascension $\alpha_\textrm{X}$ and declination $\delta_\textrm{X}$ is concerned, the bounds of the orbital elements can be translated into corresponding ranges for the Celestial coordinates by means of \rfrs{lX}{lZ} and ${\bds{\hat{l}}}^{'}$. In particular, we consider $\alpha_\textrm{X},~\delta_\textrm{X}$ as functions, in general, of some of the orbital elements $I_\textrm{X},~\Omega_\textrm{X},~\omega_\textrm{X}$ and of $f_\textrm{X}$, assumed to vary within the ranges both provided by the dynamical simulations by \citet{BaBroAJ2016} ($I_\textrm{X},~\Omega_\textrm{X},~\omega_\textrm{X}$) and our method ($f_\textrm{X}$). Then, we numerically extremize $\alpha_\textrm{X},~\delta_\textrm{X}$ and assume such minima and maxima as bounds of the allowed regions in the sky. In this specific case, $I_\textrm{X},~\omega_\textrm{X}$ are kept fixed to their values of Table~\ref{kepelemsX}, while $\Omega_\textrm{X}$ is allowed to vary within $\Omega_\textrm{X}=113\pm 13~\textrm{deg}$. It turns out that the left panel in the lower row of Figure~\ref{nodo_peri} displays the allowed portion of Celestial Sphere delimited by $31.1~\textrm{deg} \lesssim\alpha_\textrm{X}\lesssim 160.2~\textrm{deg},~-18.9~\textrm{deg}\lesssim\delta_\textrm{X}\lesssim 32.6~\textrm{deg}$ corresponding to the left panel in the upper row of Figure~\ref{nodo_peri}; the right panel in the lower row of Figure~\ref{nodo_peri}, displaying the region $40.1~\textrm{deg} \lesssim\alpha_\textrm{X}\lesssim 101.8~\textrm{deg},~-13.9~\textrm{deg} \lesssim\delta_\textrm{X}\lesssim 18.4~\textrm{deg}$, corresponds to the right panel in the upper row of Figure~\ref{nodo_peri} for $100~\textrm{deg}\lesssim\Omega_\textrm{X}\lesssim 110~\textrm{deg}$.
\subsection{Relaxing the orbital configuration of PX}
In this Section, we will look at the more recent findings by \citet{BroBaAJ2016} who explored the parameter space of PX in more depth by allowing for ranges of possible values of its orbital and physical parameters (see Tables~\ref{largo}~to~\ref{uffa}). As stressed by \citet{BroBaAJ2016}  themselves, their suggested constraints on the orbital configuration of PX, based on numerical simulations of the evolutionary patterns of certain populations of trans-Neptunian objects under the putative action of the hypothesized distant planet, should be necessarily considered just as preliminary because of the need to cover a substantial part of the possible phase space, the exploratory nature of the simulations themselves, sparsely populated,  and the small number of objects actually observed so far.

As described in Section~\ref{BaBro}, also in this Section we will compare our theoretical expressions of \rfrs{nodocazzo}{Enne} and \rfrs{pericazzo}{Hacca} for the Kronian orbital precessions  with the observational intervals of admissible values quoted in Table~\ref{perihelia}; then, we will take the intersection of the resulting allowed regions of some selected portions of the parameter space of PX.
\subsubsection{Working with partially fixed orientation in space}
Let us begin with a fixed spatial orientation of the orbit of PX  following the scenario characterized by the relation
\eqi 200 + 30\ton{\rp{m_\textrm{X}}{m_{\oplus}}}\leq \rp{a_\textrm{X}}{\textrm{au}}\leq 600 + 20\ton{\rp{m_\textrm{X}}{m_{\oplus}}},~e_\textrm{X}=0.75 - \grf{\rp{\qua{250 + 20\ton{\rp{m_\textrm{X}}{m_{\oplus}}}}~\textrm{au}}{a_\textrm{X}}}^8,\lb{condicio}\eqf
presented in Section 4 of \citet{BroBaAJ2016}.

\begin{figure*}
\centerline{
\vbox{
\begin{tabular}{cc}
\epsfysize= 7.5 cm\epsfbox{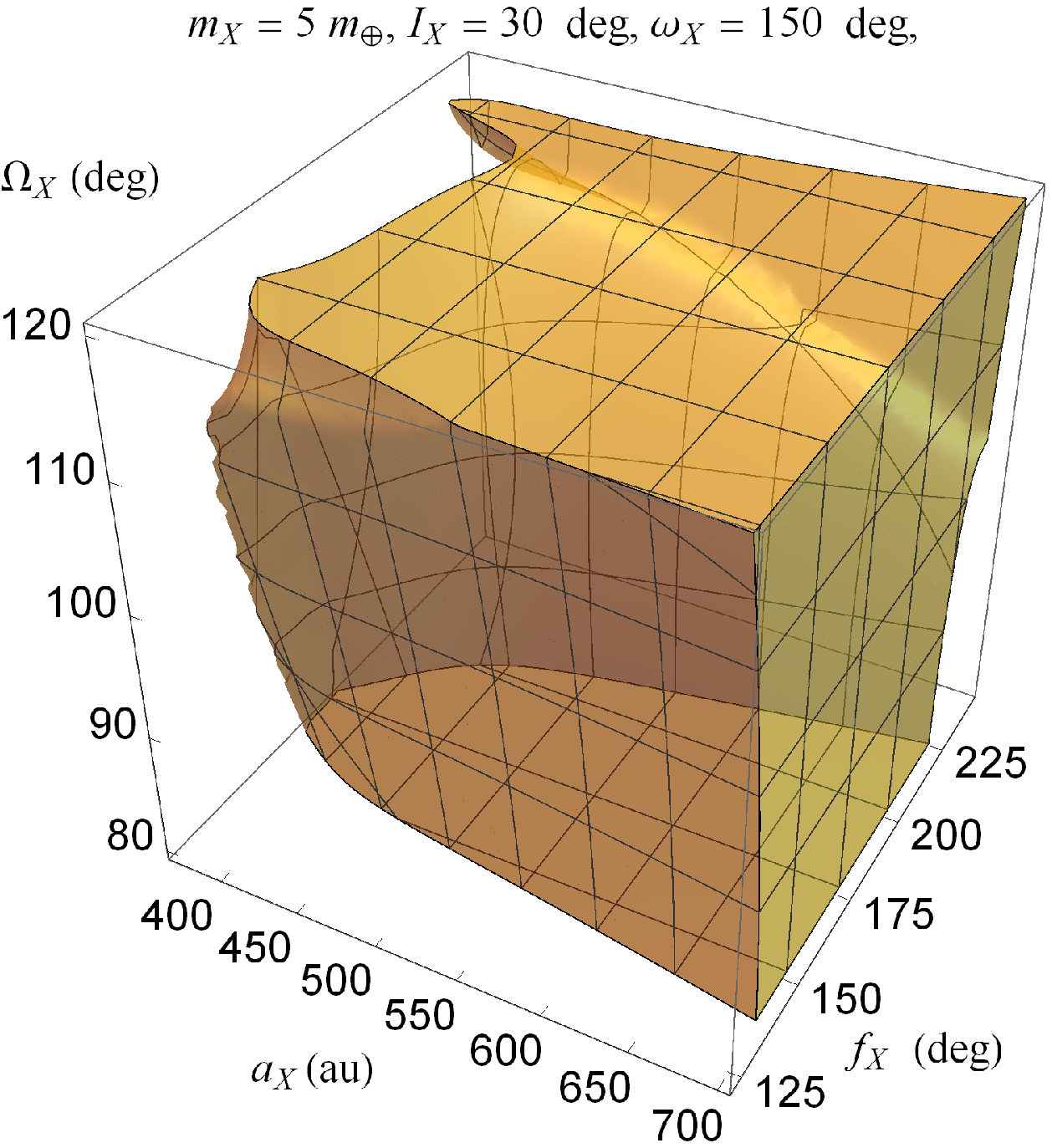} & \epsfysize= 7.5 cm\epsfbox{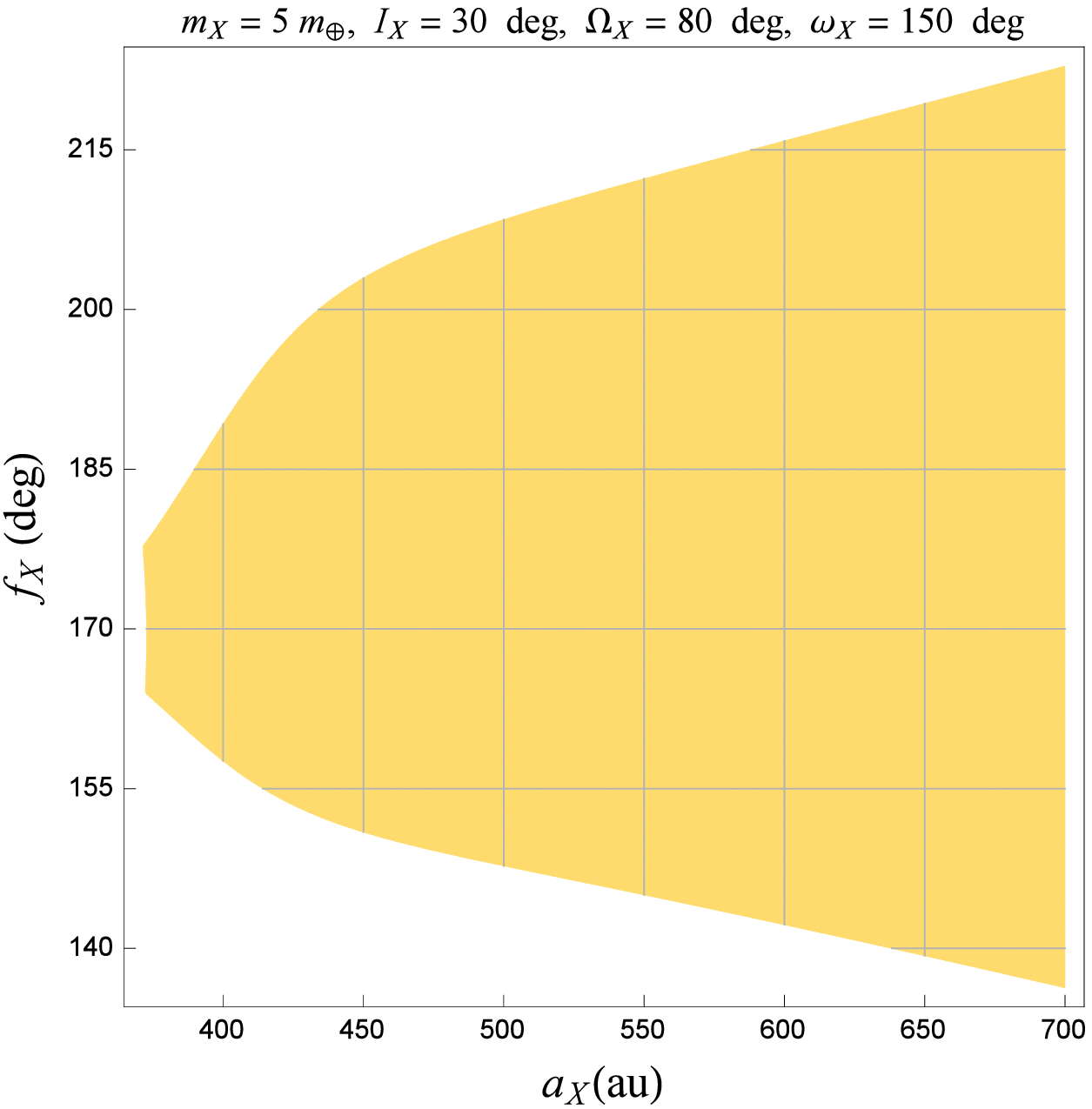} \\
\epsfysize= 7.5 cm\epsfbox{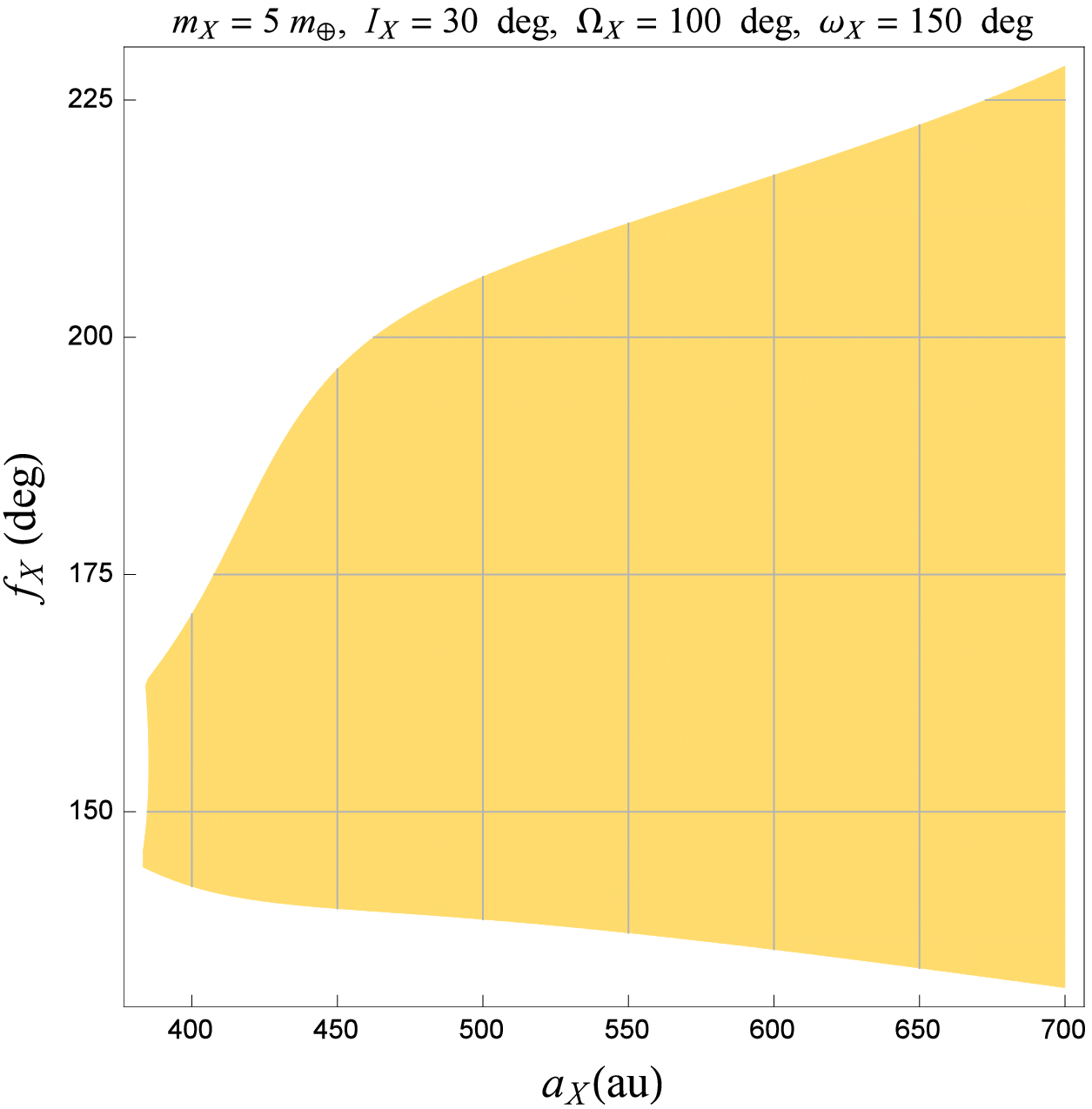} & \epsfysize= 7.5 cm\epsfbox{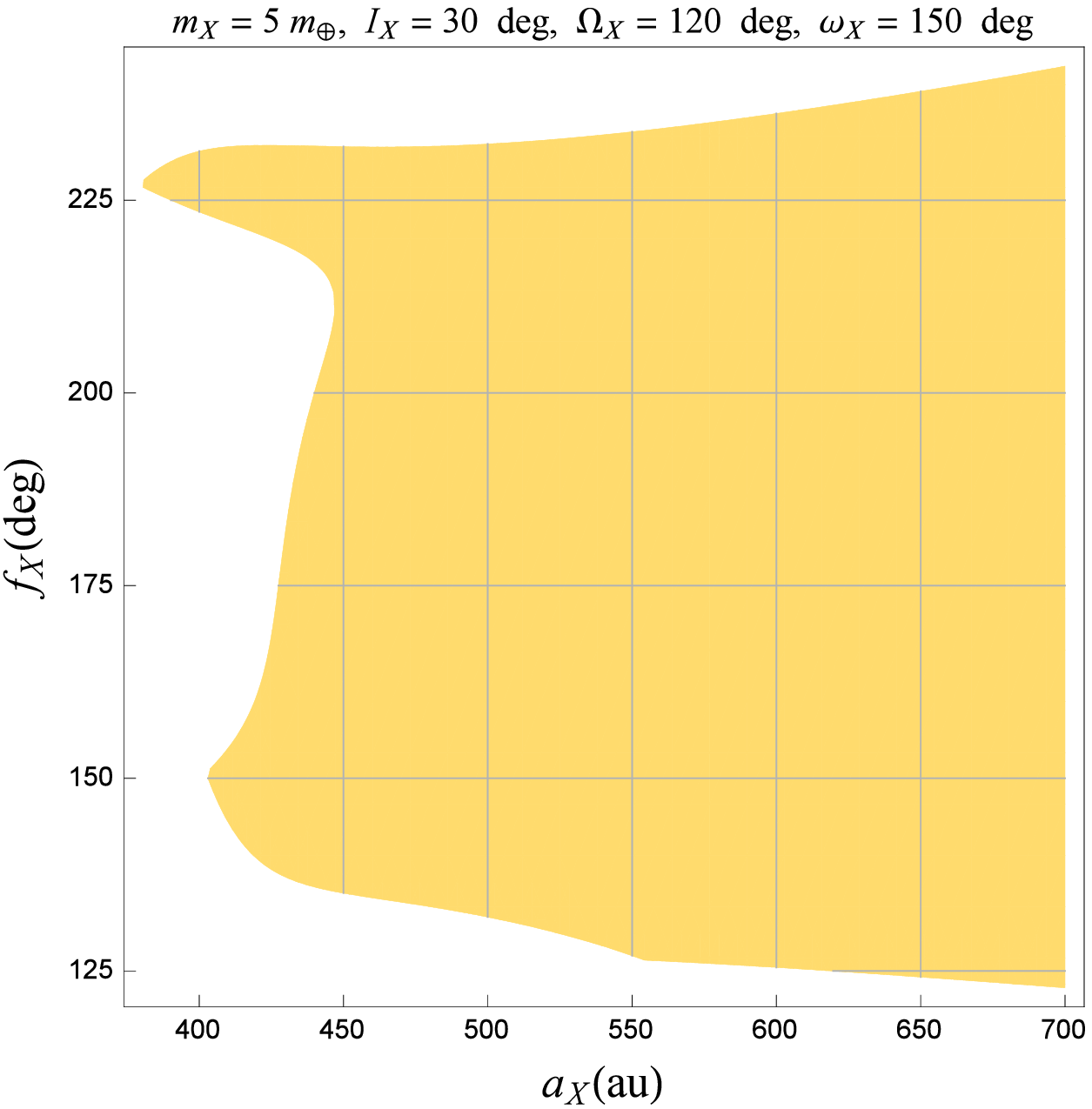} \\
\end{tabular}
}
}
\caption{Allowed regions in the parameter space of PX as determined by the INPOP10a ephemerides \citep{2011CeMDA.111..363F} through a comparison of \rfrs{nodocazzo}{Enne} and \rfrs{pericazzo}{Hacca}, computed with \rfrs{lX}{lZ} and the physical and orbital parameters of PX  for the $m_\textrm{X}=5~m_\oplus,~350~\textrm{au}\leq a_\textrm{X}\leq 700~\textrm{au},~e_\textrm{X}=0.75 - \ton{350~\textrm{au}~a^{-1}_\textrm{X}}^8$ scenario, which is just a particular case  of \rfr{condicio} and Table~\ref{largo}. While the inclination and the argument of perihelion are kept fixed to $I_\textrm{X}=30~\textrm{deg},~\omega_\textrm{X}=150~\textrm{deg}$ \citep{BaBroAJ2016,BroBaAJ2016}, the longitude of ascending node is allowed to vary within   $80~\textrm{deg}\leq \Omega_\textrm{X}\leq 120~\textrm{deg}$  \citep{BroBaAJ2016, 2016AJ....152..126B}. In the left upper corner, the allowed region in the  $\grf{a_\textrm{X},~f_\textrm{X},~\Omega_\textrm{X}}$ volume  inferred from the simultaneous overlapping of the allowed regions by both the node and the perihelion of Saturn is depicted. The other pictures represent selected sections of it in the  $\grf{a_\textrm{X},~f_\textrm{X}}$ plane obtained for fixed values of $\Omega_\textrm{X}$. The widest allowed range of allowed values for the true anomaly of PX is $125~\textrm{deg}\lesssim f_\textrm{X} \lesssim 230~\textrm{deg}$; it is almost independent of $\Omega_\textrm{X}$ and occurs for $a_\textrm{X} = 700~\textrm{au}$. A stronger dependence on $\Omega_\textrm{X}$ can be observed for lower values of $a_\textrm{X}$.
}\label{cubo0}
\end{figure*}

The  $350~\textrm{au}\leq a_\textrm{X}\leq 700~\textrm{au},~e_\textrm{X}=0.75 - \ton{350~\textrm{au}~a^{-1}_\textrm{X}}^8$ scenario, which is just a particular case of \rfr{condicio} for $m_\textrm{X} = 5~m_\oplus$, is illustrated in Figure~\ref{cubo0}.
The maximum allowed range for the true anomaly of PX, i.e. $125~\textrm{deg}\lesssim f_\textrm{X} \lesssim 230~\textrm{deg}$, occurs for the largest value of its semimajor axis considered, i.e. $a_\textrm{X} = 700~\textrm{au}$, independently of $\Omega_\textrm{X}$. It gets much narrower for $a_\textrm{X} = 350-400~\textrm{au}$ reducing to about $160~\textrm{deg}\lesssim f_\textrm{X} \lesssim 175-180~\textrm{deg}~(\Omega_\textrm{X} = 80~\textrm{deg}),~145~\textrm{deg}\lesssim f_\textrm{X} \lesssim 165-170~\textrm{deg}~(\Omega_\textrm{X} = 100~\textrm{deg}),~ f_\textrm{X} \approx 225~\textrm{deg}~(\Omega_\textrm{X} = 120~\textrm{deg})$; a more marked sensitivity to $\Omega_\textrm{X}$ can be noted.

\begin{figure*}
\centerline{
\vbox{
\begin{tabular}{cc}
\epsfysize= 7.5 cm\epsfbox{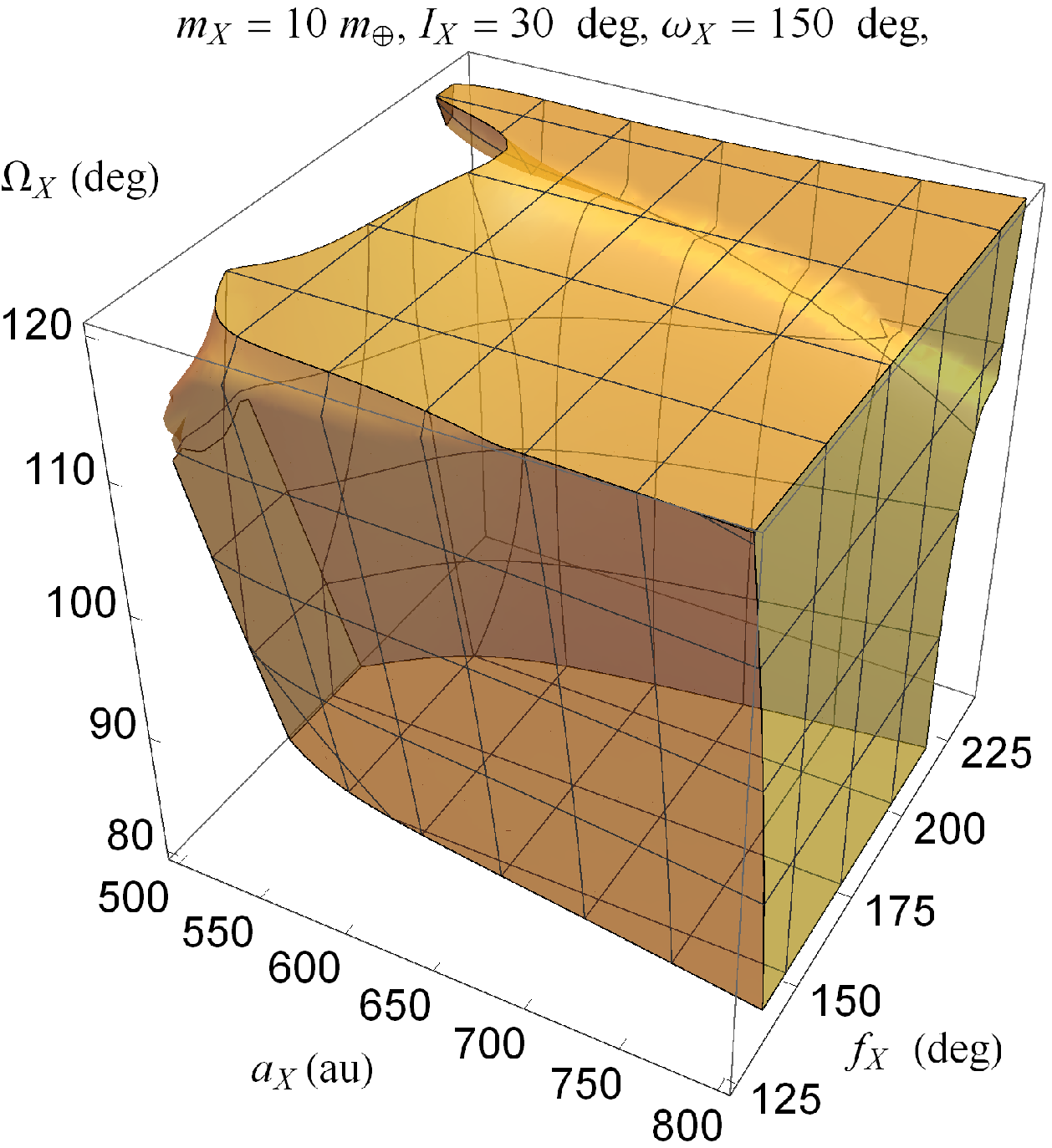} & \epsfysize= 7.5 cm\epsfbox{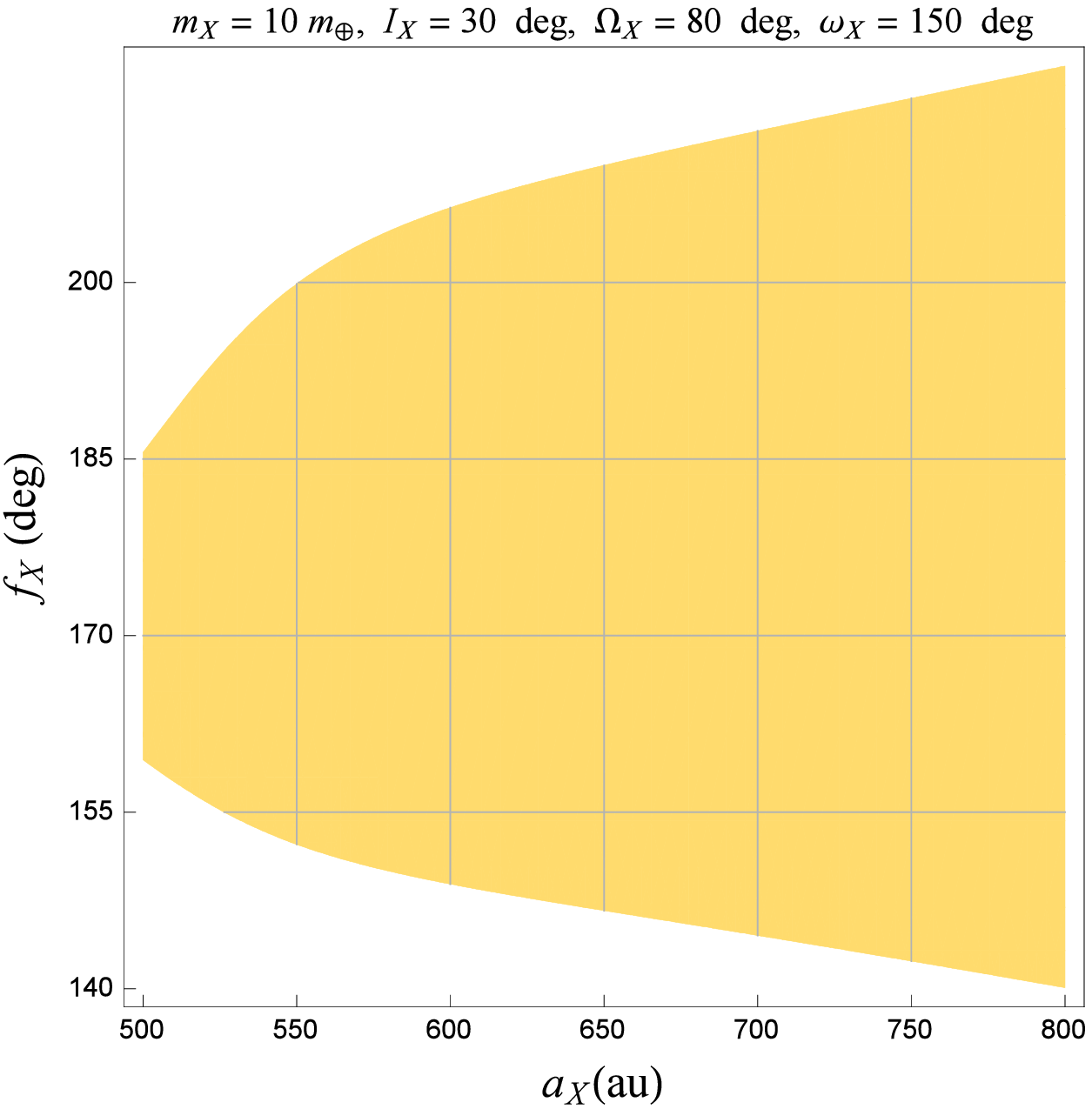} \\
\epsfysize= 7.5 cm\epsfbox{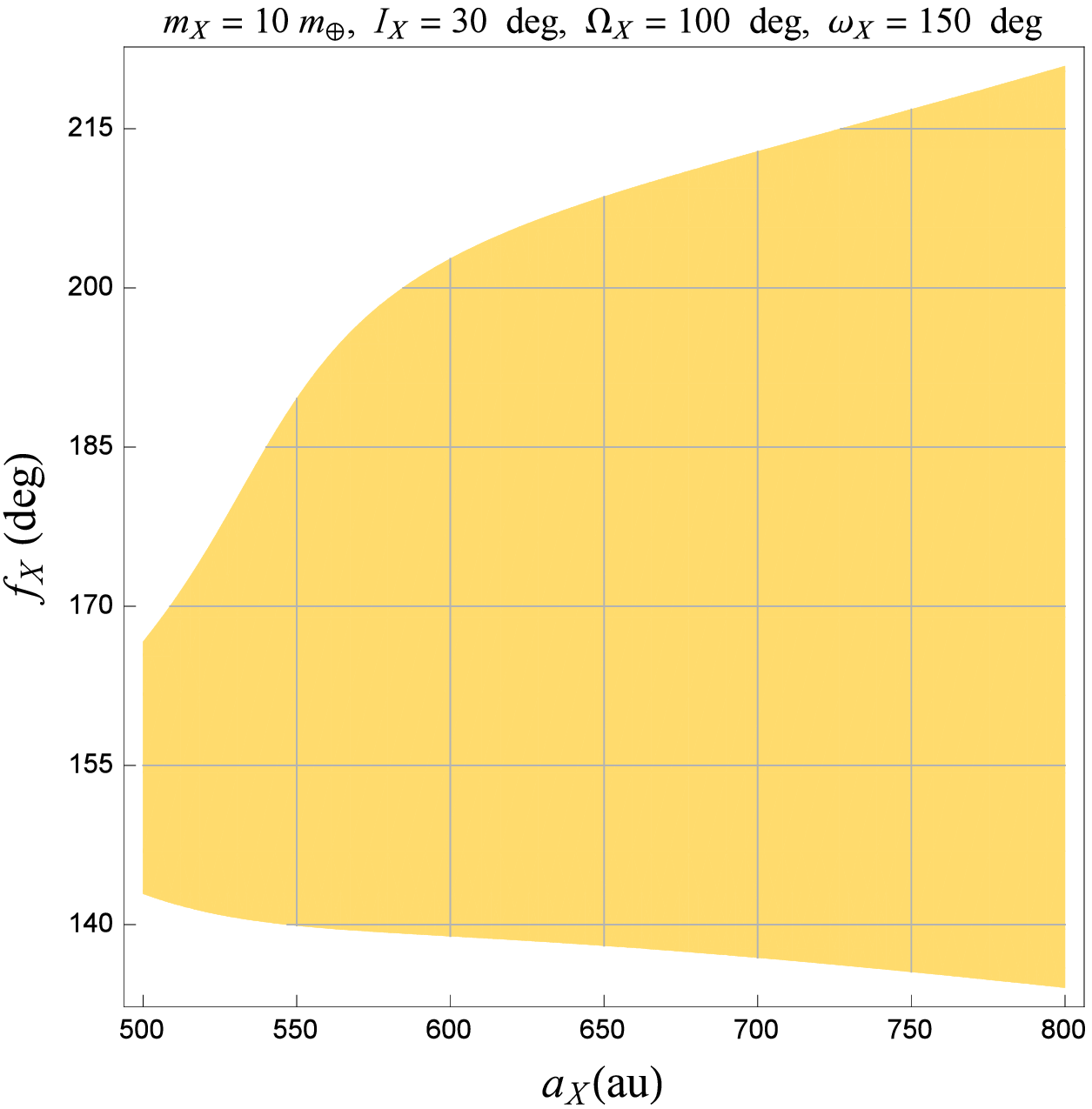} & \epsfysize= 7.5 cm\epsfbox{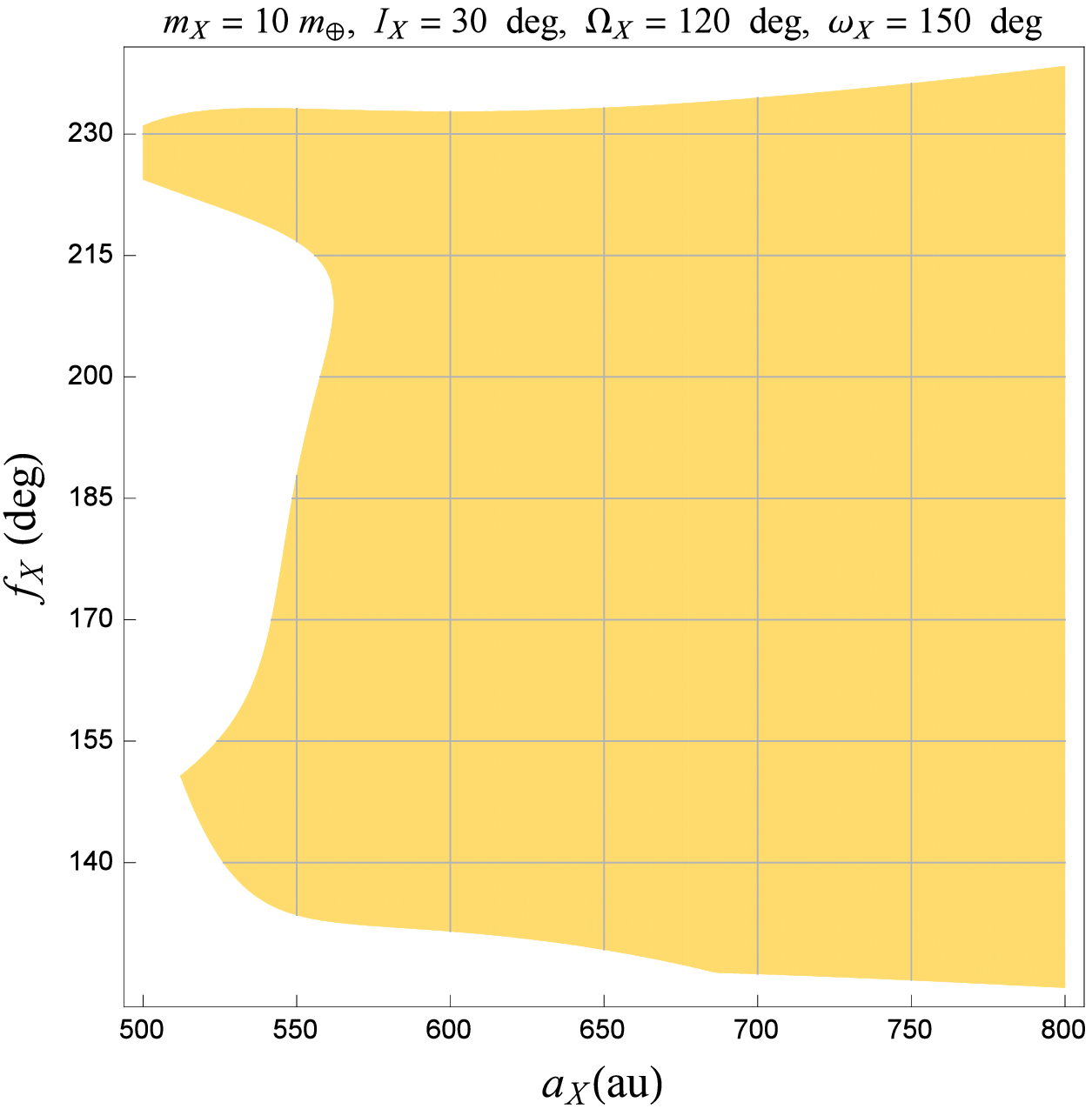} \\
\end{tabular}
}
}
\caption{Allowed regions in the parameter space of PX as determined by the INPOP10a ephemerides \citep{2011CeMDA.111..363F} through a comparison of \rfrs{nodocazzo}{Enne} and \rfrs{pericazzo}{Hacca}, computed with \rfrs{lX}{lZ} and the physical and orbital parameters of PX as  for the $m_\textrm{X}=10~m_\oplus,~500~\textrm{au}\leq a_\textrm{X}\leq 800~\textrm{au},~e_\textrm{X}=0.75 - \ton{450~\textrm{au}~a^{-1}_\textrm{X}}^8$ scenario (it is a particular case for $m_\textrm{X} = 10~m_{\oplus}$ of \rfr{condicio} in Section 2 of \citet{BroBaAJ2016}. While the inclination and the argument of perihelion are kept fixed to $I_\textrm{X}=30~\textrm{deg},~\omega_\textrm{X}=150~\textrm{deg}$ \citep{BaBroAJ2016,BroBaAJ2016}, the longitude of ascending node is allowed to vary within   $80~\textrm{deg}\leq \Omega_\textrm{X}\leq 120~\textrm{deg}$  \citep{BroBaAJ2016, 2016AJ....152..126B}. In the left upper corner, the allowed region in the  $\grf{a_\textrm{X},~f_\textrm{X},~\Omega_\textrm{X}}$ volume  inferred from the simultaneous overlapping of the allowed regions by both the node and the perihelion of Saturn is depicted. The other pictures represent selected sections of it in the  $\grf{a_\textrm{X},~f_\textrm{X}}$ plane obtained for given values of $\Omega_\textrm{X}$. The uncertainty in $\Omega_\textrm{X}$ does not substantially affect the allowed range of values of $f_\textrm{X}$, which is roughly confined in the range $130~\textrm{deg}\lesssim f_\textrm{X}\lesssim 230~\textrm{deg}$. It reduces to roughly $160~\textrm{deg}\lesssim f_\textrm{X}\lesssim 190~\textrm{deg}$ for $\Omega_\textrm{X}\approx 80~\textrm{deg}, a_\textrm{X}\approx 500~\textrm{au}$.}\label{cubo1}
\end{figure*}

Figure~\ref{cubo1} deals with the case  $m_\textrm{X} = 10~m_\oplus,~500~\textrm{au}\leq a_\textrm{X}\leq 800~\textrm{au},~e_\textrm{X}=0.75 - \ton{450~\textrm{au}~a^{-1}_\textrm{X}}^8$, another particular case of \rfr{condicio} which is explicitly considered in Section 2 of \citet{BroBaAJ2016}.
It can be noticed that the uncertainty in $\Omega_\textrm{X}$ does not affect too much the allowed range of values of $f_\textrm{X}$, which is substantially limited to roughly $130~\textrm{deg}\lesssim f_\textrm{X}\lesssim 230~\textrm{deg}$ for most of the largest values of either $\Omega_\textrm{X}$ and $a_\textrm{X}$; for $\Omega_\textrm{X}\approx 80~\textrm{deg}$ and $a_\textrm{X}\approx 500~\textrm{au}$, the range for the true anomaly of PX gets narrower amounting to $160~\textrm{deg}\lesssim f_\textrm{X}\lesssim 190~\textrm{deg}$.

\begin{figure*}
\centerline{
\vbox{
\begin{tabular}{cc}
\epsfysize= 7.5 cm\epsfbox{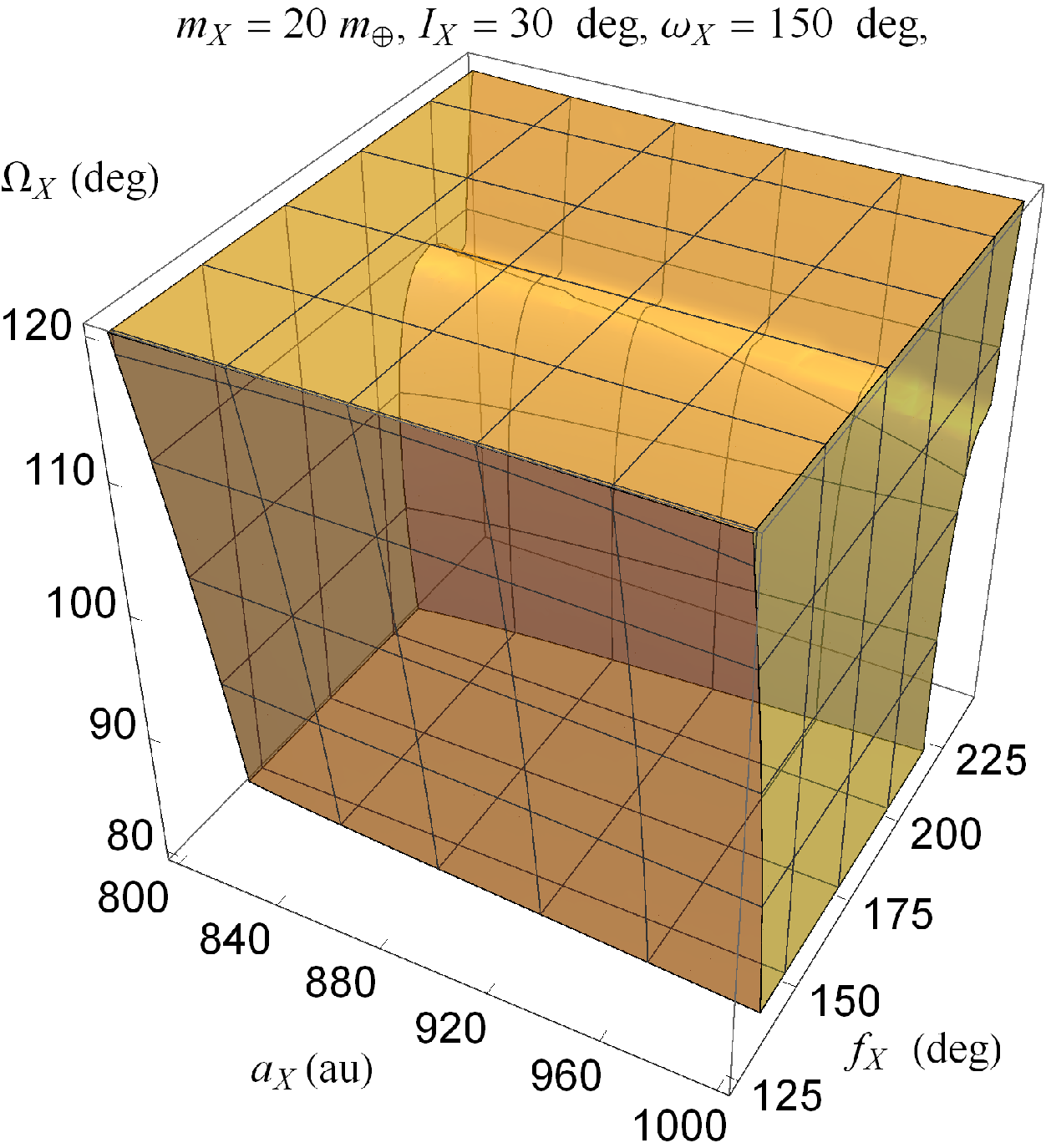} & \epsfysize= 7.5 cm\epsfbox{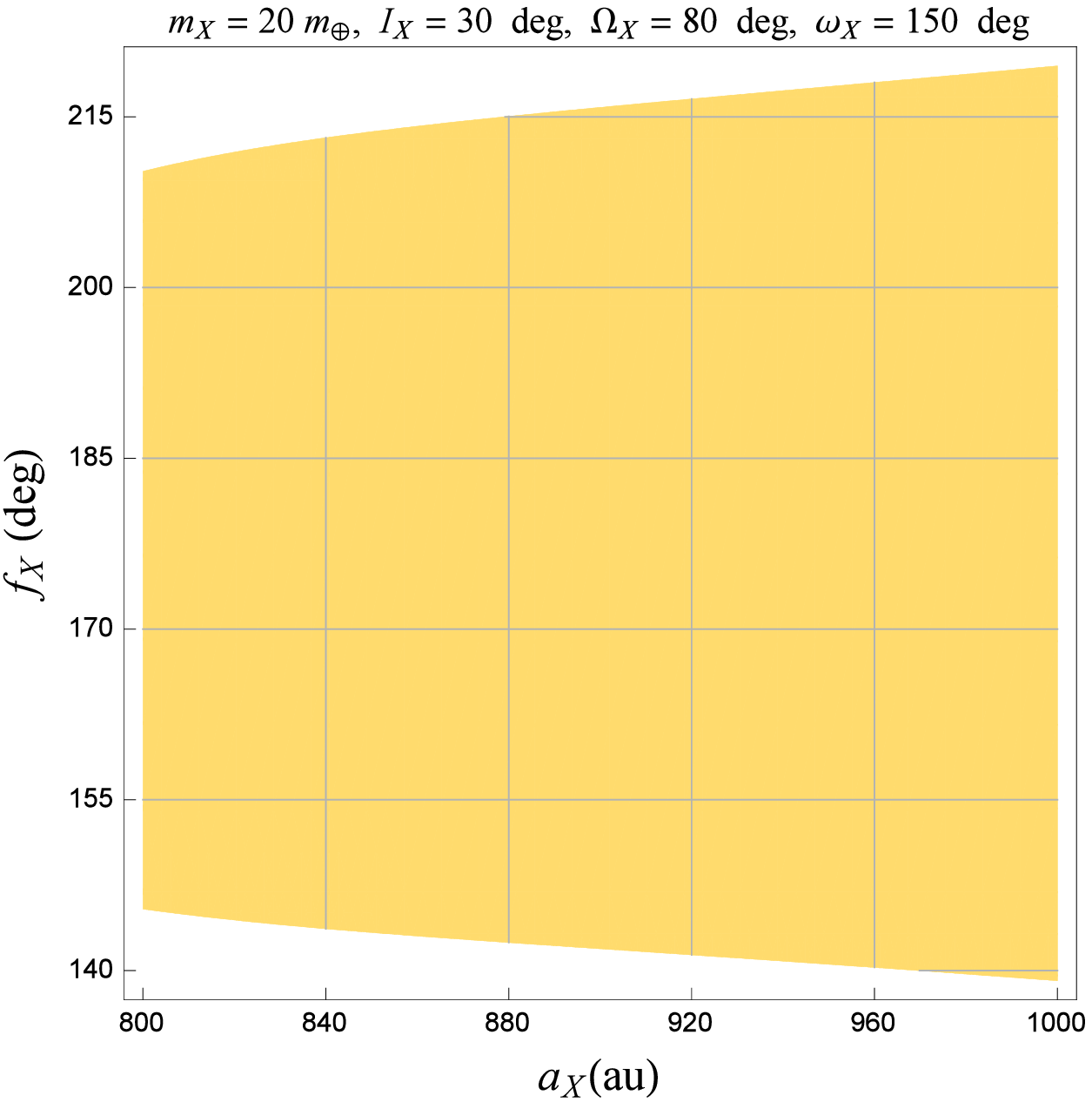} \\
\epsfysize= 7.5 cm\epsfbox{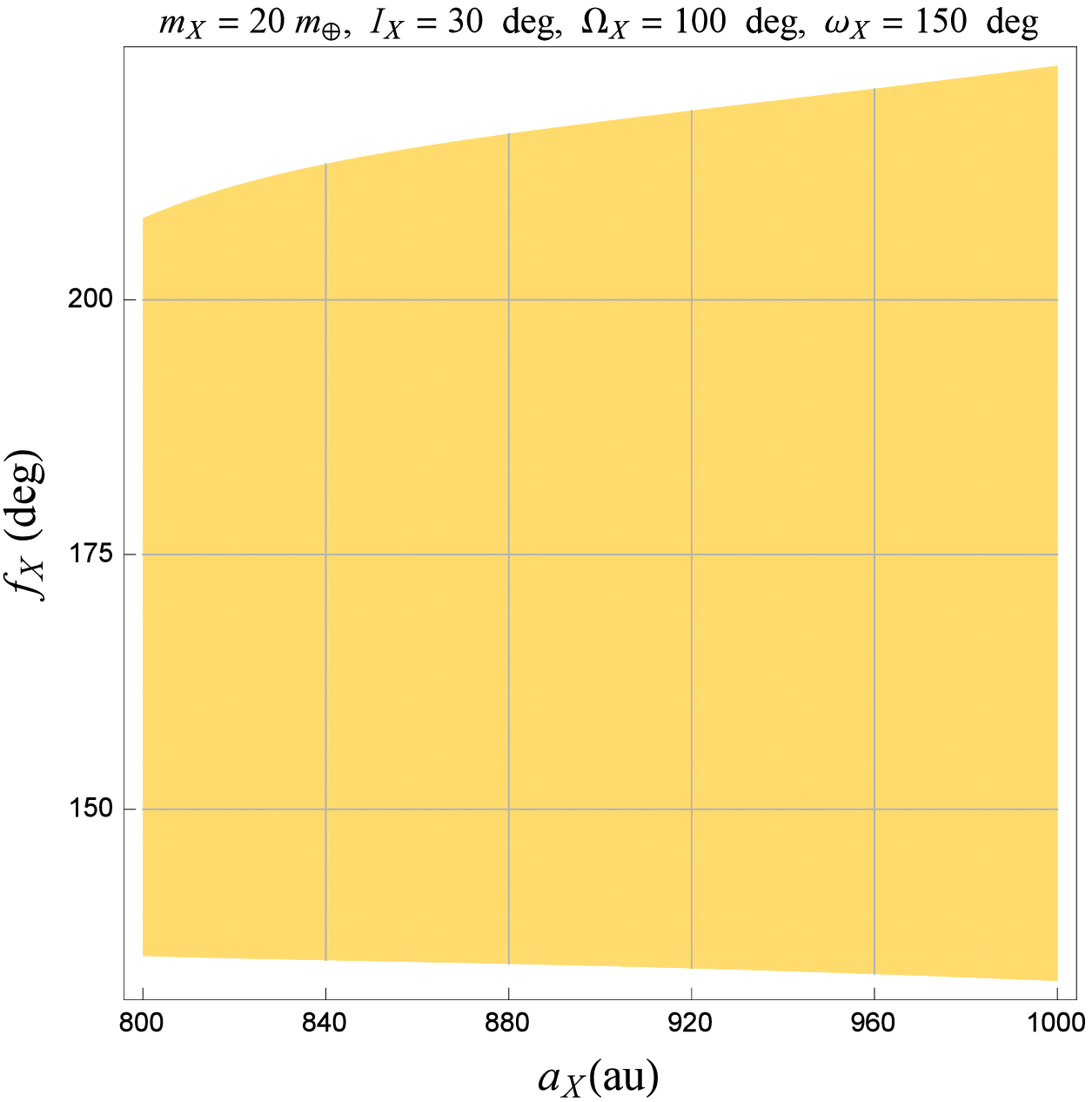} & \epsfysize= 7.5 cm\epsfbox{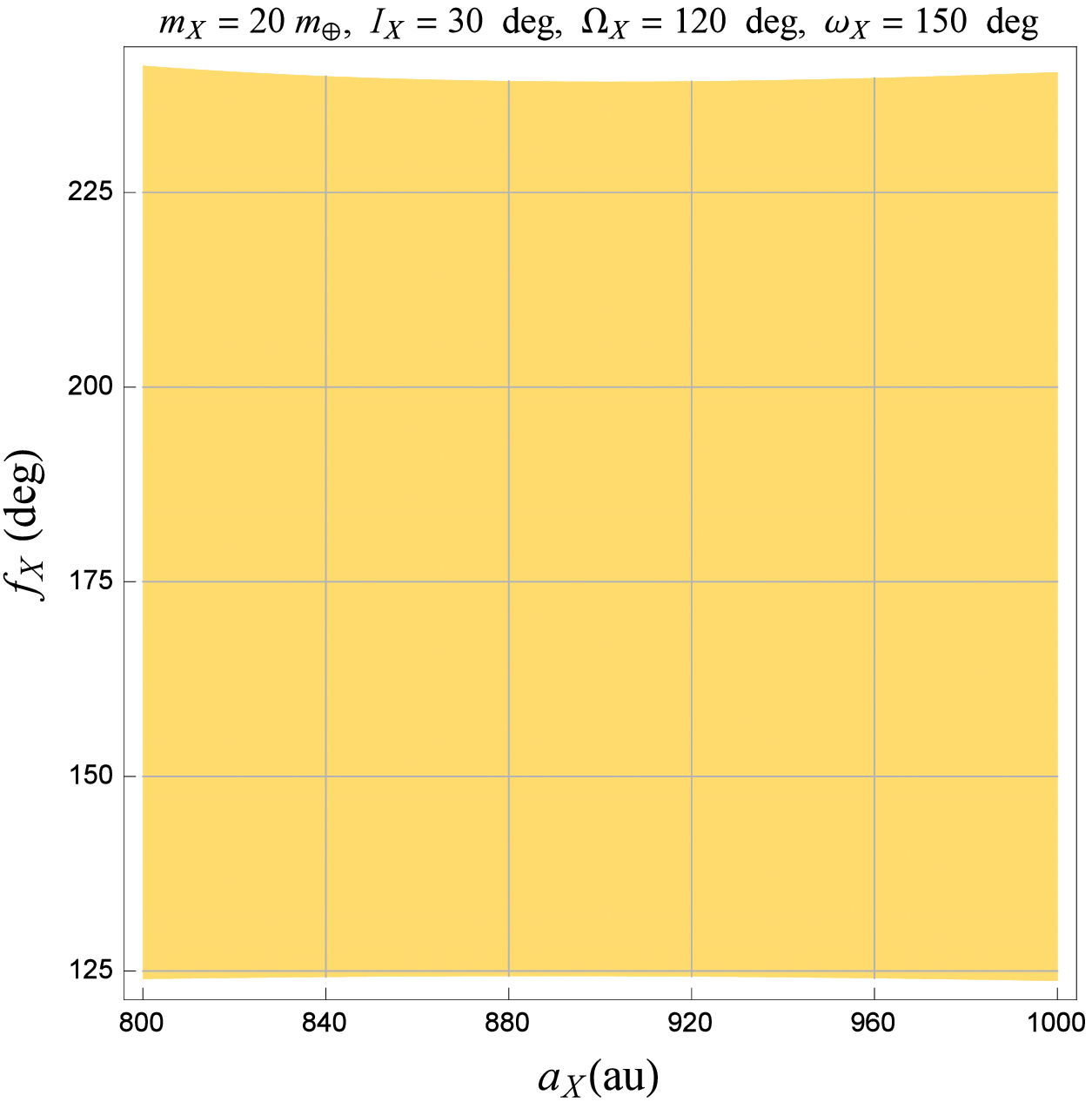} \\
\end{tabular}
}
}
\caption{Allowed regions in the parameter space of PX as determined by the INPOP10a ephemerides \citep{2011CeMDA.111..363F} through a comparison of \rfrs{nodocazzo}{Enne} and \rfrs{pericazzo}{Hacca}, computed with \rfrs{lX}{lZ} and the physical and orbital parameters of PX as  for the $m_\textrm{X}=20~m_\oplus,~800~\textrm{au}\leq a_\textrm{X}\leq 1000~\textrm{au},~e_\textrm{X}=0.75 - \ton{650~\textrm{au}~a^{-1}_\textrm{X}}^8$ (it is a particular case for $m_\textrm{X} = 20~m_{\oplus}$ of \rfr{condicio} in Section 2 of \citet{BroBaAJ2016}. While the inclination and the argument of perihelion are kept fixed to $I_\textrm{X}=30~\textrm{deg},~\omega_\textrm{X}=150~\textrm{deg}$ \citep{BaBroAJ2016,BroBaAJ2016}, the longitude of ascending node is allowed to vary within   $80~\textrm{deg}\leq \Omega_\textrm{X}\leq 120~\textrm{deg}$  \citep{BroBaAJ2016, 2016AJ....152..126B}. In the left upper corner, the allowed region in the  $\grf{a_\textrm{X},~f_\textrm{X},~\Omega_\textrm{X}}$ volume  inferred from the simultaneous overlapping of the allowed regions by both the node and the perihelion of Saturn is depicted. The other pictures represent selected sections of it in the  $\grf{a_\textrm{X},~f_\textrm{X}}$ plane obtained for given values of $\Omega_\textrm{X}$. The uncertainty in $\Omega_\textrm{X}$ does not substantially affect the allowed range of values of $f_\textrm{X}$, which is roughly confined in the range $125~\textrm{deg}\lesssim f_\textrm{X}\lesssim 250~\textrm{deg}$. For $\Omega_\textrm{X}\approx 80~\textrm{deg}, a_\textrm{X}\approx 800~\textrm{au}$, it slightly reduces down to about $150~\textrm{deg}\lesssim f_\textrm{X}\lesssim 210~\textrm{deg}$.}
\label{cubo2}
\end{figure*}

Figure~\ref{cubo2} illustrates the other scenario explicitly proposed by \citet{BroBaAJ2016} in their Section 2 characterized by  $m_\textrm{X}=20~m_\oplus,~800~\textrm{au}\leq a_\textrm{X}\leq 1000~\textrm{au},~e_\textrm{X}=0.75 - \ton{650~\textrm{au}~a^{-1}_\textrm{X}}^8$; it is a further particular case of \rfr{condicio}.
Now, the dependence from $\Omega_\textrm{X}$ and $a_\textrm{X}$ seems somewhat less pronounced than in Figure~\ref{cubo1}; the range of allowed values for the PX's position along its orbits is naively $145~\textrm{deg}\lesssim f_\textrm{X}\lesssim 210~\textrm{deg}$, reaching $125~\textrm{deg}\lesssim f_\textrm{X}\lesssim 225~\textrm{deg}$ for $\Omega_\textrm{X}=120~\textrm{deg}$.

For the sake of definiteness, in all the three scenarios previously considered we adopted $I_\textrm{X}=30~\textrm{deg},~\omega_\textrm{X}=150~\textrm{deg}$, which are the preferred values in \citet{BaBroAJ2016} and are compatible with the ranges explored in \citet{BroBaAJ2016}, while we allowed $\Omega_\textrm{X}$ to vary within the range $80~\textrm{deg}\leq\Omega_\textrm{X}\leq 120~\textrm{deg}$ preferred by \citet{BroBaAJ2016,2016AJ....152..126B}.

\begin{figure*}
\centerline{
\vbox{
\begin{tabular}{cc}
 \epsfysize= 7.5 cm\epsfbox{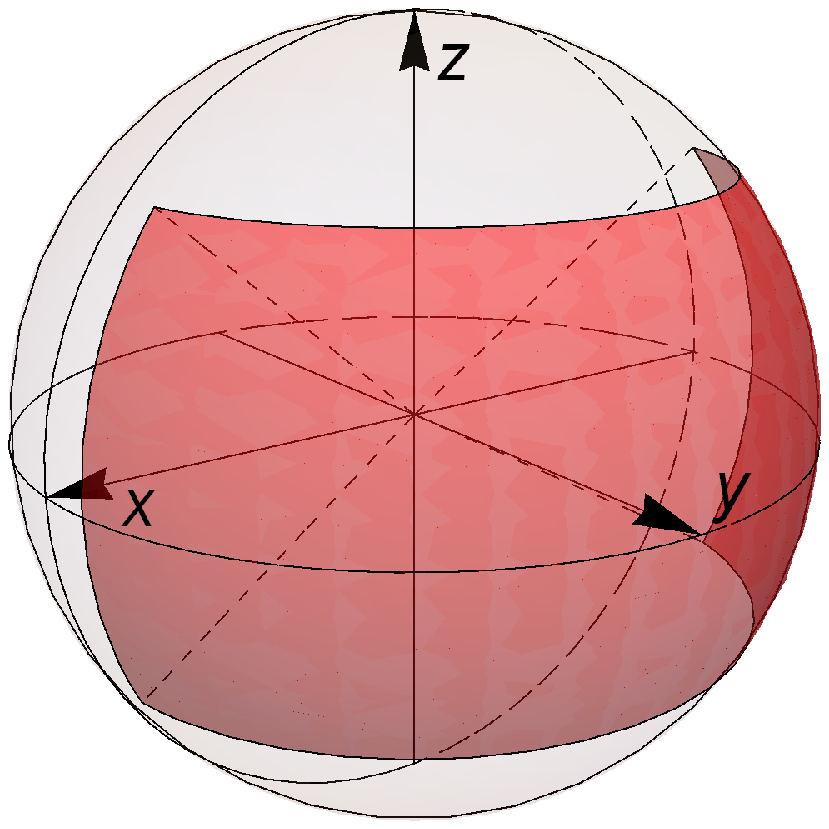} &  \epsfysize= 7.5 cm\epsfbox{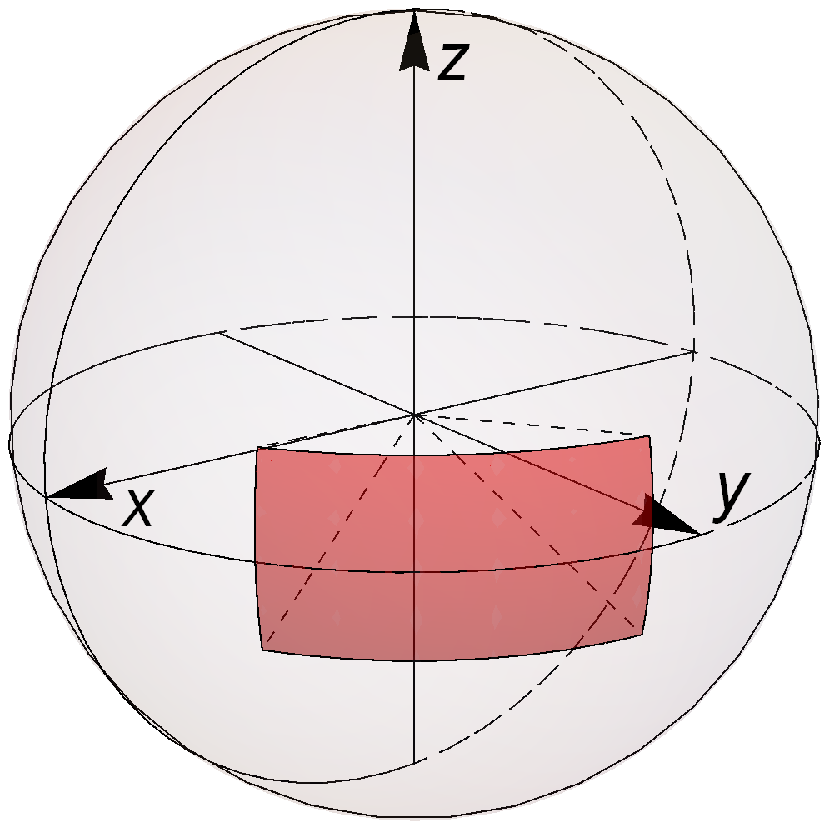}\\
\end{tabular}
}
}
\caption{Left panel: allowed region $9.1~\textrm{deg}\leq \alpha_\textrm{X}\leq 165.4~\textrm{deg},~-28.8~\textrm{deg}\leq \delta_\textrm{X}\leq 39.3~\textrm{deg}$ onto the Celestial Sphere. It yields an overall representation of most of the bounds in Figure~\ref{cubo0}~to~\ref{cubo2}. Right panel: more restricted area $35.4~\textrm{deg}\leq \alpha_\textrm{X}\leq 82.4~\textrm{deg},~-11.5~\textrm{deg}\leq \delta_\textrm{X}\leq 13.4~\textrm{deg}$ corresponding to $80~\textrm{deg}\lesssim \Omega_\textrm{X}\lesssim 100~\textrm{deg},~155~\textrm{deg}\lesssim f_\textrm{X}\lesssim 190~\textrm{deg}$ in Figures~\ref{cubo0}~to~\ref{cubo2}. The $\grf{x,~y}$ plane is the Celestial Equator, while the $z$ axis points towards the Celestial North Pole.}
\label{totalRADEC}
\end{figure*}

Figure~\ref{totalRADEC} depicts the allowed region $9.1~\textrm{deg}\leq \alpha_\textrm{X}\leq 165.4~\textrm{deg},~-28.8~\textrm{deg}\leq \delta_\textrm{X}\leq 39.3~\textrm{deg}$ onto the Celestial Sphere. It is somewhat representative of all our previous findings since it was obtained by considering the allowed regions for $f_\textrm{X}$ inferred in Figures~\ref{cubo0}~to~\ref{cubo2}, and the range of variation for $\Omega_\textrm{X}$ proposed by \citet{BroBaAJ2016,2016AJ....152..126B}. The narrower region $35.4~\textrm{deg}\leq \alpha_\textrm{X}\leq 82.4~\textrm{deg},~-11.5~\textrm{deg}\leq \delta_\textrm{X}\leq 13.4~\textrm{deg}$, corresponding approximately to $80~\textrm{deg}\lesssim \Omega_\textrm{X}\lesssim 100~\textrm{deg},~155~\textrm{deg}\lesssim f_\textrm{X}\lesssim 190~\textrm{deg}$ in Figures~\ref{cubo0}~to~\ref{cubo2} is depicted as well.
\subsubsection{Working with fixed orbital shape and size}
\begin{figure*}
\centerline{
\vbox{
\begin{tabular}{cc}
\epsfysize= 7.5 cm\epsfbox{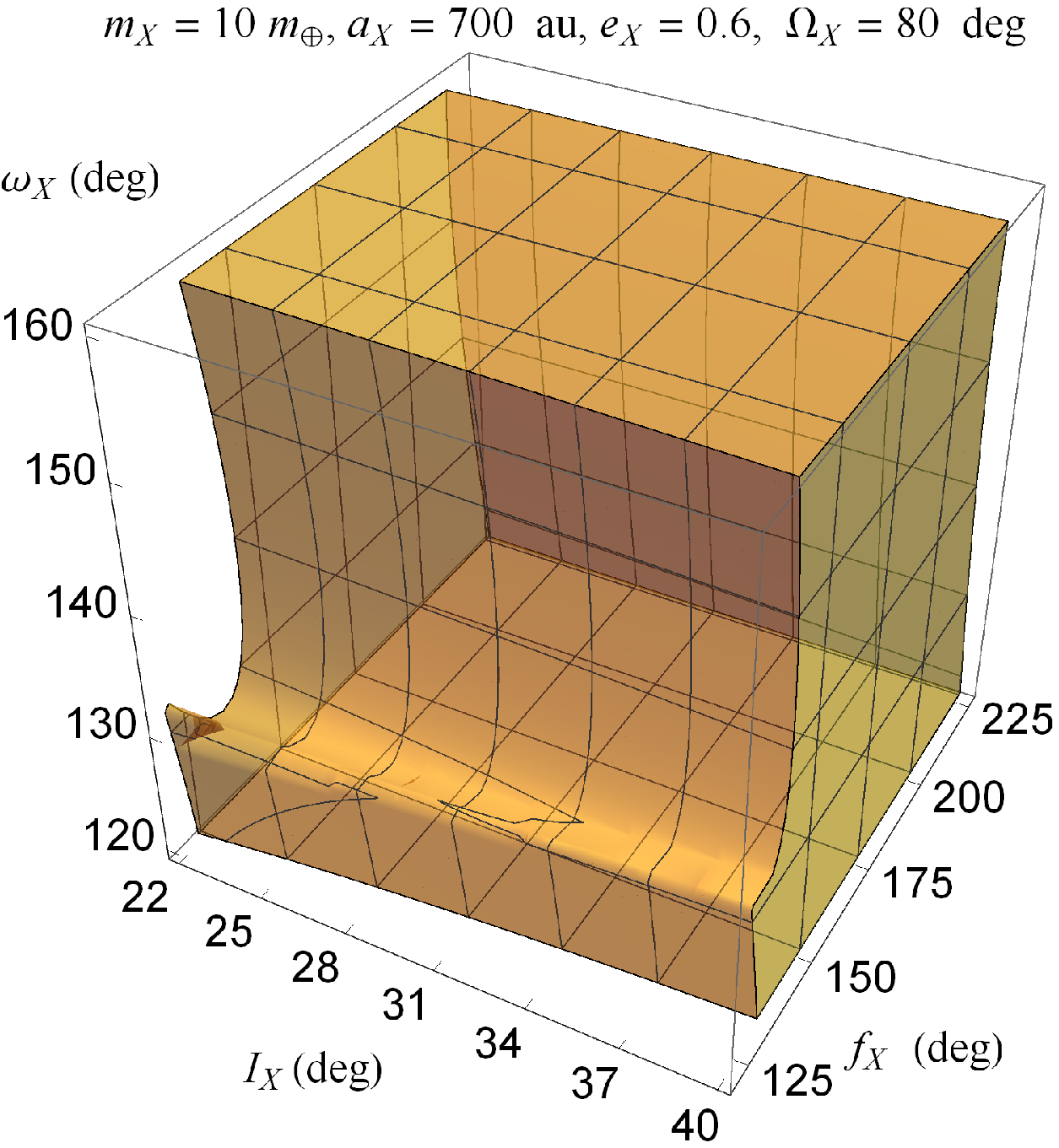} & \epsfysize= 7.5 cm\epsfbox{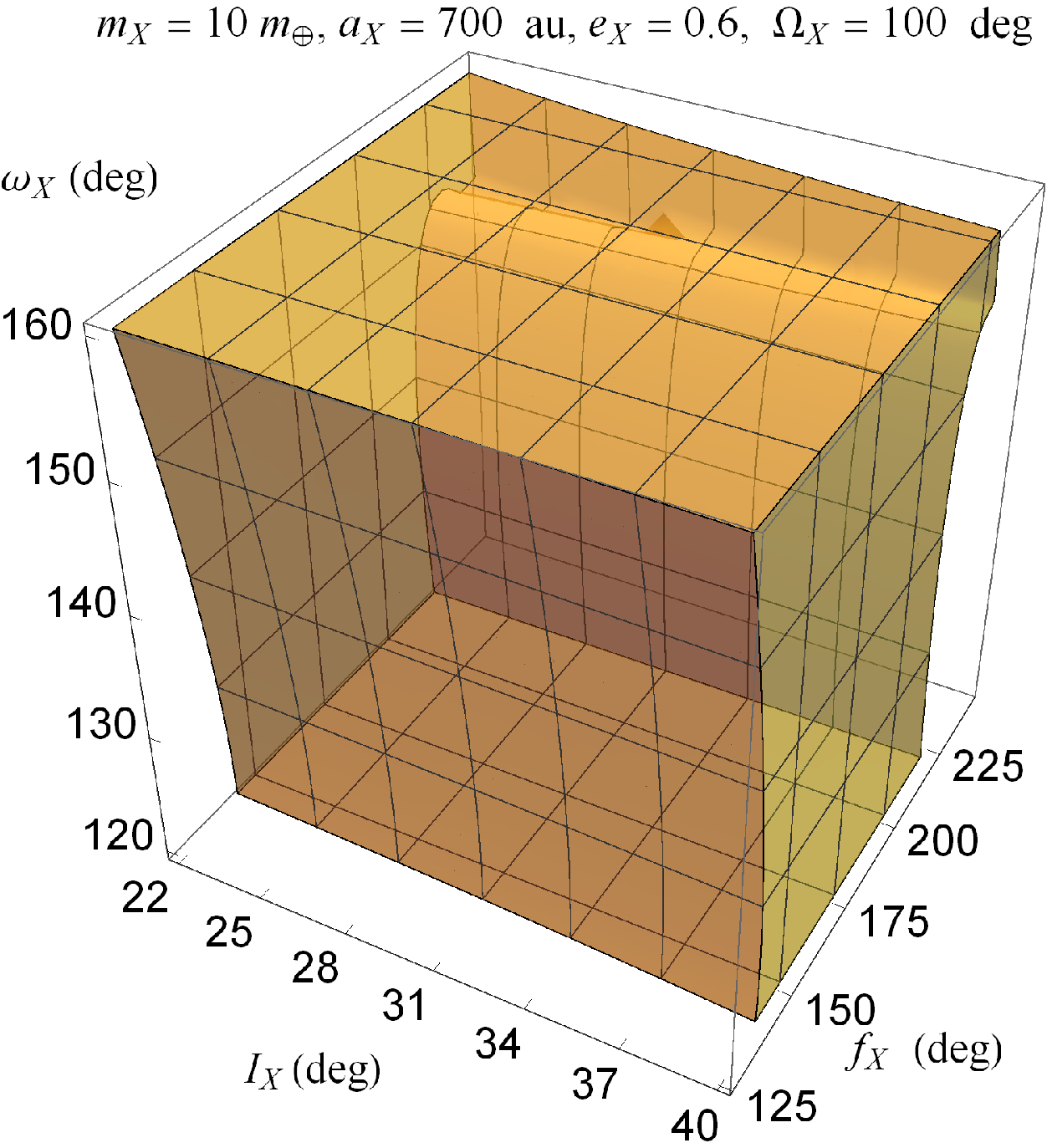} \\
\epsfysize= 7.5 cm\epsfbox{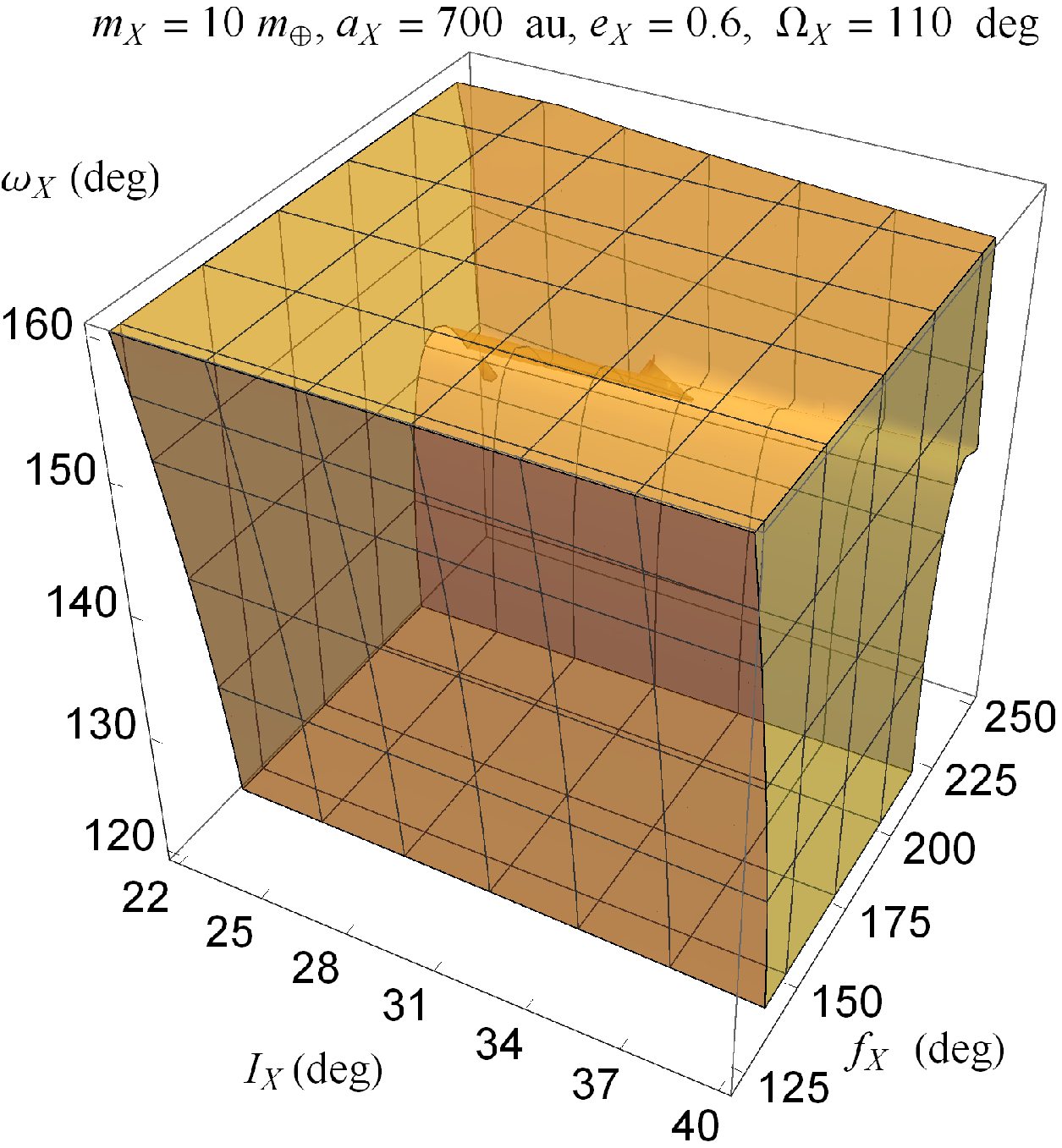} & \epsfysize= 7.5 cm\epsfbox{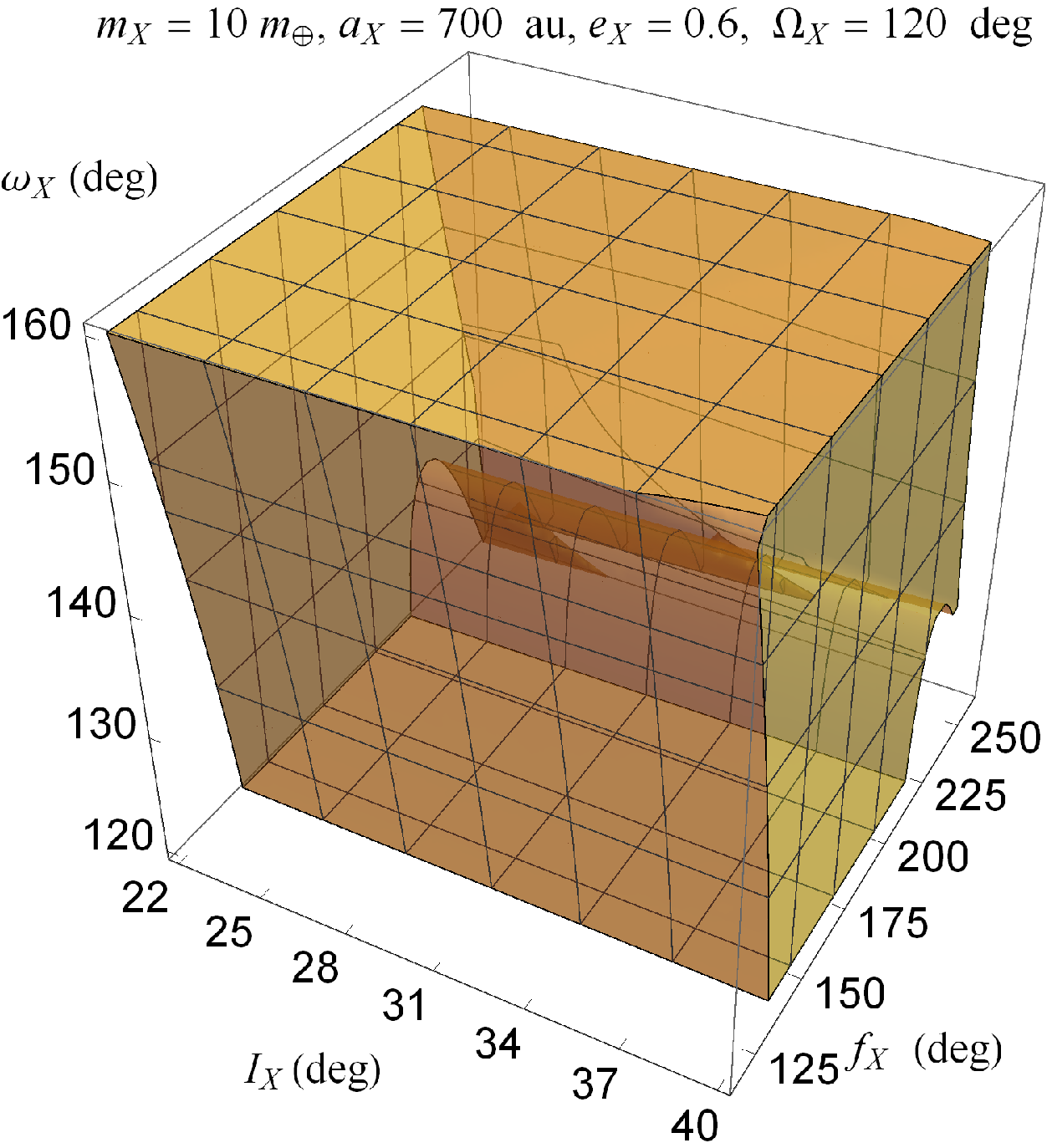} \\
\end{tabular}
}
}
\caption{Allowed regions  in the $\grf{I_\textrm{X},~f_\textrm{X},~\omega_\textrm{X}}$ volume for PX as determined by the INPOP10a ephemerides \citep{2011CeMDA.111..363F} through a comparison of \rfrs{nodocazzo}{Enne} and \rfrs{pericazzo}{Hacca}, computed with \rfrs{lX}{lZ} and the physical and orbital parameters of PX as in Section 3 of \citet{BroBaAJ2016} for the $m_\textrm{X}=10~m_\oplus,~a_\textrm{X}=700~\textrm{au},~e_\textrm{X}=0.6$ scenario, i.e. $22~\textrm{deg}\leq I_\textrm{X}\leq 40~\textrm{deg},~120~\textrm{deg}\leq \omega_\textrm{X}\leq 160~\textrm{deg}$.  The longitude of ascending node $\Omega_\textrm{X}$ is kept fixed to some values within   $80~\textrm{deg}\leq \Omega_\textrm{X}\leq 120~\textrm{deg}$  \citep{BroBaAJ2016, 2016AJ....152..126B}.
%
%
%
%
It must be recalled that $I_\textrm{X} = 30~\textrm{deg},~\omega_\textrm{X} = 150~\textrm{deg}$  are the favored values, as per Figure 7 of  \citet{BroBaAJ2016}. The maximum allowed region for the true anomaly of PX ranges from a minimum of $150~\textrm{deg}\lesssim f_\textrm{X}\lesssim ~210~\textrm{deg}$ to a maximum of $125~\textrm{deg}\lesssim f_\textrm{X}\lesssim ~250~\textrm{deg}$.
}
\label{cubo3}
\end{figure*}

\begin{figure*}
\centerline{
\vbox{
\begin{center}
 \epsfysize= 10 cm\epsfbox{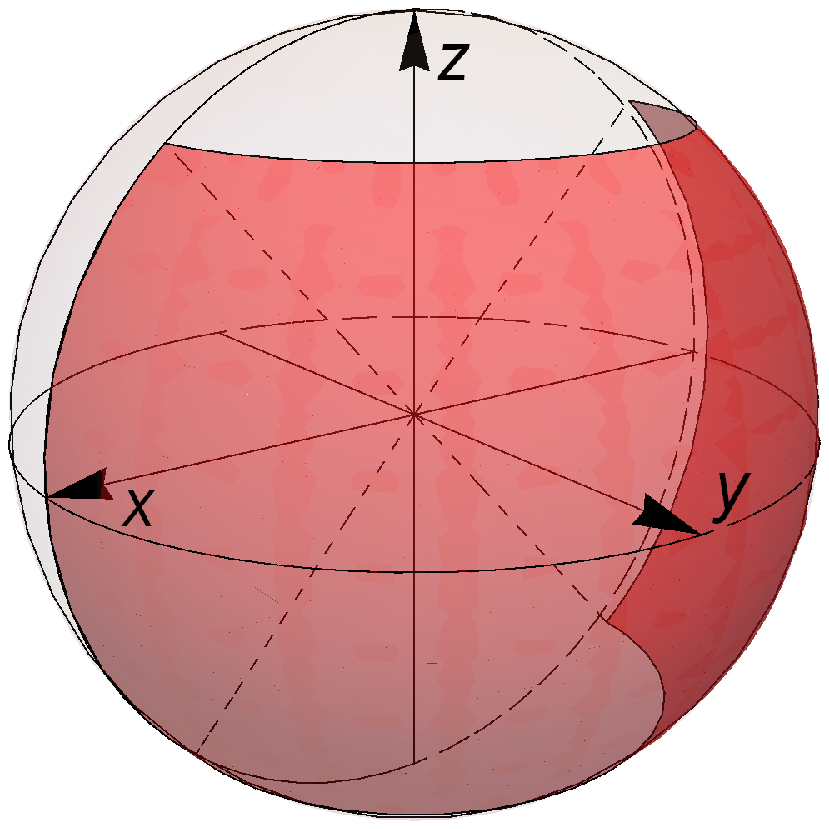} \\
\end{center}
}
}
\caption{Globally allowed region in the sky, characterized by $0~\textrm{deg} \lesssim\alpha_\textrm{X}\lesssim 176.8~\textrm{deg},~-48.4~\textrm{deg} \lesssim\delta_\textrm{X}\lesssim 48.0~\textrm{deg}$, according to the allowed values for $f_\textrm{X}$ of Figure~\ref{cubo3} and the ranges of variation for
$I_\textrm{X},~\Omega_\textrm{X},~\omega_\textrm{X}$ as per Section 3 of \citet{BroBaAJ2016}. The $\grf{x,~y}$ plane is the Celestial Equator, while the $z$ axis points towards the Celestial North Pole.}
\label{megaRADEC}
\end{figure*}

\begin{figure*}
\centerline{
\vbox{
\begin{center}
\epsfysize= 9 cm\epsfbox{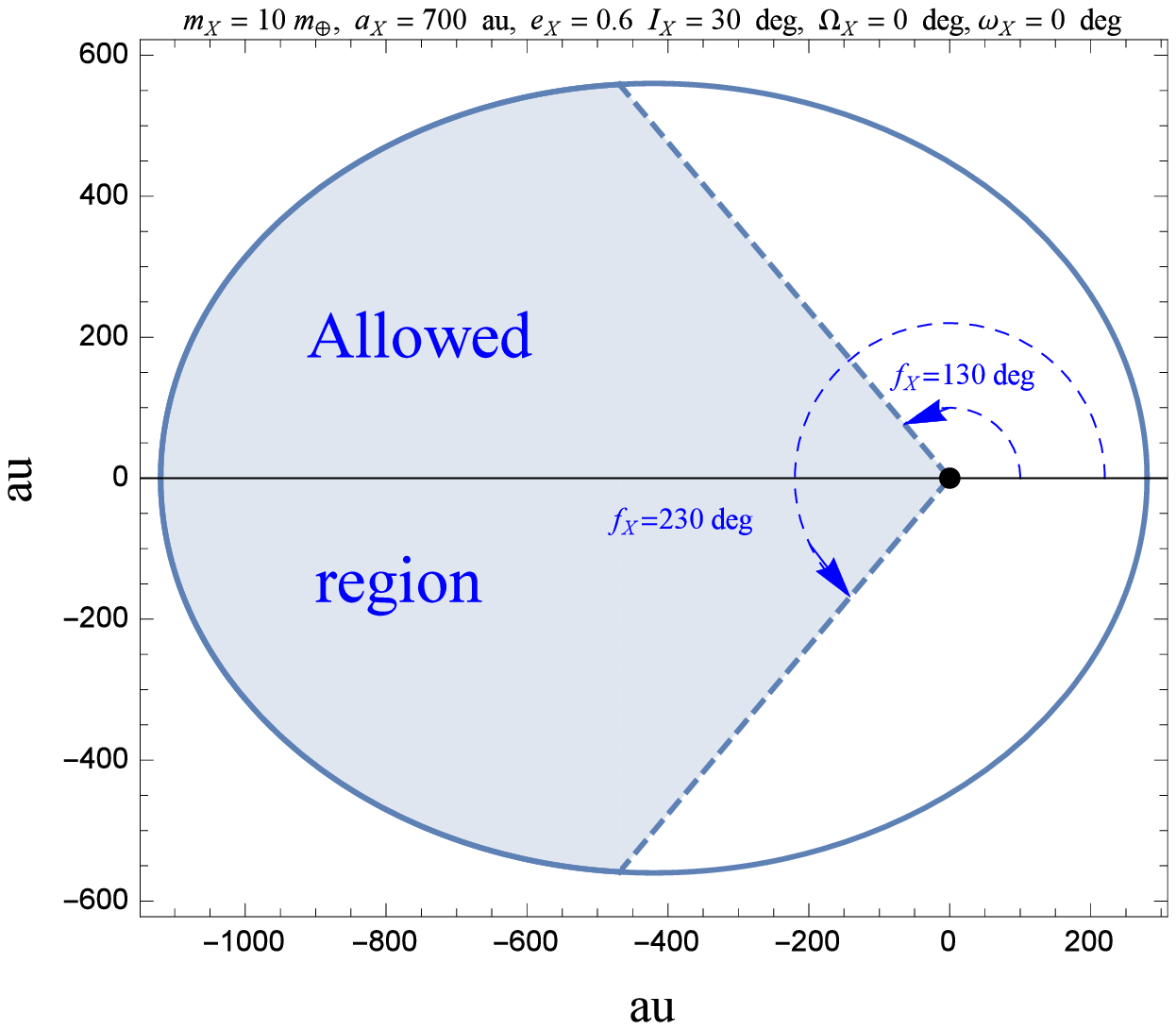} \\
\epsfysize= 9 cm\epsfbox{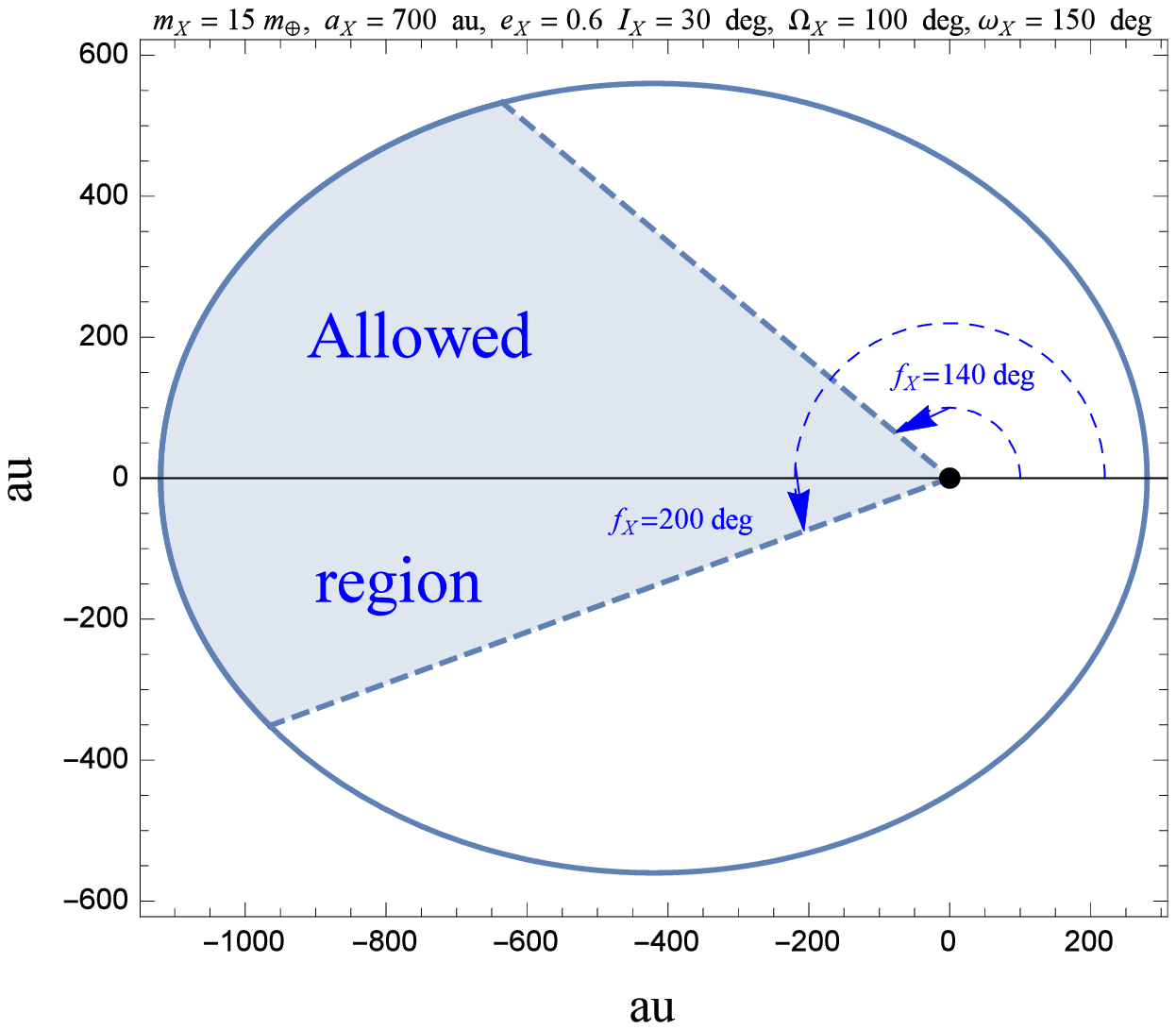} \\
\end{center}
}
}
\caption{Upper panel: allowed region (shaded area delimited by $130~\textrm{deg}\leq f_\textrm{X}\leq 230~\textrm{deg}$) for PX as determined by the INPOP10a ephemerides \citep{2011CeMDA.111..363F} through a comparison of \rfrs{nodocazzo}{Enne} and \rfrs{pericazzo}{Hacca}, computed with \rfrs{lX}{lZ} and the physical and orbital parameters of PX as in Table~\ref{uffa}. In terms of right ascension $\alpha_\textrm{X}$ and declination $\delta_\textrm{X}$, it corresponds to $144.6~\textrm{deg}\leq \alpha_\textrm{X}\leq 215.4~\textrm{deg},~-37.9~\textrm{deg}\leq \delta_\textrm{X}\leq 37.9~\textrm{deg}$; see Figure~\ref{RADEC1}. The aforementioned range for $f_\textrm{X}$ substantially holds also for the different orbital configuration of PX in Table~\ref{kepelemsX}; cfr. the left upper panel of Figure~\ref{nodo_peri}.  Lower panel: allowed region for the $m_\textrm{X} = 15~m_\oplus,~\Omega_\textrm{X}=100~\textrm{deg}$ case of the right upper panel of Figure~\ref{nodo_peri}.}
\label{ellis}
\end{figure*}

In their Section 3, \citet{BroBaAJ2016} attempt to investigate the possible spatial orientation of the orbit of PX for $m_\textrm{X} = 10~m_\oplus,~a_\textrm{X} = 700~\textrm{au},~e_\textrm{X}=0.6$; they find that the preferred regions for the inclination and the argument of perihelion are $22~\textrm{deg}\leq I_\textrm{X}\leq 40~\textrm{deg},~120~\textrm{deg}\leq\omega_\textrm{X}\leq 160~\textrm{deg}$. Figure~\ref{cubo3} explores such a portion of the parameter space of the distant perturber for fixed values of $\Omega_\textrm{X}$ within the previously adopted range i.e. $80~\textrm{deg}\leq\Omega_\textrm{X}\leq 120~\textrm{deg}$ \citet{BroBaAJ2016,2016AJ....152..126B}. It can be noticed that the true anomaly of PX ranges from $150~\textrm{deg}\lesssim f_\textrm{X}\lesssim ~210~\textrm{deg}$ ($\Omega_\textrm{X} \approx 110-120~\textrm{deg},~\omega_\textrm{X} \approx 120-130~\textrm{deg}$) to a wider interval as large as about $125~\textrm{deg}\lesssim f_\textrm{X}\lesssim ~250~\textrm{deg}$ ($\Omega_\textrm{X} \approx 80~\textrm{deg},~\omega_\textrm{X}= 120~\textrm{deg}$). By considering $\alpha_\textrm{X},~\delta_\textrm{X}$ as functions of $I_\textrm{X},~\Omega_\textrm{X},~\omega_\textrm{X},~f_\textrm{X}$ within their ranges of variation as per Section 3 of \citet{BroBaAJ2016} ($I_\textrm{X},~\Omega_\textrm{X},~\omega_\textrm{X}$) and Figure~\ref{cubo3} ($f_\textrm{X}$), it can be inferred an allowed region onto the Celestial Sphere as large as $0~\textrm{deg} \lesssim\alpha_\textrm{X}\lesssim 176.8~\textrm{deg},~-48.4~\textrm{deg} \lesssim\delta_\textrm{X}\lesssim 48.0~\textrm{deg}$; it is depicted in Figure~\ref{megaRADEC}.

Finally, let us consider the particular orbital configuration of PX selected at the end of Section 3 of \citet{BroBaAJ2016} as reported in Table~\ref{uffa}.
It turns out that our method yields an allowed range $130~\textrm{deg}\leq f_\textrm{X}\leq 230~\textrm{deg}$, which is illustrated in Figure~\ref{ellis}.
It corresponds to a position in the sky in terms of right ascension $\alpha_\textrm{X}$ and declination $\delta_\textrm{X}$  delimited by $144.6~\textrm{deg}\leq \alpha_\textrm{X}\leq 215.4~\textrm{deg},~-37.9~\textrm{deg}\leq \delta_\textrm{X}\leq 37.9~\textrm{deg}$, as depicted in Figure~\ref{RADEC1}.

\subsection{Perspectives on the method adopted}
Unfortunately, the supplementary node precessions are currently available only for the INPOP10a ephemerides \citep{2011CeMDA.111..363F}; if and when they will be determined by the astronomers also with their most recent ephemerides (see, e.g., \citet{2016arXiv160100947F, 2013MNRAS.432.3431P}), it will be possible to further strengthen our preliminary constraints on the location of PX and remove the degeneracy in $\Omega_\textrm{X}$, at least to a certain extent. Unfortunately, no more accurate draconitic extra-precessions have been determined so far with the most recent versions of the INPOP and EPM planetary ephemerides; in this respect, it would also be of the utmost importance if the astronomers will accurately determine the supplementary secular rates of change of more planetary orbital elements since they are all theoretically impacted by a distant perturber, apart from the semimajor axis \citep{2012CeMDA.112..117I}.
\section{Discussion of other approaches and results in the literature}\lb{lite}
\subsection{General consideration of the method presently adopted}\lb{difesa}
In principle, the approach followed in Section~\ref{costra} may be criticized because of a certain lack of rigor mainly in view of the need of reprocessing the same set of observations by explicitly modeling the dynamical effect one is looking for and solving for it; otherwise, it could not be excluded that part of the signal of interest, if real, may be somewhat removed, being absorbed in the estimation of, e.g., the planetary initial conditions. As a consequence, the resulting constraints might turn out to be too tight, excluding more or less larger admissible portions of the parameter space.
Thus, a complementary dedicated covariance analysis of the available data records by explicitly modeling the dynamical action of PX by the teams routinely engaged in the production of the planetary ephemerides is certainly desirable. On the other hand, it must be stressed that also such a potentially more rigourous approach may have its own drawbacks. Indeed, a selection of the dynamical effects to be modeled
and of the parameters entering them which can be practically estimated is unavoidably always made in real data reduction. Thus, any other sort of unmodeled/mismodeled forces, both standard and exotic, may impact the solved-for parameter(s); the effect of any sort of \virg{Russell teapots} may
well creep into the  full covariance analysis by somewhat biasing the estimated values one is interested to.

Furthermore, the following considerations are in order.
Generally speaking, the opportunistic approach consisting of using data originally processed for different scopes to constrain a wide variety of exotic physical effects arising from several theoretical scenarios aiming to modify the currently known laws of the gravitational interaction at various scales  is of common usage in the literature \citep{1987ApJ...320..871N, 2012EPJP..127..155A, 2013PhRvD..87h4041L, 2013CQGra..30p5020S, 2013MNRAS.433.3584X, 2013ApJ...774...65C, 2014Galax...2..466A, 2014RAA....14..139L, 2014Ap&SS.350..103D, 2014MNRAS.444.1639C, 2014NewA...31...51W, 2014RAA....14.1019L, 2014MNRAS.438.1832X, 2015IJMPD..2450060R, 2015IJMPD..2450066F, 2015Univ....1..422A, 2015NewA...35...36D, 2015AnPhy.361...62D, 2015PhRvD..92f4033Z, 2015IJTP...54.1739D, 2015ForPh..63..711K, 2015APh....68....1H, 2015EPJC...75..539D,2015PhRvD..91j4014R, 2015PhRvD..92f4049H, 2015MNRAS.451.3034N, 2016CQGra..33g5005C,2016arXiv160100947F,2016arXiv160903613R,2016arXiv160605263C,2016MPLA...3150021D} since, e.g., the early days of general relativity; suffice it to recall that \citet{LeVer1859} fit purely Newtonian dynamical and measurement models to the Hermean observations which yielded the celebrated anomalous perihelion precession of Mercury of $42.98$ arcsec cty$^{-1}$ \citep{1915SPAW...47..831E,1986Natur.320...39N,2003Ap&SS.284.1159P}.

More specifically, a peculiar form of the MOdified Newtonian Dynamics (MOND) theory predicts certain subtle effects on the planetary motions which are able to mimic just those of a far pointlike pertuber located in a specific position \citep{2009MNRAS.399..474M, 2011MNRAS.412.2530B}. Now, the bounds on the MOND-type parameter inferred analytically in \citep{2010OAJ.....3....1I} in the same way as done here with older supplementary perihelion advances turned out to be compatible with those later determined in \citet{2014PhRvD..89j2002H} by explicitly modeling it in the data reduction procedure of the Cassini telemetry. Furthermore,  both \citet{2011MNRAS.412.2530B} and \citet{2016arXiv160100947F}  used a direct comparison of a theoretically predicted orbital precession $\dot\varpi_{\rm theo}$ with supplementary precessions $\Delta\dot\varpi_{\rm obs}$ determined without modeling the PX-type MOND effect to infer constraints on the characteristic parameter at hand. In particular, \citet{2016arXiv160100947F} wrote about the usage of the supplementary perihelion precessions determined with INPOP15a in Table~\ref{perihelia}: \virg{In terms of comparison to alternative theories of gravitation and in particular to MOND (Ref. 2 [i.e. \citep{2011MNRAS.412.2530B}]), the new estimated values [i.e. $\Delta\varpi_{\rm obs}$] are still competitive for selecting possible MOND functions. Following the Ref. 2 notations, only $\mu_{20(y)}$ is indeed the only possible functions regarding the impact of the quadrupole term $Q_2$ on the advance of planet perihelia.}

Last but not least, another notable case in which both the approaches returned essentially the same outcome is the rejection of the hypothesis that the Pioneer anomaly \citep{1998PhRvL..81.2858A, 2002PhRvD..65h2004A,2010LRR....13....4T} was due to an anomalous extra-acceleration of gravitational origin acting at least on all the major bodies of the Solar System orbiting in the same region where the alleged anomaly manifested itself \citep{2006NewA...11..600I,2007PhRvD..76d2005T,2007FoPh...37..897I,2008AIPC..977..254S,2010IAUS..261..179S,PitPio,2009sf2a.conf..105F,2012sf2a.conf...25F,Folkner09}.
It came before that its likely mundane origin rooted in standard non-gravitational physics was later demonstrated \citep{2011AnP...523..439R,2012PhRvL.108x1101T,2012PhLB..711..337F,2014PhRvD..90b2004M}.
\subsection{Comparing the present results for the originally proposed configuration of PX with other ones in the literature}
A first step in the complementary direction mentioned in Section~\ref{difesa} was meritoriously  taken in the recent work by \citet{2016A&A...587L...8F}. They produced new post-fit residuals $\Delta\rho_\textrm{X}$ of the geocentric range $\rho$ of Saturn (the red curves in their Figure 1) by explicitly including the specific version of PX of Table~\ref{kepelemsX} \citep{BaBroAJ2016} with $m_\textrm{X}=10~m_\oplus$ and $\Omega_\textrm{X}=113$ deg in the dynamical models of the Solar System used to reprocess a decadal record of ranging data from Cassini extending from April 2004 to April 2014 for selected values of\footnote{The parameterization used by \citet{2016A&A...587L...8F} for $f_\textrm{X}$ is different from ours here since they assumed $-180~\textrm{deg}\leq f_\textrm{X}\leq 180~\textrm{deg}$.} $f_\textrm{X}$. Such post-fit residuals were subsequently contrasted to the \virg{nominal} post-fit ones $\Delta\rho_0$ obtained by reducing the same data set without modeling PX at all; the set of the estimated parameters is the same for both the residuals, apart from those of the Moon which were not included in the solution for $\Delta\rho_\textrm{X}$. As an outcome, certain regions of admissible values for $f_\textrm{X}$ were inferred $(85~\textrm{deg}\lesssim f_\textrm{X}\lesssim 235~\textrm{deg}$ and $260~\textrm{deg}\lesssim f_\textrm{X}\lesssim 295~\textrm{deg})$ along with a claimed most probable interval of values for it amounting to $f_\textrm{X} = 117.8^{+11}_{-10}~\textrm{deg}$.
The results by \citet{2016A&A...587L...8F} can be compared with our findings in Section~\ref{BaBro}, Figure~\ref{nodo_peri}, from which it turns out that our allowed region for the true anomaly of PX is $130~\textrm{deg}\lesssim f_\textrm{X}\lesssim 240~\textrm{deg}$. It overlaps with the largest allowed region claimed by \citet{2016A&A...587L...8F} to a large extent; on the other hand, their claimed most probable interval of values is incompatible with our bounds. Cfr. the upper panel of our Figure~\ref{miaellisse}, extracted from Figure~\ref{nodo_peri}, with Figure 6 of \citet{2016A&A...587L...8F}, adapted in the lower panel of Figure~\ref{miaellisse} in order to display just the allowed regions inferred from the analysis of the existing 2004-2014 Cassini data.  %
%
%
%
%

%
%

As far as the range-based study by \citet{2016AJ....152...94H} is concerned, they looked at the  configuration for PX originally proposed by \citet{BaBroAJ2016} with $\Omega_\textrm{X}=113~\textrm{deg}$ in their Section 4 and Figure 3. \citet{2016AJ....152...94H} first claimed that they were able to reproduce the results by \citet{2016A&A...587L...8F}. Then, after having adopted a different statistical criterion and having kept also outlying data points, \citet{2016AJ....152...94H} stated that there would be \virg{significant room for further investigation because excluding outlying data points and using a standard $\chi^2$ statistic allow for more nuanced model evaluation}. Nonetheless, no explicit new intervals of allowed values for $f_\textrm{X}$ were provided by \citet{2016AJ....152...94H}; a mere visual inspection of Figure 3 of \citet{2016AJ....152...94H} does not allow to infer them in a
clear way.
\subsection{Comparing our results for different orbital and physical configurations with other studies in the literature}
\begin{figure*}
\centerline{
\vbox{
\begin{center}
\epsfysize= 9 cm\epsfbox{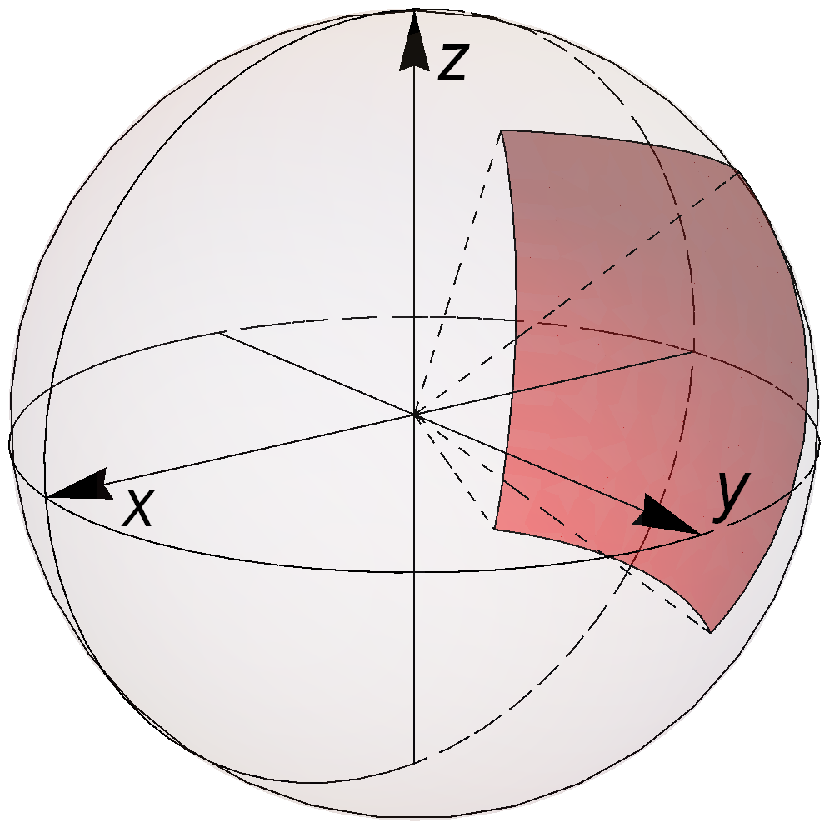} \\
\end{center}
}
}
\caption{Allowed portion of the Celestial Sphere, determined by $144.6~\textrm{deg}\lesssim \alpha_\textrm{X}\lesssim 215.4~\textrm{deg},~-37.9~\textrm{deg}\lesssim \delta_\textrm{X}\lesssim 37.9~\textrm{deg}$, for the physical and orbital configuration of PX of Table~\ref{uffa} and the upper panel of Figure~\ref{ellis}. The $\grf{x,~y}$ plane is the Celestial Equator, while the $z$ axis points towards the Celestial North Pole.}
\label{RADEC1}
\end{figure*}

\begin{figure*}
\centerline{
\vbox{
\begin{center}
\epsfysize= 10.0 cm\epsfbox{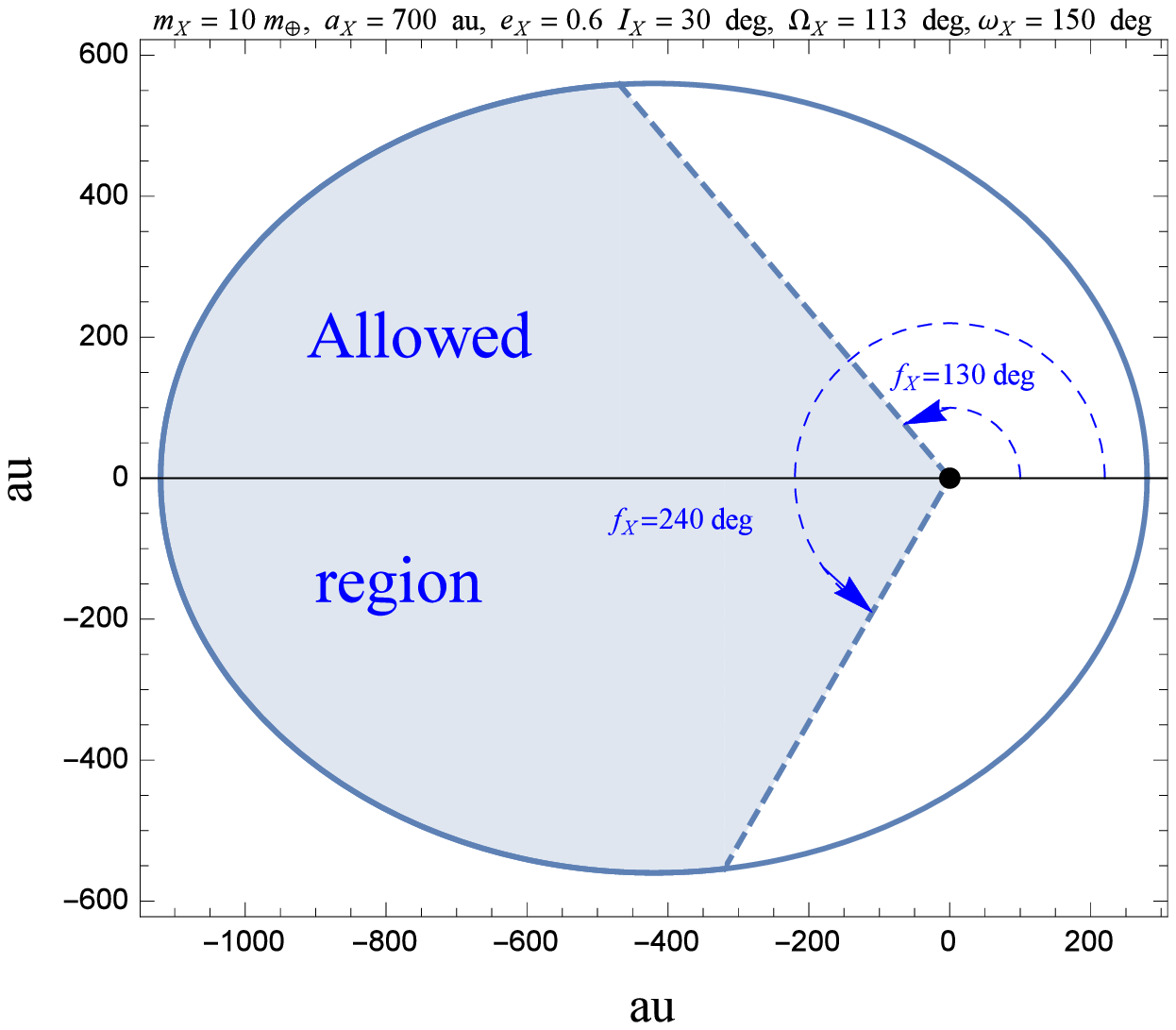} \\
\epsfysize= 10.0 cm\epsfbox{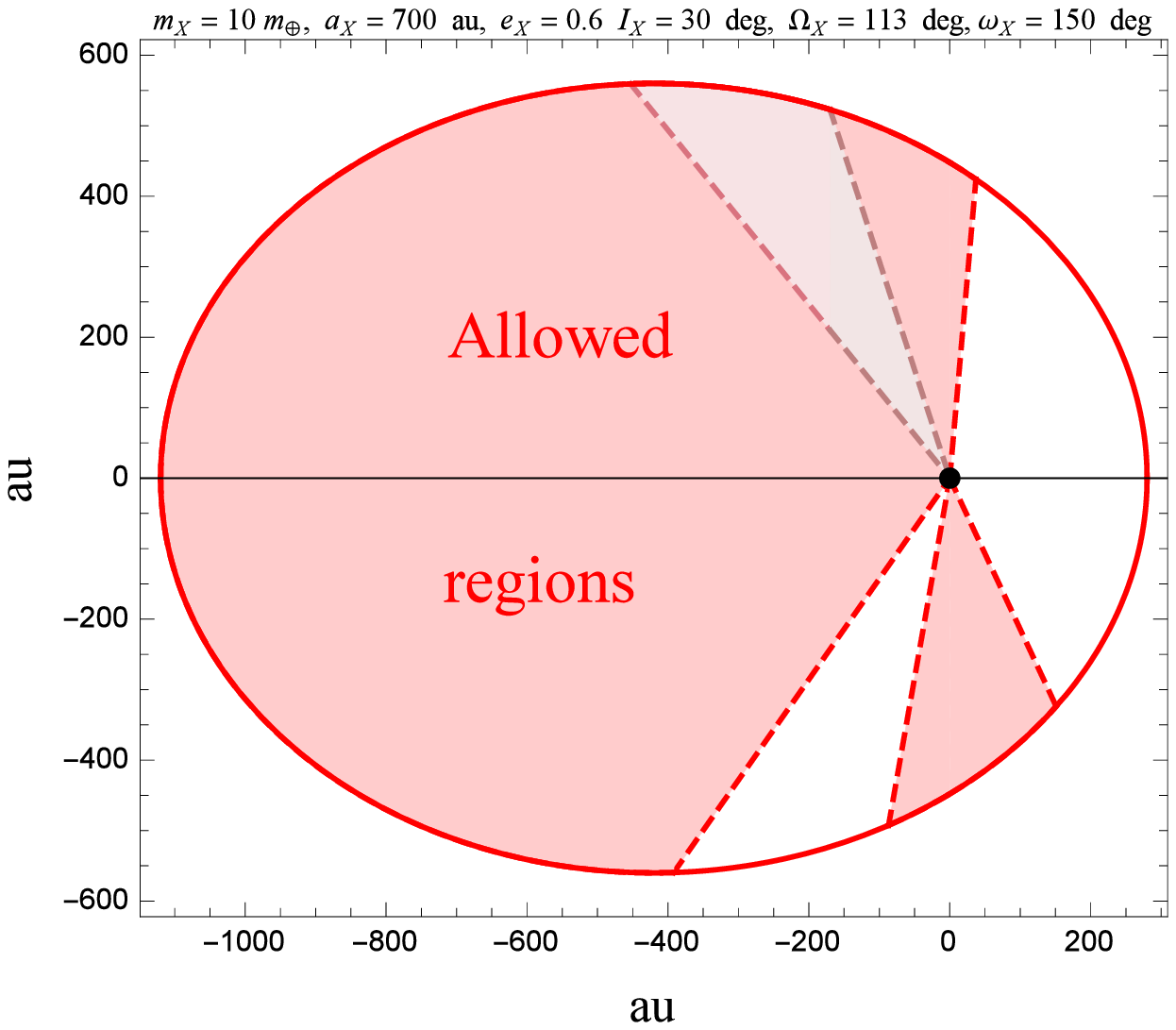}\\
\end{center}
}
}
\caption{Upper panel: allowed region for Table \ref{kepelemsX} extracted from our Figure~\ref{nodo_peri} for $m_\textrm{X}=10~m_\oplus,~\Omega_\textrm{X} = 113~\textrm{deg}$ \citep{BaBroAJ2016}. Lower panel: allowed regions as per \citet{2016A&A...587L...8F} for the same scenario of PX. According to  \citet{2016A&A...587L...8F}, the allowed regions are $85~\textrm{deg}\lesssim f_\textrm{X}\lesssim 235~\textrm{deg}$, which includes also the claimed most probable interval of values  $108~\textrm{deg}\leq f_\textrm{X}\leq 129~\textrm{deg}$, here depicted as the darker shaded area, and the disjointed area $260~\textrm{deg}\lesssim f_\textrm{X}\lesssim 295~\textrm{deg}$.}
\label{miaellisse}
\end{figure*}

In Figure 10 of their Section 4, \citet{BroBaAJ2016}, using all constraints on the orbital and physical parameters of the hypothesized distant planet arising from their dynamical simulations of selected populations of KBOs, predict its possible orbital path across the sky. The result is a continuous, sinusoidal strip in the $\grf{\alpha_\textrm{X},~\delta_\textrm{X}}$ plane  limited by $-40~\textrm{deg}\leq\delta_\textrm{X}\leq 40~\textrm{deg}$; no intervals in the $\alpha_\textrm{X}$ axis  seem to be forbidden. It is neither straightforward nor easy to make a direct comparison with our results for the allowed regions onto the Celestial Sphere in Section~\ref{costra} since quantitative details about the values of the orbital and physical parameters of PX used  are missing in \citet{BroBaAJ2016}; furthermore, only a visual inspection of their Figure 10 is possible. Nonetheless, let us try a comparison with, say, our Figures~\ref{nodo_peri} and \ref{RADEC1}, both corresponding to $a_\textrm{X}=700~\textrm{au}$. From the second panel from the the top of Figure 10 of \citet{BroBaAJ2016}, it seems that such a value of the perturber's semimajor axis is possible only for roughly $0~\textrm{deg}\lesssim \alpha_\textrm{X}\lesssim 135-140~\textrm{deg}$. In turn, the first panel from the top of Figure 10 of \citet{BroBaAJ2016} tells us that the corresponding range in declination is approximately $-40~\textrm{deg}\lesssim \delta_\textrm{X}\lesssim 40~\textrm{deg}$; actually, a range of values for $\delta_\textrm{X}$ corresponds to each value of $\alpha_\textrm{X}$. Unfortunately, it is impossible to associate specific values of both the mass of PX and its other orbital parameters to such naive guesses based only on looking at Figure 10 of \citet{BroBaAJ2016}. Now, depending on both $m_\textrm{X}$ and $\Omega_\textrm{X}$, our Figure~\ref{nodo_peri} and \ref{RADEC1} tell us that that the allowed regions onto the Celestial Sphere can be rather different, especially as far as $\alpha_\textrm{X}$ is concerned; cfr. Figure~\ref{RADEC1} ($144.6~\textrm{deg}\leq \alpha_\textrm{X}\leq 215.4~\textrm{deg},~-37.9~\textrm{deg}\leq \delta_\textrm{X}\leq 37.9~\textrm{deg}$) and the lower right panel of Figure~\ref{nodo_peri} ($40.1~\textrm{deg} \lesssim\alpha_\textrm{X}\lesssim 101.8~\textrm{deg},~-13.9~\textrm{deg} \leq\delta_\textrm{X}\leq 18.4~\textrm{deg}$).

In their Section 5.2, \citet{2016AJ....152...94H} adopted a Markov Chain Monte Carlo (MCMC) approach to generate a distribution of  realizations of PX with specific masses and orbital elements. They, then, used them to generate associated preferred locations for PX from Cassini range data alone in the $\grf{\alpha_\textrm{X},~\delta_\textrm{X}}$ plane, depicted in their Figure 6. As a result, they obtained two narrow bands with curvilinear contours  essentially perpendicular to the orbit of Saturn. From a visual inspection of Figure 6 of \citet{2016AJ....152...94H}, it seems that they are approximately centered in $\alpha_\textrm{X} \approx 30~\textrm{deg}$ and $\alpha_\textrm{X} \approx 200~\textrm{deg}$, being as large as about $-60~\textrm{deg}\lesssim\delta_\textrm{X}\lesssim 60~\textrm{deg}$ in declination. Then, \citet{2016AJ....152...94H} considered the intersections of such allowed regions with the swath of possible positions in the sky of Figure 10 by \citet{BroBaAJ2016} finding two much narrower domains quantitatively characterized in Table 1 of \citet{2016AJ....152...94H}. It contains the median and $1\sigma$ ranges for the best-fit quantities arising from the combined MCMC analysis of the Cassini data by \citet{2016AJ....152...94H} with the dynamical constraints from \citet{BroBaAJ2016}. While one of such two regions, corresponding to $m_\textrm{X}< 5~m_\oplus$ and denoted as Zone-1, has to be discarded, the other one, characterized, among other things, by\footnote{\citet{2016AJ....152...94H} used the mean anomaly ${\mathcal{M}}$ as positional angle along the orbit of PX. As such, the range for $f_\textrm{X}$ quoted here is calculated via the Kepler equation for $e_\textrm{X} = 0.5$.} $142~\textrm{deg}\lesssim f_\textrm{X}\lesssim 164~\textrm{deg}$  and called Zone-2, is considered by \citet{2016AJ....152...94H} as the preferred region on the sky to find PX. We reproduce part of it in Table~\ref{HoPa}.
It turns out that such an interval of values for $f_\textrm{X}$ is compatible with our bounds obtained in Section~\ref{costra} for many of the orbital and physical scenarios of PX, retrieved from \citet{BaBroAJ2016,BroBaAJ2016},  which fall within the ranges reported in Table 1 of \citet{2016AJ....152...94H} for Zone-2. Conversely, it can be demonstrated that, if we apply our approach of Section~\ref{costra} based on the node and the perihelion precessions of Saturn to the ranges of parameters of PX quoted in the Zone-2 of Table 1 of \citet{2016AJ....152...94H}, it is possible to obtain just $142~\textrm{deg}\lesssim f_\textrm{X}\lesssim 164~\textrm{deg}$.  As far as the position in the sky is concerned, it should be noted that if we apply our method, described in Section~\ref{BaBro}, to extract the maximum and the minimum values of $\alpha_\textrm{X},~\delta_\textrm{X}$ from suitable variations of the orbital elements of PX, the allowed region in the sky corresponding to the range of values reported for the preferred region in Table 1 of \citet{2016AJ....152...94H}, is $3~\textrm{deg}\lesssim \alpha_\textrm{X}\lesssim 78~\textrm{deg},~ -38~\textrm{deg}\lesssim \delta_\textrm{X}\lesssim 5.3~\textrm{deg}$, which is larger than $\alpha_\textrm{X}= 39.5^{+5.5}_{-5.3}~\textrm{deg},~\delta_\textrm{X}= -15.4^{+6.7}_{-6.6}~\textrm{deg}$ quoted by \citet{2016AJ....152...94H} in their Table 1.
\section{Constraining the mass of PX}
\begin{figure*}
\centerline{
\vbox{
\begin{tabular}{cc}
\epsfysize= 6.2 cm\epsfbox{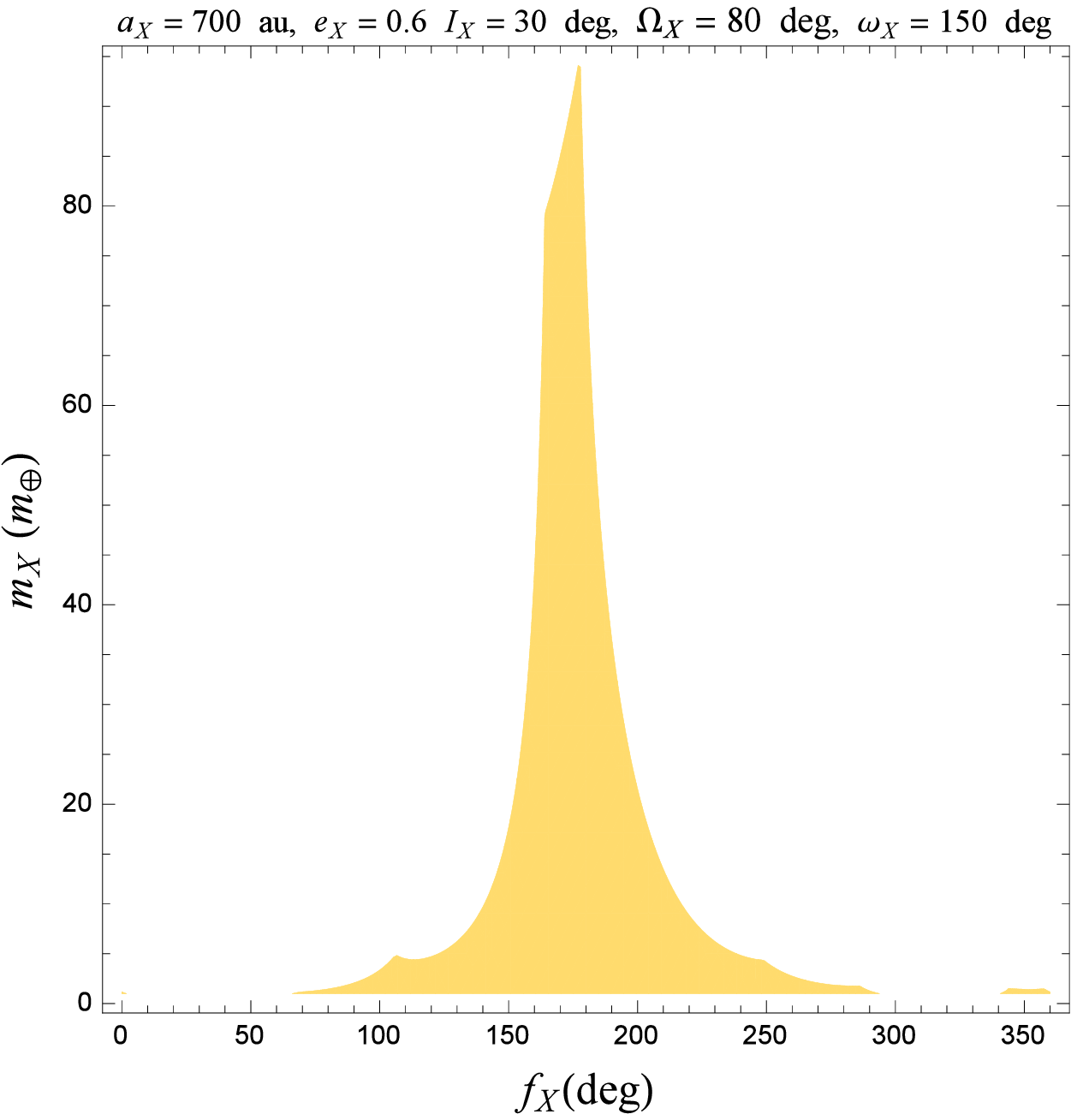} & \epsfysize= 6.2 cm\epsfbox{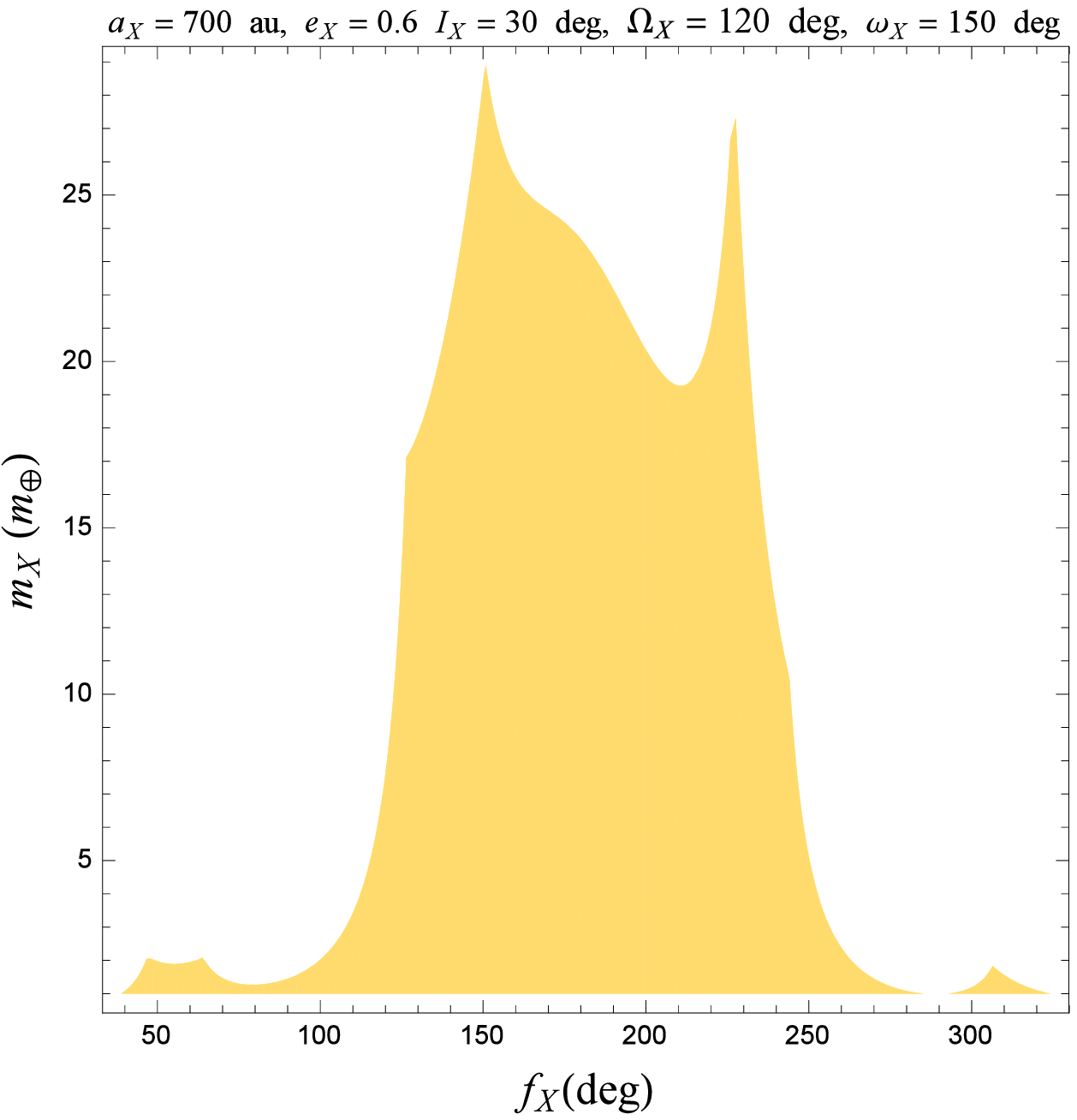} \\
\epsfysize= 6.2 cm\epsfbox{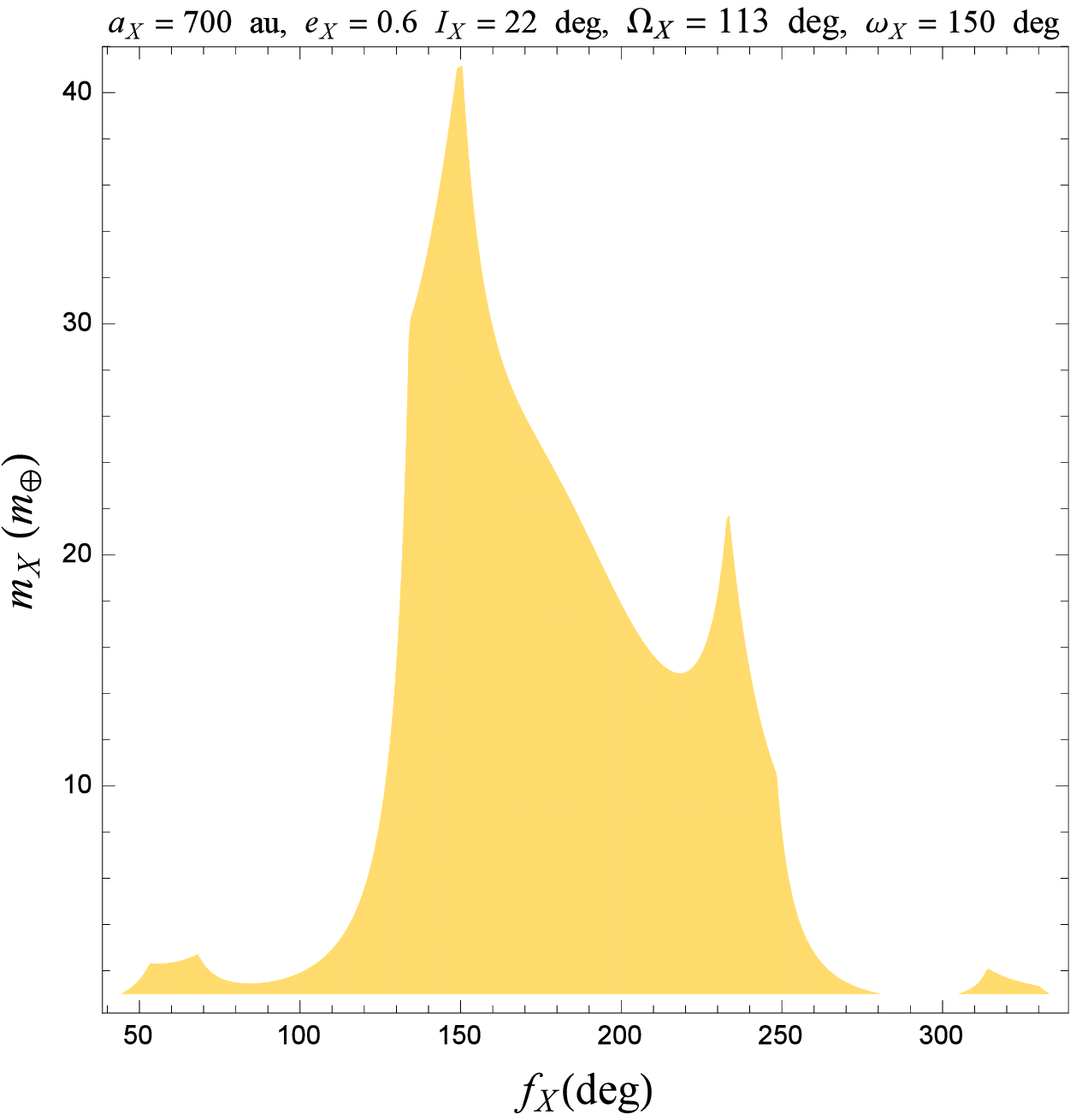} & \epsfysize= 6.2 cm\epsfbox{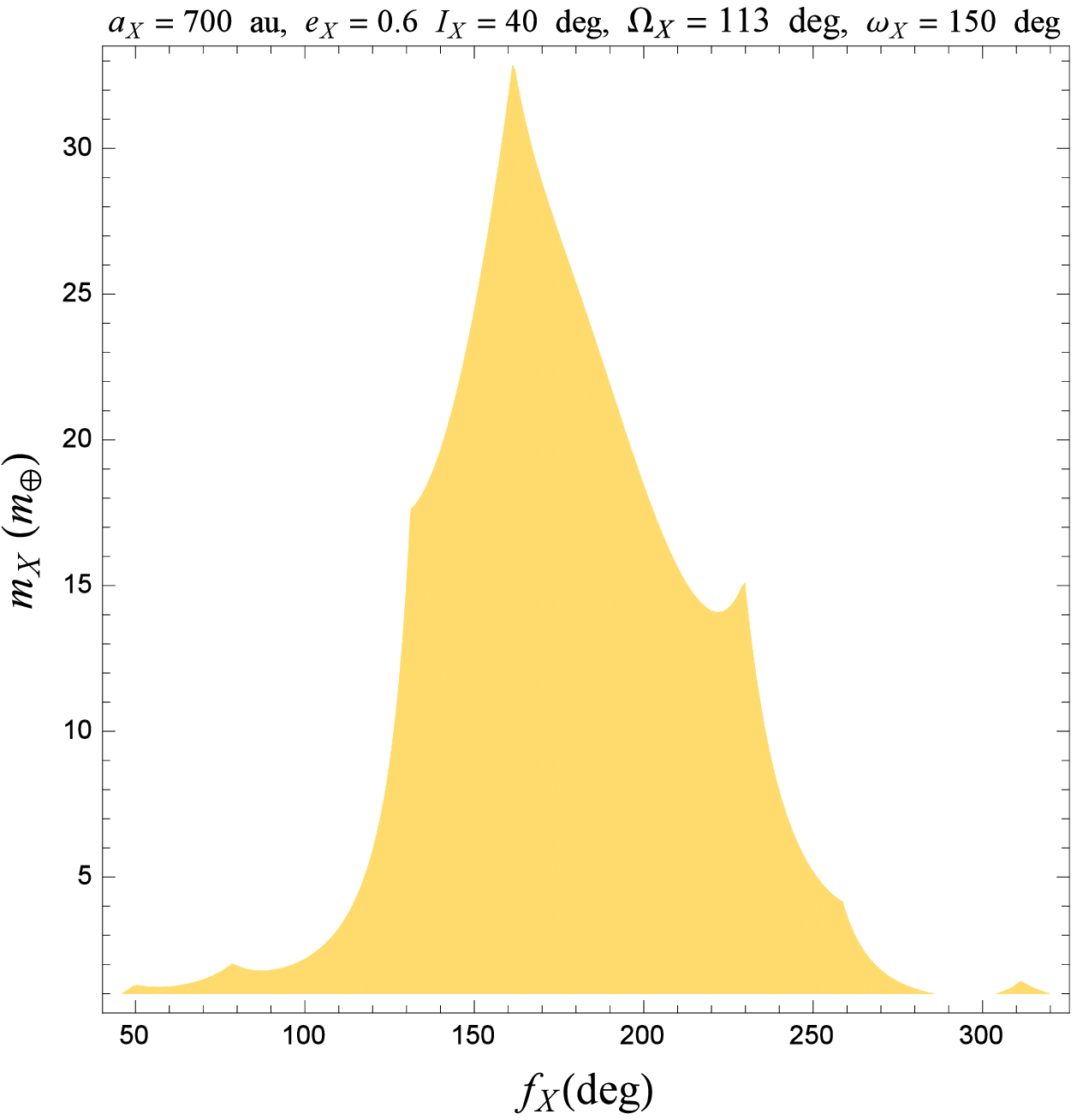} \\
\epsfysize= 6.2 cm\epsfbox{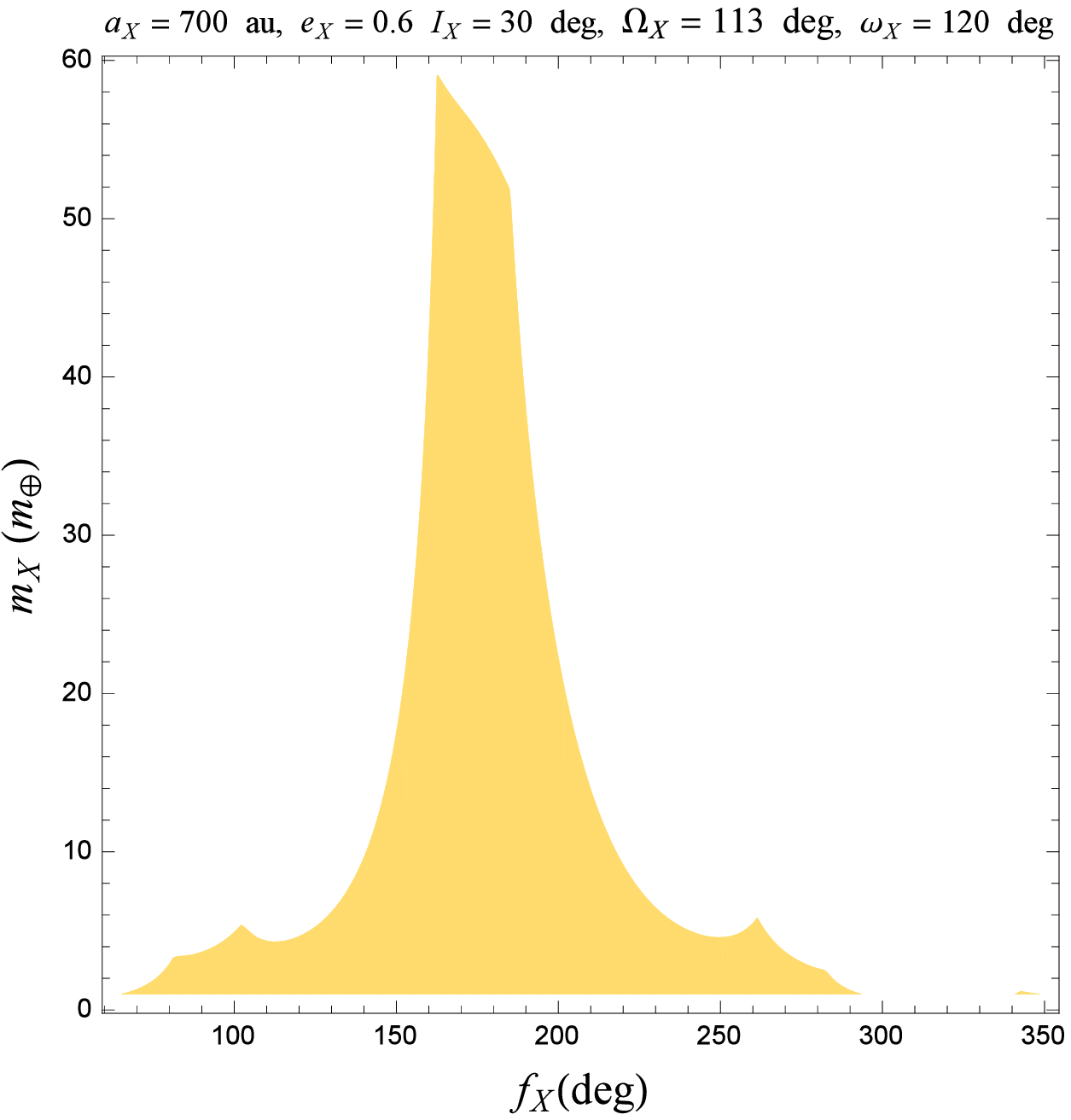} & \epsfysize= 6.2 cm\epsfbox{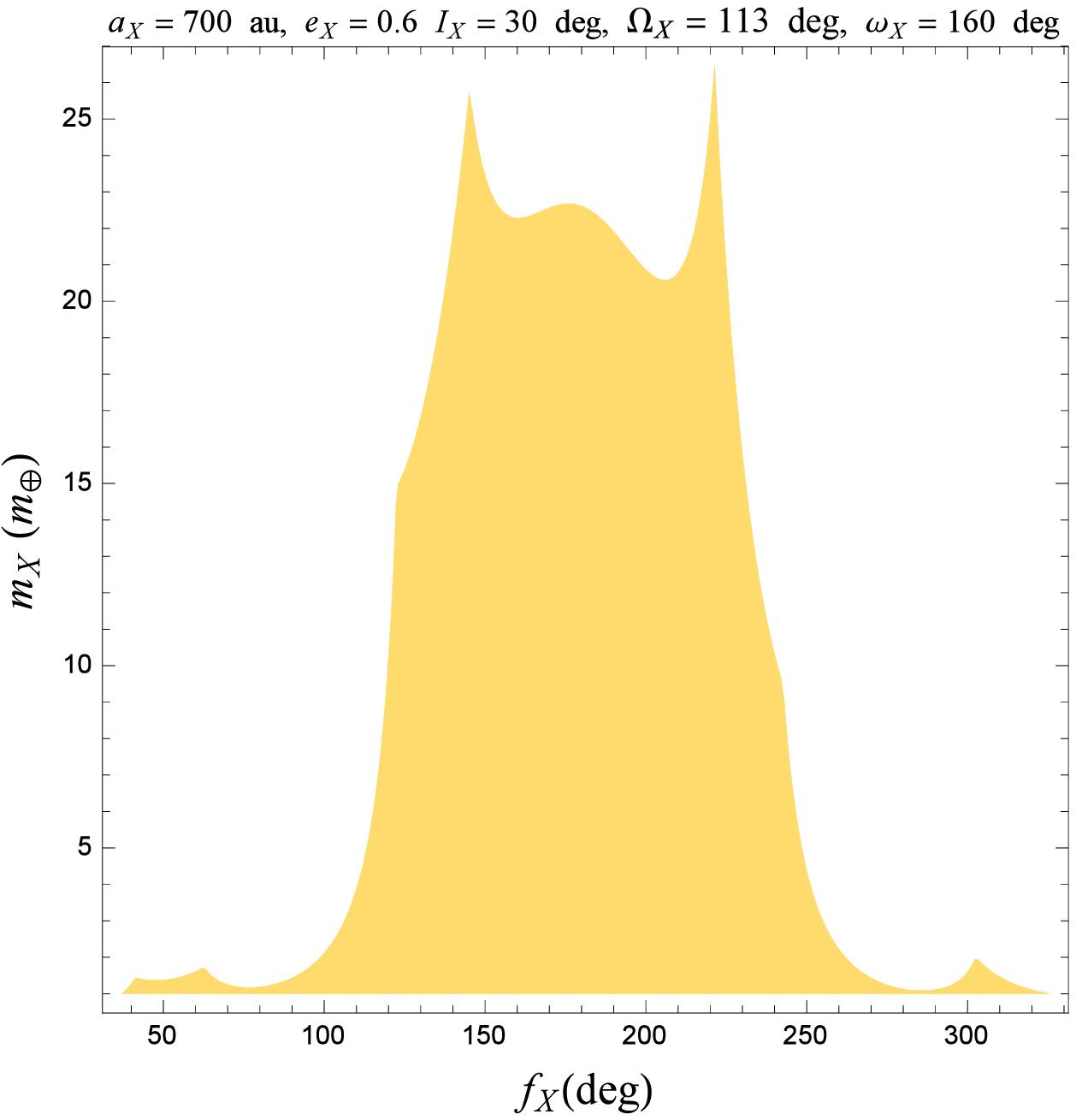} \\
\end{tabular}
}
}
\caption{Dynamical constraints on the mass of PX inferred from the orbital precessions of Saturn as in Section~\ref{costra}. No apriori limitations on the values of $m_\textrm{X}$ were applied. The range of variations of the Euler-type angular orbital parameters $I_\textrm{X},~\Omega_\textrm{X},~\omega_\textrm{X}$ of PX were taken from the relaxed parameter space of Table~\ref{largo}.}\label{masse1}
\end{figure*}

\begin{figure*}
\centerline{
\vbox{
\begin{tabular}{cc}
\epsfysize= 6.2 cm\epsfbox{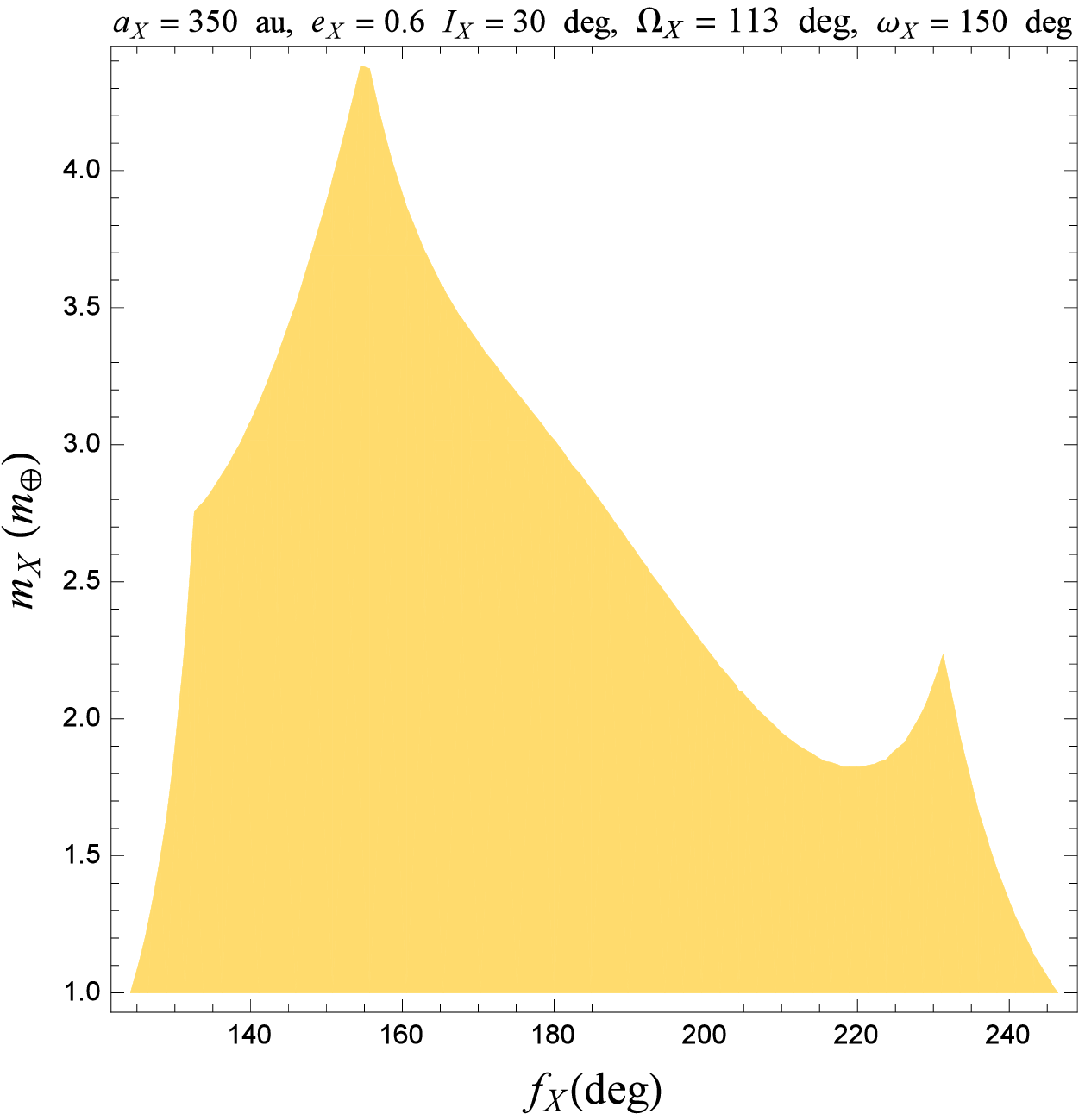} & \epsfysize= 6.2 cm\epsfbox{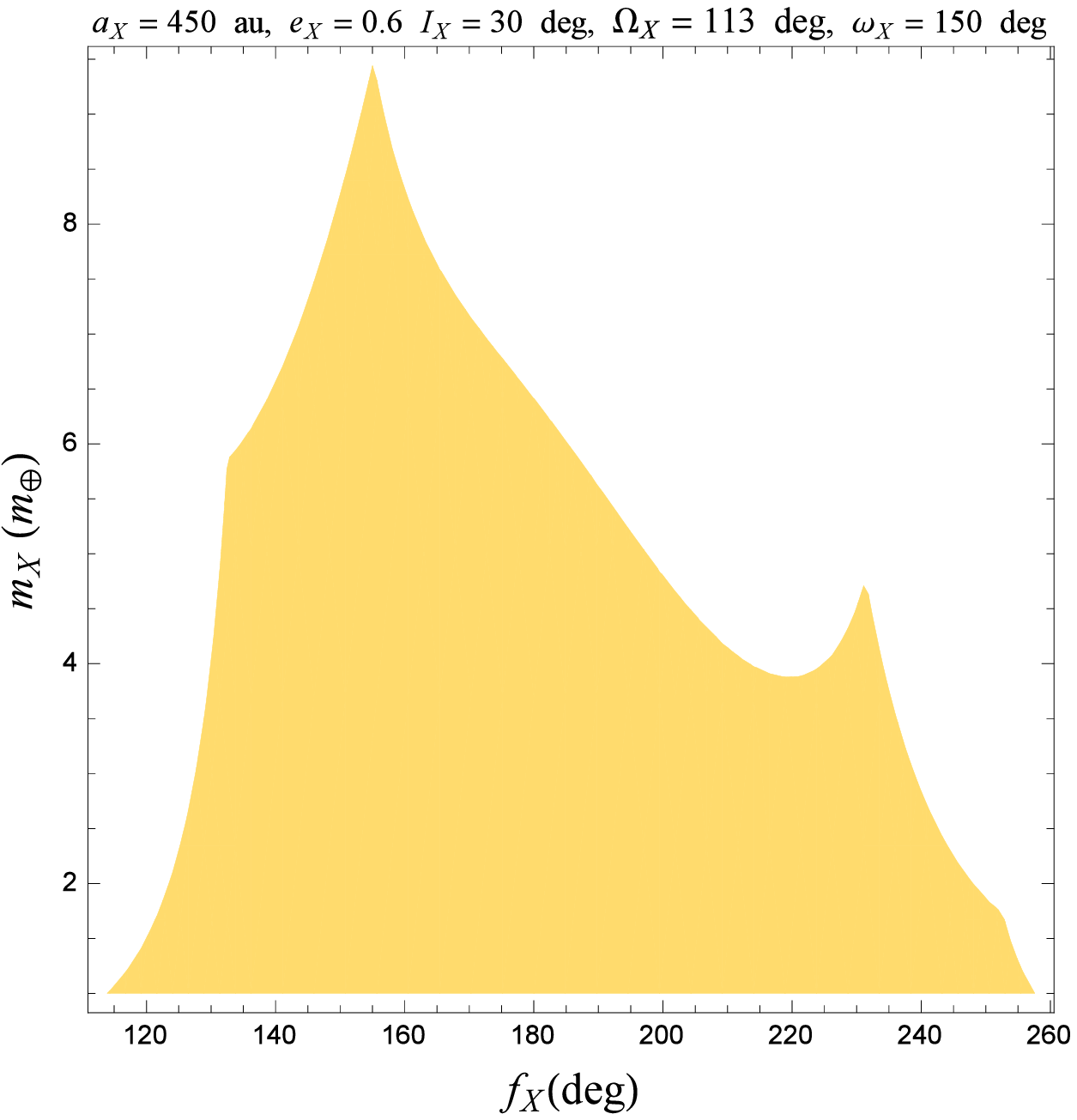} \\
\epsfysize= 6.2 cm\epsfbox{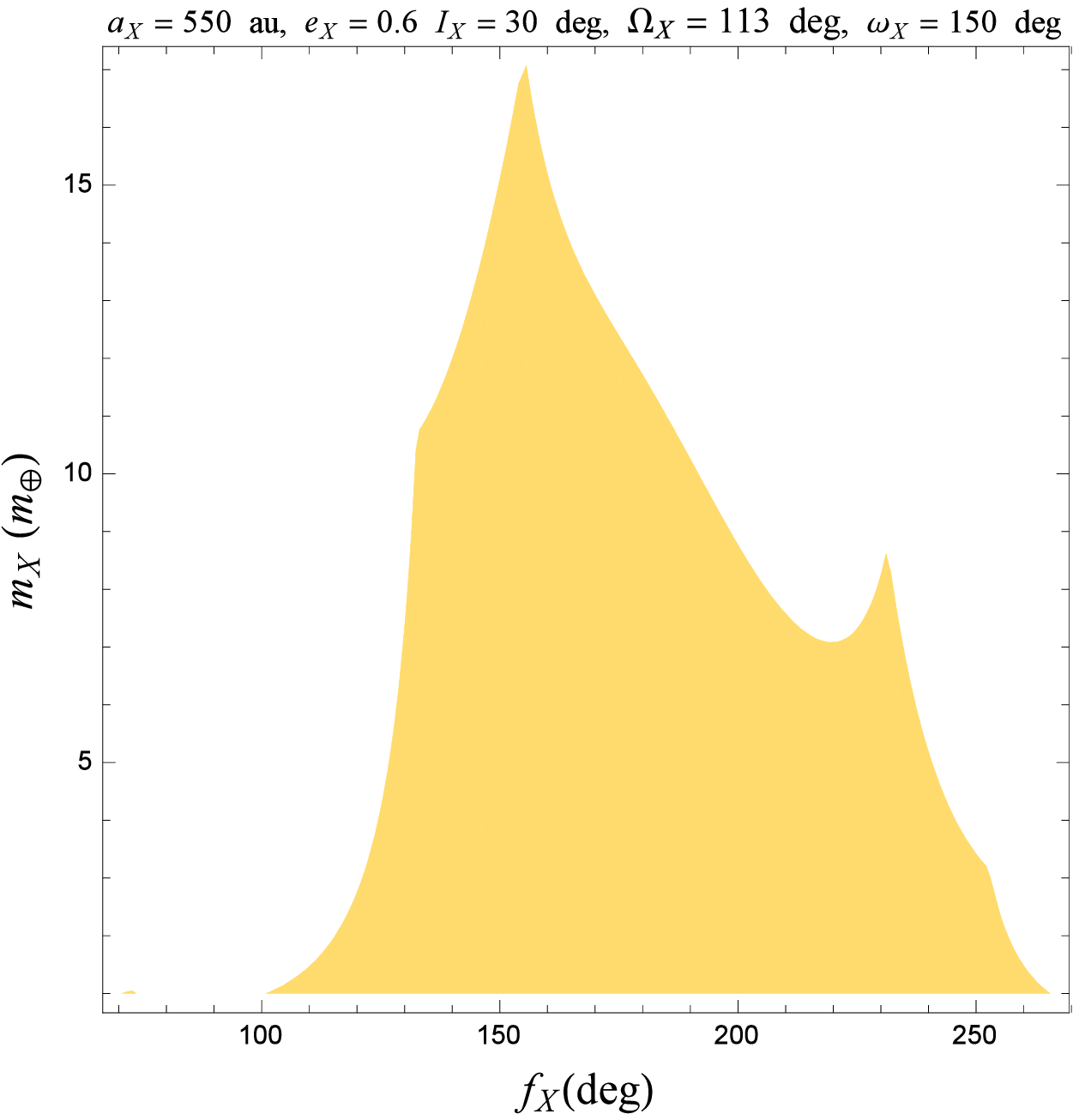} & \epsfysize= 6.2 cm\epsfbox{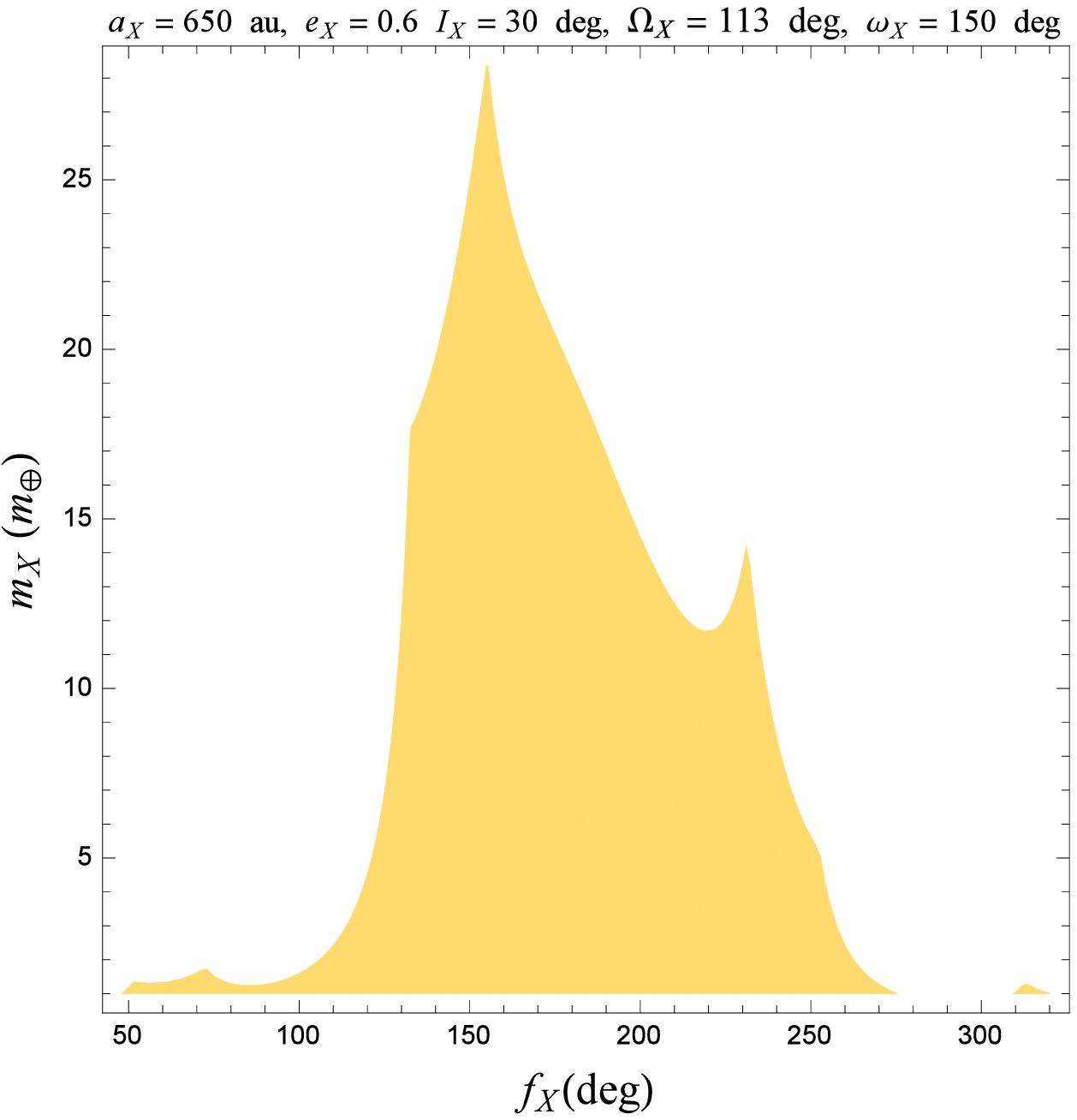} \\
\epsfysize= 6.2 cm\epsfbox{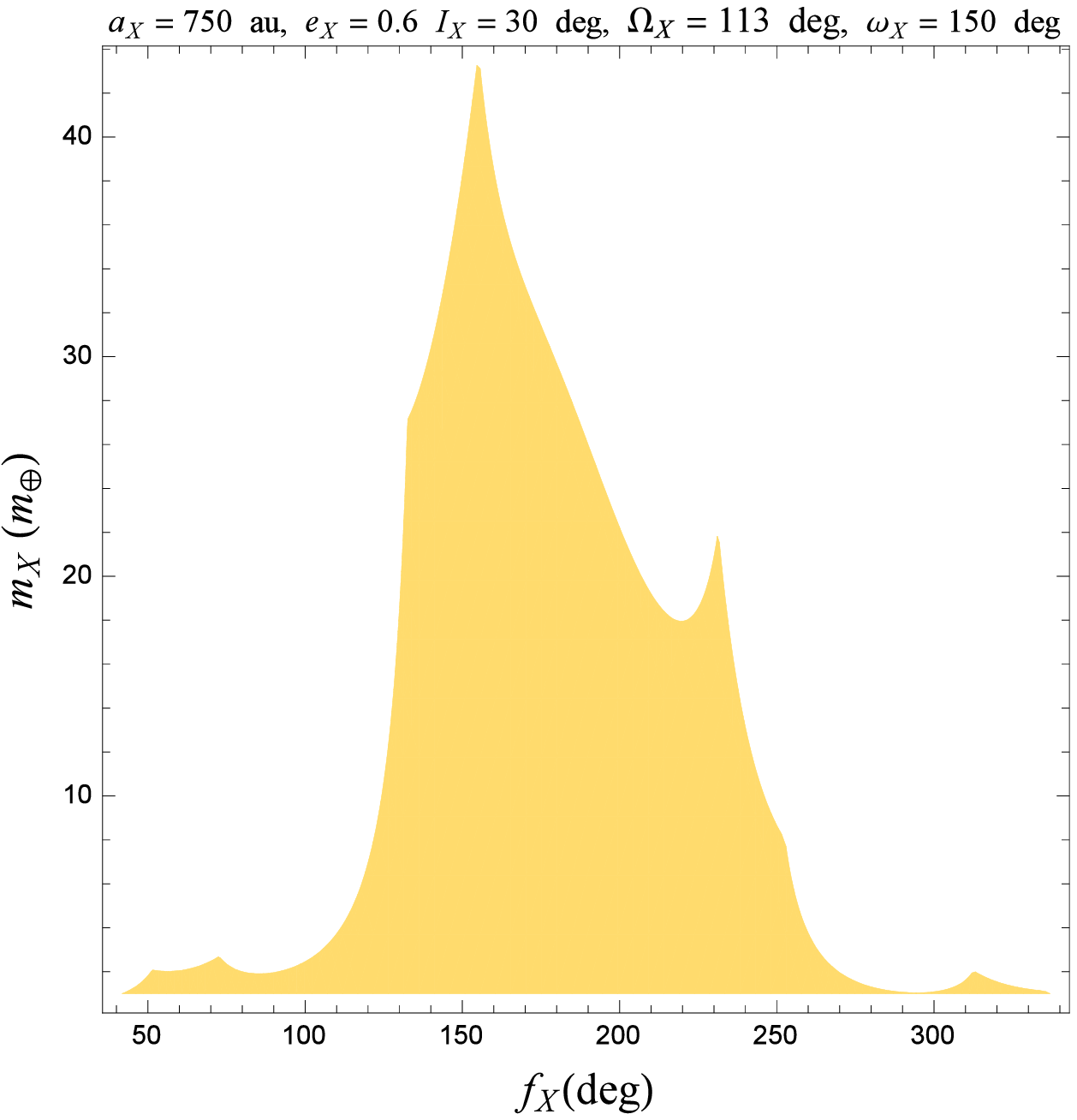} & \epsfysize= 6.2 cm\epsfbox{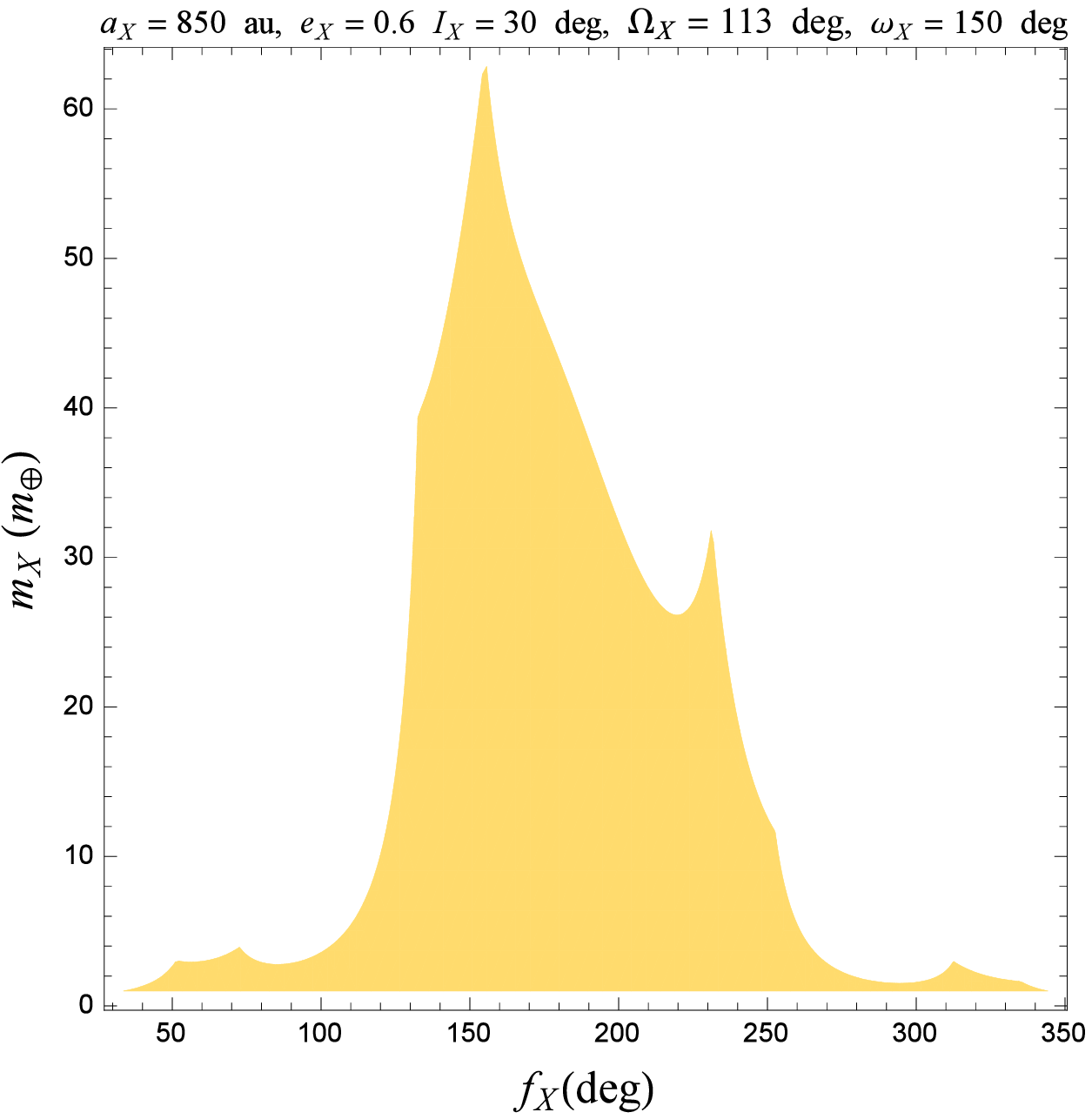} \\
\end{tabular}
}
}
\caption{Dynamical constraints on the mass of PX inferred from the orbital precessions of Saturn as in Section~\ref{costra}. No apriori limitations on the values of $m_\textrm{X}$ were applied. The inclination of the orbit of PX in space, its orientation in its plane and its shape  were taken from Table~\ref{kepelemsX} and $\Omega_\textrm{X}=113~\textrm{deg}$.}\label{masse2}
\end{figure*}

\begin{figure*}
\centerline{
\vbox{
\begin{center}
\epsfysize= 7 cm\epsfbox{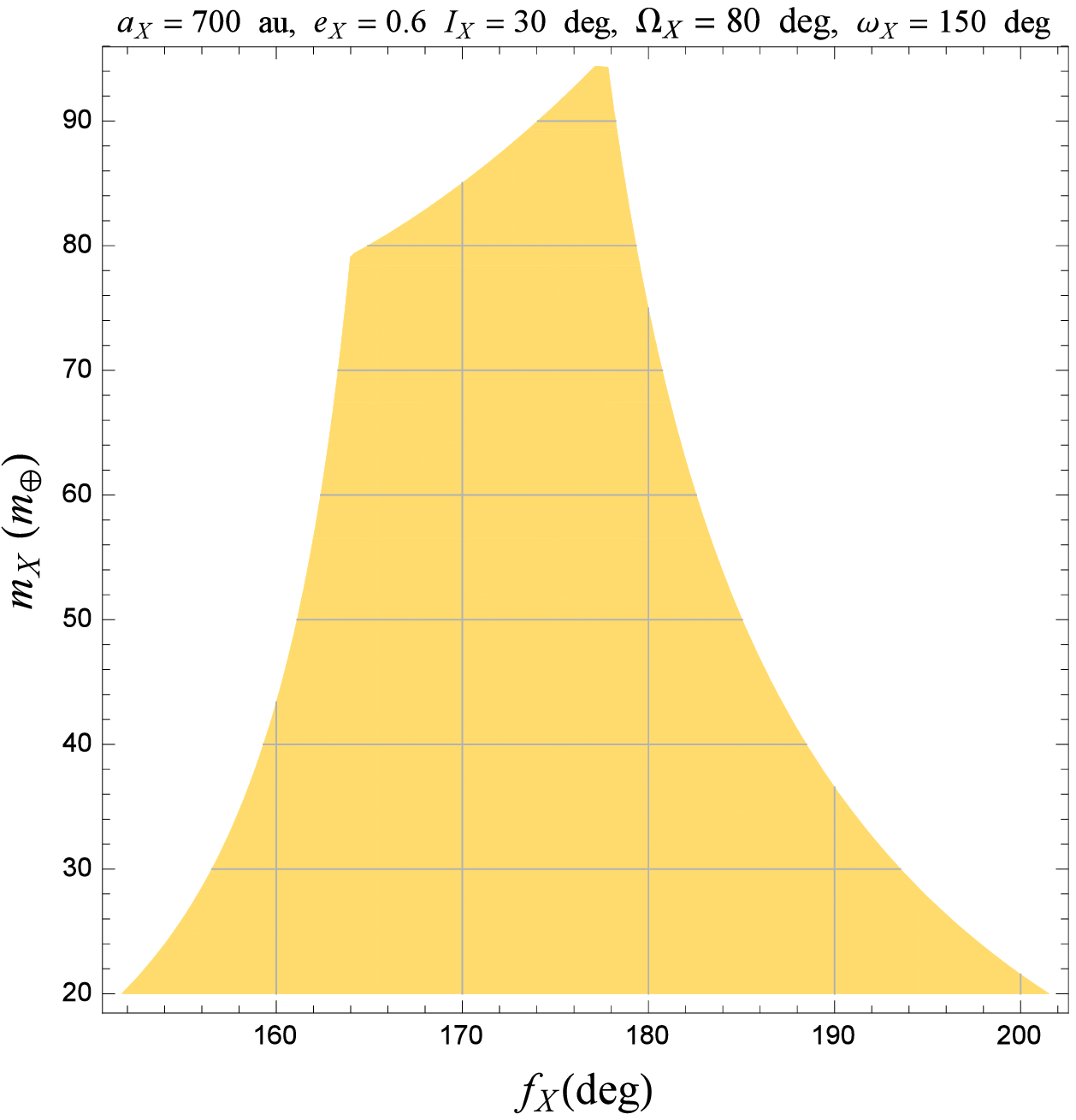}  \\
\epsfysize= 7 cm\epsfbox{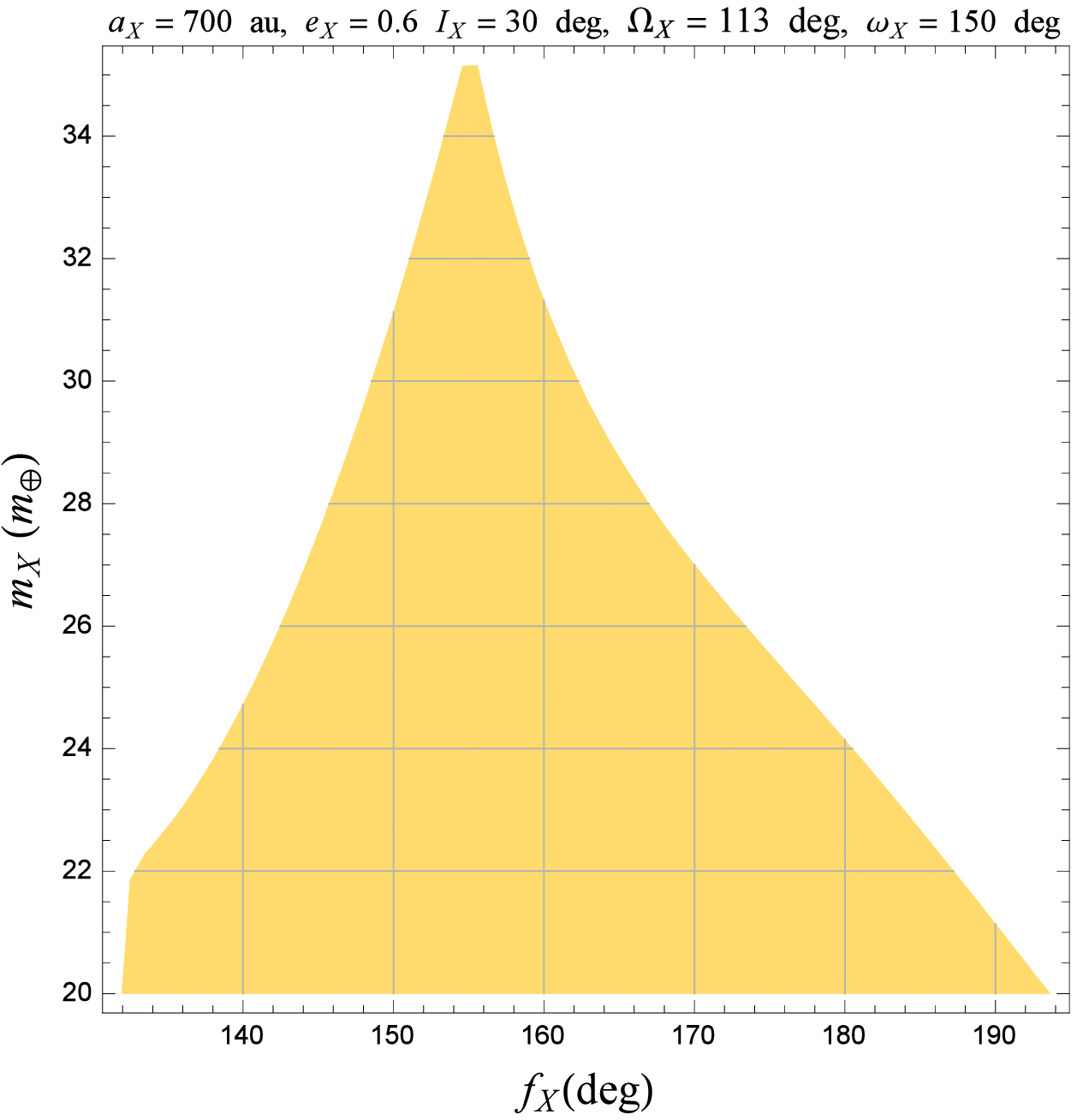}  \\
\epsfysize= 7 cm\epsfbox{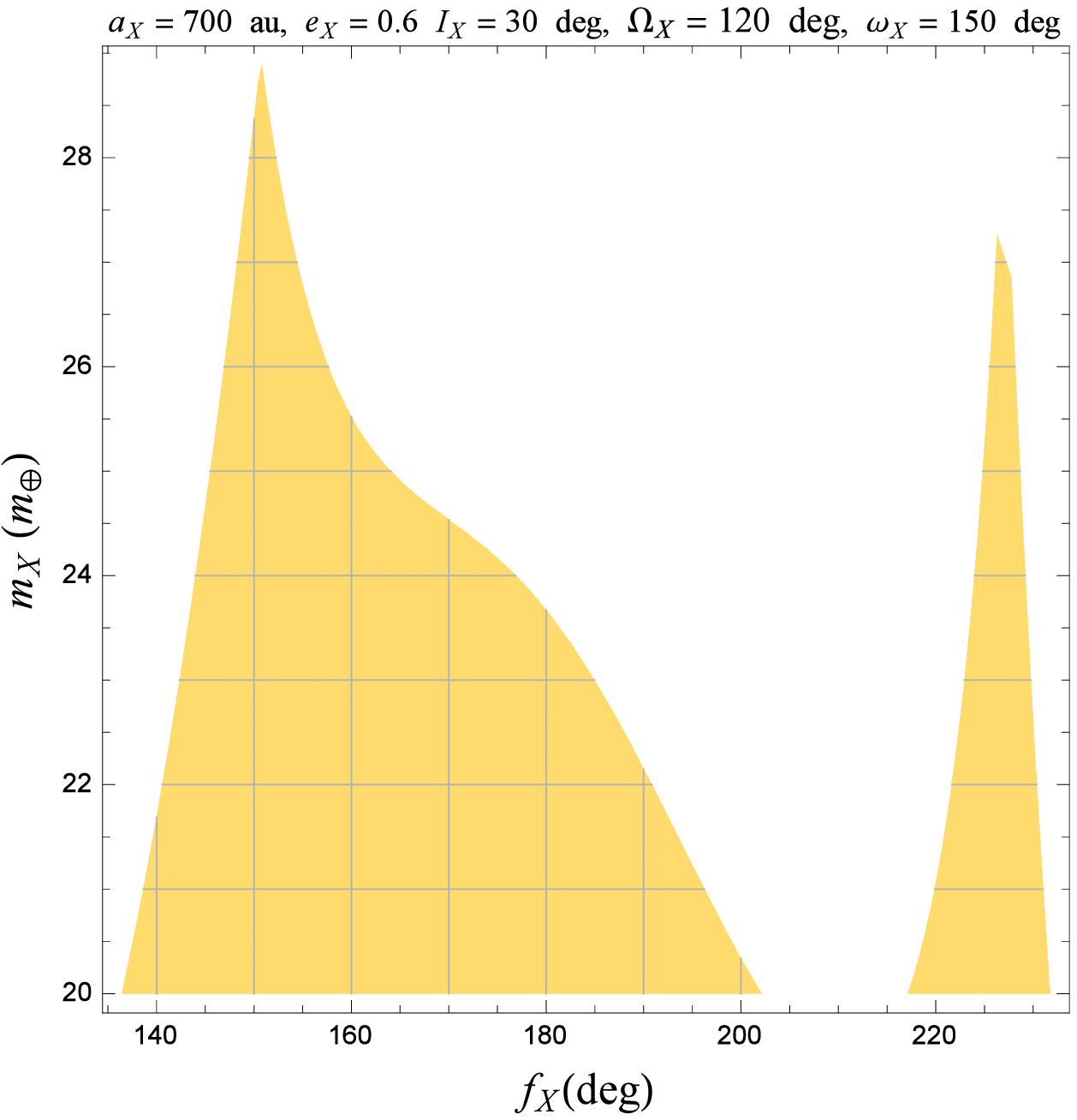}
\end{center}
}
}
\caption{\footnotesize{Dynamical constraints on the mass of PX inferred from the orbital precessions of Saturn as in Section~\ref{costra}. No apriori limitations on the values of $m_\textrm{X}$ were applied. The orbital configuration of Table~\ref{kepelemsX}, with $80~\textrm{deg}\leq\Omega_\textrm{X}\leq 120~\textrm{deg}$, is adopted. Values of $m_\textrm{X}$ even much larger than $20~m_\oplus$ are possible, despite they are confined in increasingly limited portions of the orbit of PX around its aphelion. They  strongly depend on the location of the orbital plane of PX in space.}}\label{massuccia}
\end{figure*}

\lb{massa}
In this Section, we address the problem of attempting to constrain $m_\textrm{X}$ treating it as a free parameter, without any a priori limitations on its possible values. To this aim, we will adopt the same strategy followed in Section~\ref{costra} in order to have preliminary bounds based solely on the orbital dynamics of Saturn. As such, we will treat  $m_\textrm{X}$ as one of the independent variables of \rfr{nodocazzo} and \rfr{pericazzo}; as far as the orbital configuration of PX is concerned, we will explore the relaxed parameter space of Table~\ref{largo}. Figures~\ref{masse1}~to~\ref{masse2} show the allowed regions  in the $\grf{f_\textrm{X},~m_\textrm{X}}$ plane for given values of the orbital elements of PX chosen within their range of variations according to  Table~\ref{largo}. It can be noted that, while there is room practically along the entire orbit for masses as little as $m_\textrm{X}\approx 1~m_\oplus$ or so, larger masses can be found, as expected, only in increasingly limited portions of the orbit, mainly concentrated around the aphelion. Moreover, the resulting upper limits for the admissible values of $m_\textrm{X}$, as provided by the Kronian orbital precessions, are often much greater than $20~m_\oplus$ envisaged by \citet{BroBaAJ2016}. For example, in the case of the preferred orbital configuration of Table~\ref{kepelemsX} with $\Omega_\textrm{X}=113~\textrm{deg}$, it turns out from Figure~\ref{massuccia} that masses larger than $20~m_\oplus$ are, indeed, quite possible, with a maximum of about $36~m_\oplus$ confined at $f_\textrm{X}\approx155~\textrm{deg}$. A strong dependence on $\Omega_\textrm{X}$ is observed, with $m_\textrm{X}^\textrm{max} = 92~m_\oplus$ at about $f_\textrm{X}\approx 178~\textrm{deg}$ for $\Omega_\textrm{X}=80~\textrm{deg}$.

Finally, it must be stressed that the aforementioned strategy is able to provide only upper limits on $m_\textrm{X}$. No physically meaningful lower limits $m_\textrm{X}^\textrm{min}$, with $m_\textrm{X}^\textrm{min} > 0$, can, instead, be obtained because of the error bars in the orbital precessions of Table \ref{perihelia} which, in principle, would allow also negative values for $m_\textrm{X}$. In other words, the allowed regions in Figures~\ref{masse1}~to~\ref{massuccia} may well extend down to negative values of $m_\textrm{X}$. 
\section{The potential impact on New Horizons: a preliminary sensitivity analysis}\lb{Niù}
After having encountered Pluto in the mid of July 2015, the spacecraft New Horizons \citep{2008SSRv..140....3S} is currently en route to the object $2014~\textrm{MU}_{69}$ of the Kuiper Belt, at about $43~\textrm{au}$, to be reached in 2019. New Horizons is endowed with a radio science apparatus, named  REX \citep{2008SSRv..140..217T}, which should allow to determine the spacecraft geocentric range $\rho$ with an accuracy better than $\sigma_\rho=10~\textrm{m}$ over 6 years after 2015, i.e. at geocentric distances to beyond $50~\textrm{au}$ \citep{2008SSRv..140...23F}.

\begin{figure*}
\centerline{
\vbox{
\begin{tabular}{cc}
\epsfysize= 5.0 cm\epsfbox{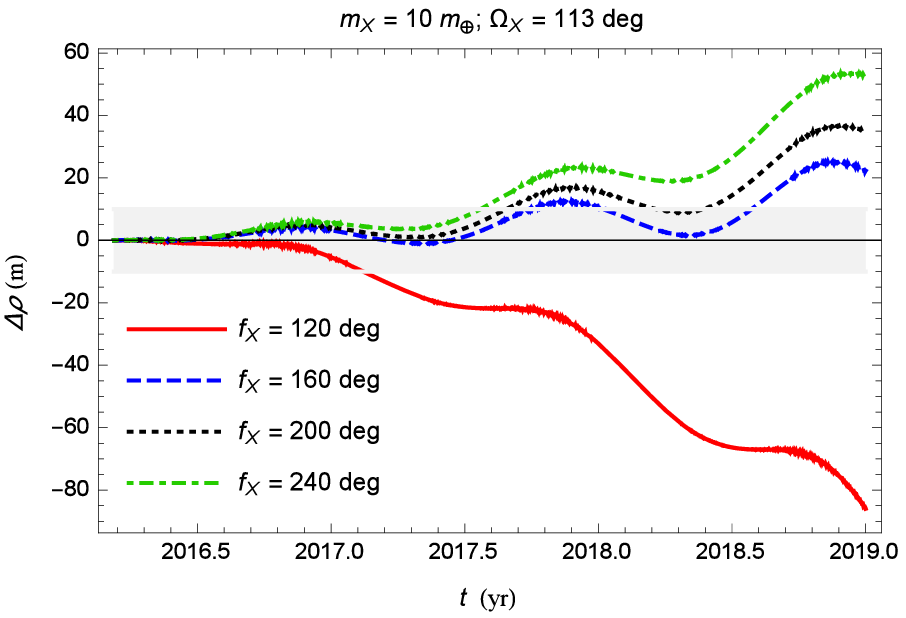} & \epsfysize= 5.0 cm\epsfbox{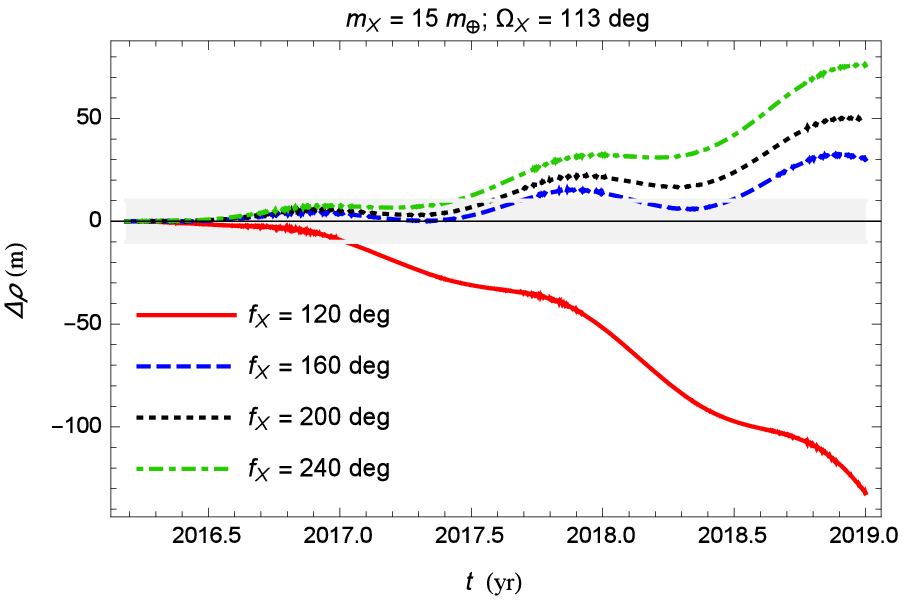} \\
\epsfysize= 5.0 cm\epsfbox{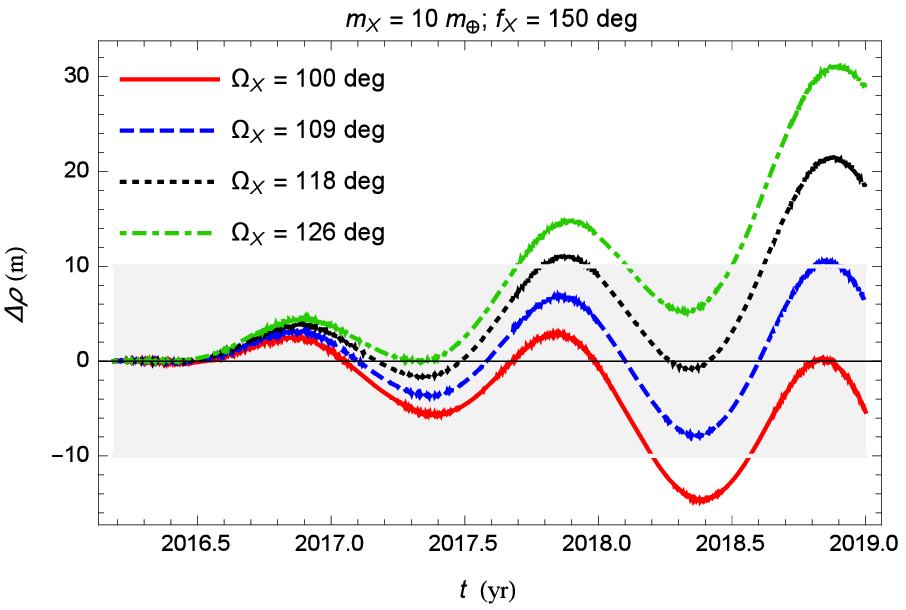} & \epsfysize= 5.0 cm\epsfbox{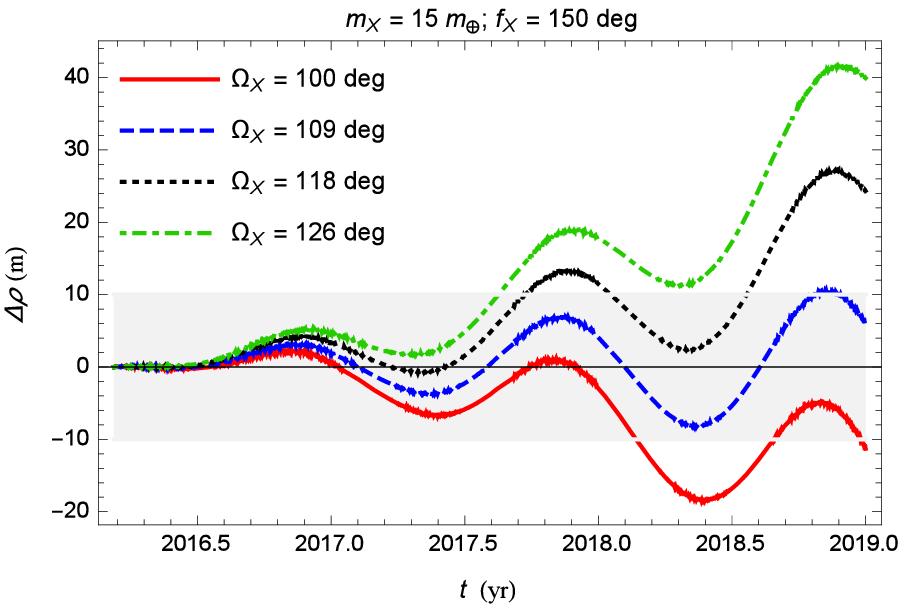} \\
\end{tabular}
}
}
\caption{Simulated Earth-New Horizons  range signature $\Delta\rho(t)$ due to a PX with \citep{BaBroAJ2016} $a_\textrm{X} =  700~\textrm{au},~e_\textrm{X}=0.6,~I_\textrm{X}=30~\textrm{deg},~\omega_\textrm{X}=150~\textrm{deg}$ for $m_\textrm{X}=10~m_\oplus~(\textrm{left column}),~m_\textrm{X}=15~m_\oplus~(\textrm{right column})$ obtained as a difference between two time series of $\rho(t)=\sqrt{\ton{x_\textrm{NH}(t) - x_\oplus(t)}^2 + \ton{y_\textrm{NH}(t) - y_\oplus(t)}^2 + \ton{z_\textrm{NH}(t) - z_\oplus(t)}^2}$ calculated by  numerically integrating the barycentric equations of motion in  Cartesian rectangular coordinates of all the major bodies of the Solar System and New Horizons over 3 years (2016--2019) with and without the perturber's potential $U_\textrm{X}$ of \rfr{hogg} \citep{1991AJ....101.2274H}.  The initial conditions, corresponding to February 1, 2016, were retrieved from the WEB interface HORIZONS by NASA JPL; they were the same for both the integrations which share also the entire standard N--body dynamical models to the first post-Newtonian level. Upper row: $f_\textrm{X}$ is assumed as a free parameter varying in the range of Figure~\ref{nodo_peri}, while the perturber's node is kept fixed at $\Omega_\textrm{X} = 113$ deg. Lower row: $\Omega_\textrm{X}$ is assumed as a free parameter varying in the range seemingly favored by \citet{BaBroAJ2016}, while the perturber's true anomaly is kept fixed at $f_\textrm{X} = 150$ deg, within the allowed regions $\mathcal{D}$ of Figure~\ref{nodo_peri}. The arrival at  $2014~\textrm{MU}_{69}$, at about $43~\textrm{au}$ from the Sun, is scheduled for the beginning of 2019. During the cruise to it, the probe should coast mainly unperturbed, apart from the putative action of PX and a relatively small number of orbital maneuvers. The expected  ranging precision is better than $\sigma_\rho=10$ m up to $50~\textrm{au}$ \citep{2008SSRv..140...23F}. It defines the light gray shaded area.}\label{NH_range}
\end{figure*}

\begin{figure*}
\centerline{
\vbox{
\begin{tabular}{cc}
\epsfysize= 5.0 cm\epsfbox{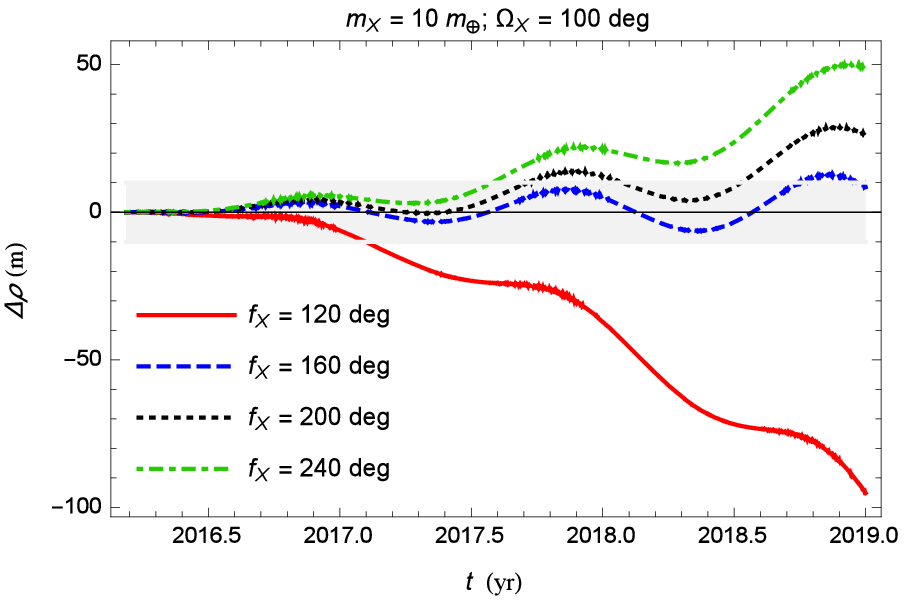} & \epsfysize= 5.0 cm\epsfbox{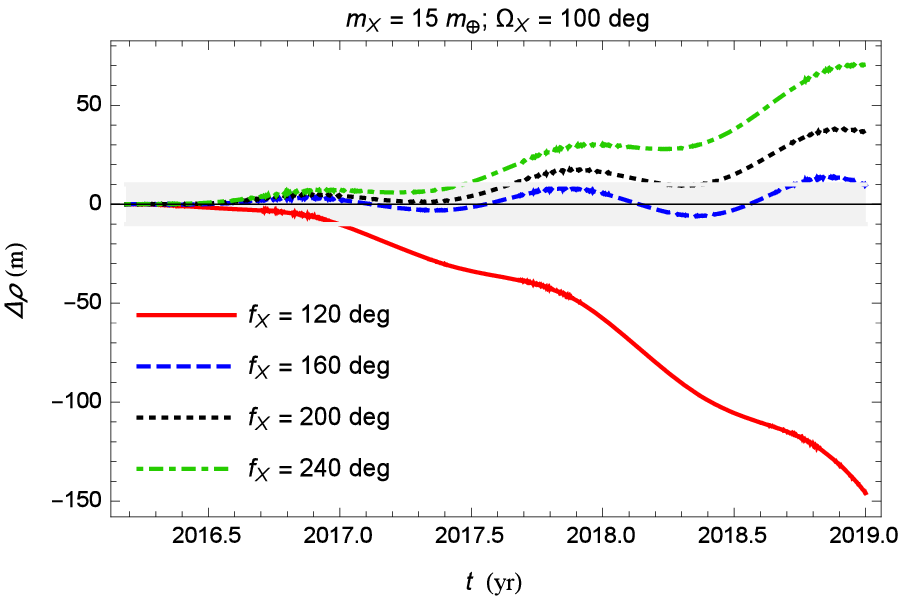} \\
\epsfysize= 5.0 cm\epsfbox{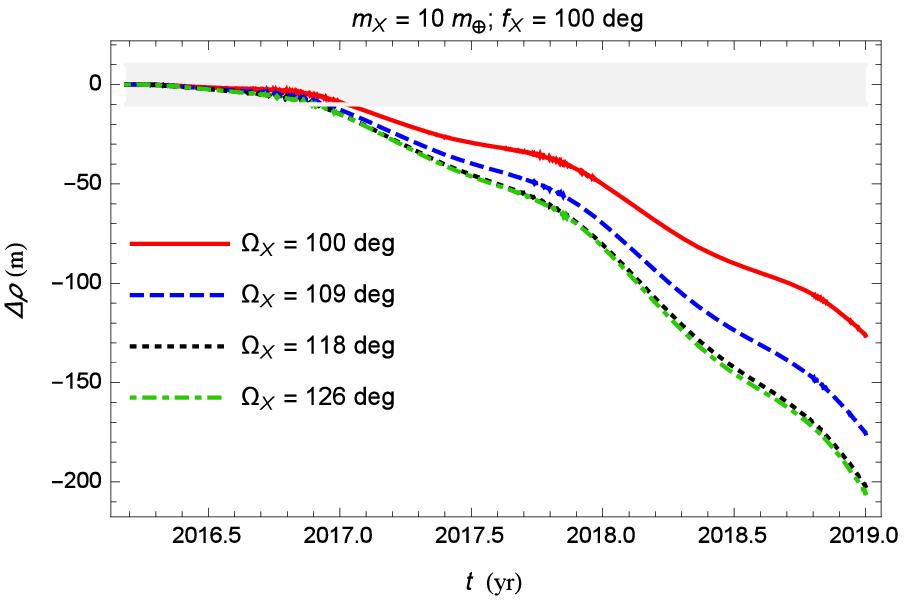} & \epsfysize= 5.0 cm\epsfbox{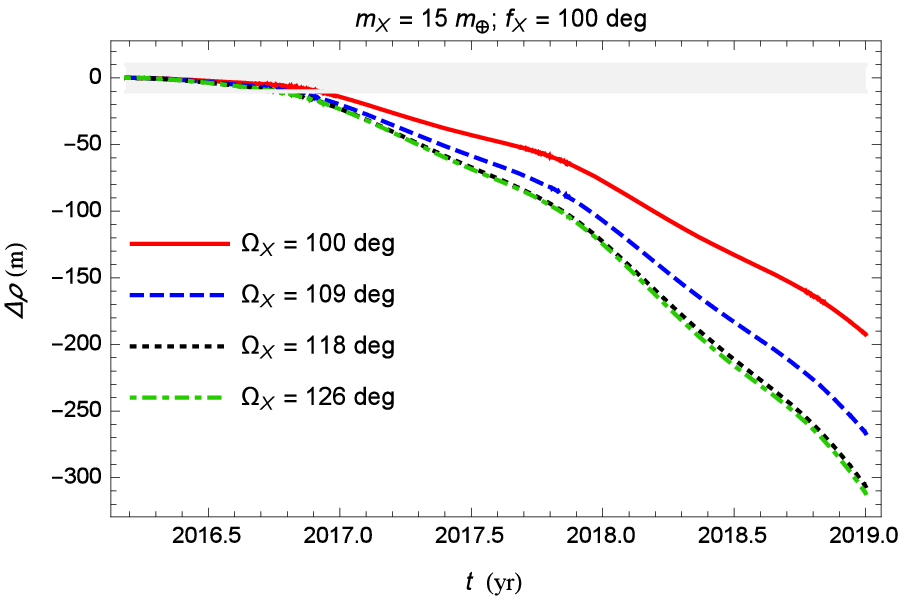} \\
\end{tabular}
}
}
\caption{As in Figure~\ref{NH_range}, but with $\Omega_\textrm{X} = 100~\textrm{deg}$ (upper row) and $f_\textrm{X} = 100~\textrm{deg}$ (lower row).}\label{NH_range2}
\end{figure*}

Following an idea put forth in \citet{2013CeMDA.116..357I}, its range residuals may, in principle, be used to put tighter constraints on the admissible locations of PX in the near--mid future; later, \citet{2015PhRvD..92j4048B} proposed a dedicated deep-space mission aimed to test the currently accepted gravitational laws up  to $100~\textrm{au}$. For the sake of simplicity, we will investigate only the preferred orbital configuration by \citet{BaBroAJ2016}.
Figures~\ref{NH_range}~to~\ref{NH_range2} depict the simulated range perturbation induced by a hypothetical distant perturber with the physical and orbital features of Table~\ref{kepelemsX} within approximately the ranges of admissible values for $f_\textrm{X},~\Omega_\textrm{X}$ set by Figure~\ref{nodo_peri}. Incidentally, the observable probed here,i.e. the Earth--spacecraft distance, is of similar nature as the one used by \citet{2016A&A...587L...8F} with Cassini. It can be noted that the peak-to-peak amplitudes range from about $\approx 50-60~\textrm{m}$ up to $\approx 150-300~\textrm{m}$ depending on the orbital geometry of PX.
In obtaining each curve of Figures~\ref{NH_range}~to~\ref{NH_range2}, we simultaneously integrated the barycentric equations of motion of all of the currently known major bodies of the Solar System  in Cartesian rectangular coordinates over a time span 3 years long starting from February 1, 2016  with and without PX. The dynamical accelerations modeled in both the integrations include the general theory of relativity to the first post-Newtonian level, and all the major known Newtonian effects such as the Sun's oblateness, pointlike mutual perturbations by the eight planets and the three largest asteroids, two massive rings accounting for the minor asteroids  and the Kuiper Belt's objects. Then, from the so--obtained solutions of the perturbed and unperturbed equations of motion, we numerically produced differential time series $\Delta\rho\ton{t}$ of the Earth--spacecraft range $\rho\ton{t}$. The amplitudes of such simulated signals can be compared to $\sigma_\rho$ in order to gain insight about the potential ability of New Horizons to constrain/detect its existence in future. It turns out that, even over a limited time span as that required to reach $2014~\textrm{MU}_{69}$, the expected range signatures due to the hypothesized PX could be as large as some hundreds of meters.
On the other hand, caution is advised because of the orbital maneuvers, not taken into account in the present tentative simulations, which will require accurate modeling. Moreover, the present investigation should be considered just as a preliminary sensitivity analysis
based on the expected precision of the probe's measurements: actual overall accuracy will be finally determined by several sources of systematic uncertainties like, e.g., the heat dissipation from the Radioisotope Thermoelectric Generator (RTG) and the ability in accurately modeling the orbital maneuvers. As such, by no means, it is meant to replace a full covariance study based on extensive simulations of the ranging data from New Horizons and their reduction with accurate models including also PX itself, which is outside the scopes of this paper. Nonetheless, the possibility of using the telemetry of New Horizons seems appealing, and would deserve further, more detailed and dedicated studies which, hopefully, may be prompted by the present one; as a  potentially interesting topic for further investigations, we mention\footnote{I thank H. Beust for such a suggestion.} the possibility of jointly using the latest years of Cassini data as well to achieve a better fit.
\section{Summary and conclusions}\lb{fine}
 Combining the recent experimental determinations of the supplementary perihelion and node rates of Saturn, quoted in our Table~\ref{perihelia}, with our analytical formulas of \rfrs{nodocazzo}{Enne} and \rfrs{pericazzo}{Hacca} for the theoretical apsidal and draconitic precessions induced by the putative new planet of the Solar System under consideration  allows us to tentatively constrain its position along its orbit for selected scenarios of such a hypothesized body characterized by given values of its mass and its orbital elements. We stress that, since its gravitational tug on the known bodies of the Solar System was neither modelled nor estimated in the ephemerides which the extra-precessions of Saturn used by us come from, our analyses should be deemed as proof-of-principle investigations to encourage future, more rigorous analyses. The present study, which should be supplemented by dedicated data reductions by the international teams producing the planetary ephemerides in which the gravitational action of PX should be explicitly modeled, can be useful in either further refining the scenarios describing the shepherding action on the KBOs by the hypothesized trans-Neptunian super-Earth and better addressing further observational campaigns. In this respect, the attempt by \citet{2016A&A...587L...8F} is a first valuable step in the right direction. Nonetheless, the preliminary bounds resulting from our analysis are reasonable and, to a non-negligible extent, in agreement with other studies making use of different observables.

As far as the originally proposed configuration of PX \citep{BaBroAJ2016}, displayed in our Table~\ref{kepelemsX}, is concerned, a direct and meaningful comparison with the results by \citet{2016A&A...587L...8F} is possible. Indeed, they too considered such a specific scenario by choosing the central value of the perturber's node within the range of variation suggested in the literature for it. \citet{2016A&A...587L...8F}, who produced post-fit residuals of the Earth-Saturn range in a dedicated data reduction of real range measurements covering the decade 2004-2014 by  explicitly modeling PX, obtained disjointed allowed regions for it. They are $85~\textrm{deg}\lesssim f_\textrm{X}\lesssim 235~\textrm{deg}$, which includes also their claimed most probable interval of values $f_\textrm{X} = 117.8^{+11}_{-10}~\textrm{deg}$, and $260~\textrm{deg}\lesssim f_\textrm{X}\lesssim 295~\textrm{deg}$. For the same choice of the value of the node of the originally proposed model of PX, we obtained only one continuous allowed region given by $130~\textrm{deg}\lesssim f_\textrm{X}\lesssim 240~\textrm{deg}$, as depicted by our Figure~\ref{nodo_peri}. It is almost entirely included in the largest allowed region by \citet{2016A&A...587L...8F}, whose most probable range is, on the other hand, excluded by us: our Figure~\ref{miaellisse} clearly illustrates this fact. Instead, the location of the orbital plane in space of the hypothesized distant perturber at hand, assumed as an unconstrained parameter within its admitted range of variation in its earliest scenario, is left substantially undetermined by our analysis.
At present, the current accuracy level in planetary ephemerides does not seem to allow to remove the degeneracy in the node of PX. Future improvements, which should come from the analysis of more radiometric data from Cassini, doomed to fatally plunge into Saturn's atmosphere in late 2017, may break it, at least to a certain extent. To this aim, we remark that a future determination of the supplementary precessions of further orbital elements of Saturn by the astronomers producing the planetary ephemerides would be a quite valuable and highly desirable step to infer tighter constraints on both the location of PX along its orbit and of its orbital plane in space.

The investigation of the relaxed versions by \citet{BroBaAJ2016} of the preferred orbital configuration of PX \citep{BaBroAJ2016}, reproduced in our Tables~\ref{largo}~to~\ref{uffa}, makes more difficult to present our results in an unitary way and comparing them with other findings in the literature. A common feature of all our analyses performed by sampling the enlarged parameter space of the putative new planet and shown by Figures~\ref{cubo0}~to~\ref{RADEC1} is that the simultaneous use of both the node and the perihelion precessions of Saturn allows us to obtain always single continuous  areas or volumes of permitted parameters of PX by allowing to discard  smaller disjointed regions which the use of only one Kronian orbital precession at a time would force us to keep. Moreover, the resulting intervals of permitted values of the true anomaly of PX are always essentially centered on its aphelion, spanning roughly from $\Delta f_\textrm{X}\approx 30~\textrm{deg}$ for certain specific values characterizing some of the boundaries of the perturber's parameter space up to about $\Delta f_\textrm{X}\approx 100-125~\textrm{deg}$ for most of the remaining domain. Importantly, our strategy is able to reproduce the range $142~\textrm{deg}\lesssim f_\textrm{X}\lesssim 164~\textrm{deg}$ corresponding to the relatively restricted portion of the parameter space of PX, reproduced in our Table~\ref{HoPa}, picked up in \citet{2016AJ....152...94H} by comparing  their own allowed regions with the continuous sinusoidal strip by \citet{BroBaAJ2016} in the plane of the sky.

We also attempted to put constraints on the upper limits of the mass of PX for given orbital configurations retrieved from \citet{BaBroAJ2016,BroBaAJ2016} by treating it as a free, apriori unconstrained parameter. Our Figures~\ref{masse1}~to~\ref{massuccia} show that the nodal and apsidal precessions of Saturn actually leave room for larger values of $m_\textrm{X}$ than the upper limit of $20~m_\oplus$ suggested by \citet{BroBaAJ2016}, but they are confined in quite small orbital portions nearly around the aphelion. In particular, in the case of the preferred orbital geometry by \citet{BaBroAJ2016}, the largest allowed value for the mass of PX is $m_\textrm{X} \approx 36~m_\oplus$ which could stay only at about $f_\textrm{X}\approx 155~\textrm{deg}$ if $\Omega_\textrm{X} = 113~\textrm{deg}$. Instead, if $\Omega_\textrm{X} = 80~\textrm{deg}$, a mass as large as $m_\textrm{X} \approx 92~m_\oplus$ at $f_\textrm{X}\approx 178~\textrm{deg}$ is still admissible. Since the supplementary orbital precessions of Saturn are statistically compatible with zero, our strategy is not able to provide physically meaningful lower limits on $m_\textrm{X}$, being unphysical negative values of $m_\textrm{X}$ admissible.

Finally, we mention also the potentially appealing opportunity of using in the mid future the precise telemetry of the ongoing New Horizons mission, which is currently traveling at about 35 au and whose range would be perturbed up to some hundreds of meters by PX, or, hopefully, of some missions proposed to test the currently accepted laws of the gravitational interaction in the far reaches of the Solar System, at about 100 au. Caution is, however, advised since it remains to be seen whether it will be actually possible to deal with smooth orbital arcs long enough for New Horizons by effectively modeling the non-gravitational perturbations affecting it and its orbital maneuvers. Perhaps, also the latest data from Cassini might be jointly used to improve the fit.
\appendix
\section{Tables and Figures}

\begin{table*}
\caption{Nominal orbital elements originally proposed by \citet{BaBroAJ2016} for a distant pointlike perturber PX of mass $m_\textrm{X}=10~m_\oplus$ in order to explain reasonably well the observed clustering of  either the apsidal and nodal lines of the orbits of some of the known distant KBOs. The values for the inclination $I_\textrm{X}$ and the argument of perihelion $\omega_\textrm{X}$ are referred to the ecliptic. The  values for the semimajor axis $a_\textrm{X}$ and the eccentricity $e_\textrm{X}$ proposed by \citet{BaBroAJ2016} correspond to a perihelion distance  $q_\textrm{X}=280$ au and an aphelion distance $Q_\textrm{X}=1120$ au. As far as the longitude of the ascending node of PX is concerned, it seems that \citet{BaBroAJ2016} favor the range $\Omega_\textrm{X} =113\pm 13~\textrm{deg}$, based on the orientation of the orbital planes of the KBOs considered.  }
\label{kepelemsX}
\centering
\begin{tabular}{llll}
\noalign{\smallskip}
\hline
$a_\textrm{X}$ (au) & $e_\textrm{X}$ & $I_\textrm{X}$ (deg) & $\omega_\textrm{X}$ (deg) \\
\hline
$700$ & $0.6$ & $30$ & $150$  \\
\hline
\end{tabular}
\end{table*}
\begin{table*}
\caption{Enlarged parameter space of PX, valid for $5~m_\oplus\leq m_\textrm{X}\leq 20~m_\oplus$, as per Section 4 of \citet{BroBaAJ2016} and Section 3 of \citet{2016AJ....152..126B}. It must be remarked that, in their Section 2, \citet{BroBaAJ2016} seemingly used the displayed formulas for $a_\textrm{X},~e_\textrm{X}$ with $m_\textrm{X} = 10,20~m_\oplus$ by keeping the inclination and the argument of perihelion fixed to $I_\textrm{X} = 30~\textrm{deg},~\omega_\textrm{X} = 150~\textrm{deg}$. Instead, in their Section 3, \citet{BroBaAJ2016} adopted the displayed ranges for $I_\textrm{X},~\omega_\textrm{X}$ with the fixed values $m_\textrm{X}=10~m_\oplus,~a_\textrm{X}=700~\textrm{au},~e_\textrm{X} = 0.6$. For another PX scenario proposed by \citet{BroBaAJ2016}, see Table~\ref{uffa}.}
\label{largo}
\centering
\begin{tabular}{lllllll}
\noalign{\smallskip}
\hline
  $a_\textrm{X}~(\textrm{au})$ & $e_\textrm{X}$ & $I_\textrm{X}~(\textrm{deg})$ & $\Omega_\textrm{X}~(\textrm{deg})$ & $\omega_\textrm{X}~(\textrm{deg})$\\
\hline
  $[200 + 30~\rp{m_\textrm{X}}{m_\oplus},~600 + 20~\rp{m_\textrm{X}}{m_\oplus}]$ & $0.75 - \qua{\rp{\ton{250 + 20~\rp{m_\textrm{X}}{m_\oplus}}~\textrm{au}}{a_\textrm{X}}}^8$ & $[22,~40]$ & $[80,~120]$ & $[120,~160]$  \\
\hline
\end{tabular}
\end{table*}
\clearpage
\begin{table*}
\caption{Orbital and physical parameters of PX as proposed at the end of Section 3 of \citet{BroBaAJ2016} in the framework of the final full three-dimensional simulation with a large number of particles with randomly chosen starting angles.}
\label{uffa}
\centering
\begin{tabular}{llllll}
\noalign{\smallskip}
\hline
$m_\textrm{X}~(m_\oplus)$  & $a_\textrm{X}~\textrm{(au)}$  & $e_\textrm{X}$ & $I_\textrm{X}~\textrm{(deg)}$  & $\omega_\textrm{X}~\textrm{(deg)}$  & $\Omega_\textrm{X}~\textrm{(deg)}$ \\
\hline
$10$ & $700$ & $0.6$ & $30$ & $0$ & $0$  \\
\hline
\end{tabular}
\end{table*}
\clearpage
\begin{table*}
\footnotesize{\caption{Most recent determinations $\Delta\dot\varpi_\textrm{obs},~\Delta\dot\Omega_\textrm{obs}$ of the supplementary secular precessions  of the longitudes of the perihelion $\varpi$ and the ascending node $\Omega$ of Saturn obtained with the EPM2011 \citep{2013MNRAS.432.3431P}, INPOP10a \citep{2011CeMDA.111..363F} and INPOP15a \citep{2015CeMDA.123..325F, 2016arXiv160100947F} ephemerides by processing data records spanning about $\Delta t = 100~\textrm{yr}$ which include also accurate radiometric measurements from the Cassini spacecraft. The dynamical action of  the putative distant object hypothesized by \citet{BaBroAJ2016}  was not modeled in any of the ephemerides used. The units are milliarcseconds per century (mas cty$^{-1}$). In principle, $\Delta\dot\varpi_\textrm{obs},~\Delta\dot\Omega_\textrm{obs}$ account for any unmodelled/mismodelled dynamical effects in the ephemerides used for their determination.}
\label{perihelia}
\centering
\begin{tabular}{llll}
\noalign{\smallskip}
\hline
 & EPM2011 \citep{2013MNRAS.432.3431P} & INPOP10a \citep{2011CeMDA.111..363F} & INPOP15a \citep{2016arXiv160100947F} \\
\hline
$\Delta\dot\varpi_{\rm obs}$  & $-0.32\pm 0.47$ & $0.15\pm 0.65$ & $0.625\pm 2.6 $ \\
$\Delta\dot\Omega_{\rm obs}$  & $-$ & $-0.1\pm 0.4$ & $-$ \\
\hline
\end{tabular}}
\end{table*}
\begin{table}
\caption{Selected physical and orbital parameters of PX from Zone-2 of Table 1 in \citet{2016AJ....152...94H}. Here, ${\mathcal{M}}_\textrm{X}$ denotes the mean anomaly.}
\label{HoPa}
\centering
\begin{tabular}{lllllll}
\noalign{\smallskip}
\hline
$m_\textrm{X}~(m_\oplus)$ & $a_\textrm{X}~(\textrm{au})$ & $e_\textrm{X}$ & $I_\textrm{X}~(\textrm{deg})$ & $\Omega_\textrm{X}~(\textrm{deg})$ &  $\omega_\textrm{X}~(\textrm{deg})$ & ${\mathcal{M}}_\textrm{X}~(\textrm{deg})$\\
\hline
$17.7^{+8.4}_{-9.1}$ & $478.7^{+70.5}_{-91.7}$ & $0.5^{+0.1}_{-0.1}$ & $32.7^{+5.3}_{-6.9}$ & $97.8^{+14.8}_{-14.7}$ & $138.7^{+14.1}_{-13.0}$ & $117.2^{+23.7}_{-25.0}$  \\
\hline
\end{tabular}
\end{table}
\clearpage

\section*{Acknowledgements}
I would like to thank an anonymous referee and H. Beust for their detailed and stimulating critical remarks which greatly improved the overall quality of the manuscript.

\bibliography{PXbib}{}

\begin{thebibliography}{114}
\expandafter\ifx\csname natexlab\endcsname\relax\def\natexlab#1{#1}\fi

\bibitem[{{Acedo}(2014)}]{2014Galax...2..466A}
{Acedo} L., 2014, Galaxies, 2, 466

\bibitem[{{Acedo}(2015)}]{2015Univ....1..422A}
{Acedo} L., 2015, Universe, 1, 422

\bibitem[{{Anderson} {et~al}\mbox{.}(1998){Anderson}, {Laing}, {Lau}, {Liu},
  {Nieto}, \& {Turyshev}}]{1998PhRvL..81.2858A}
{Anderson} J.~D., {Laing} P.~A., {Lau} E.~L., {Liu} A.~S., {Nieto} M.~M.,
  {Turyshev} S.~G., 1998, Phys. Rev. Lett., 81, 2858

\bibitem[{{Anderson} {et~al}\mbox{.}(2002){Anderson}, {Laing}, {Lau}, {Liu},
  {Nieto}, \& {Turyshev}}]{2002PhRvD..65h2004A}
{Anderson} J.~D., {Laing} P.~A., {Lau} E.~L., {Liu} A.~S., {Nieto} M.~M.,
  {Turyshev} S.~G., 2002, Phys. Rev. D, 65, 082004

\bibitem[{{Avalos-Vargas} \& {Ares de Parga}(2012)}]{2012EPJP..127..155A}
{Avalos-Vargas} A., {Ares de Parga} G., 2012, European Physical Journal Plus,
  127, 155

\bibitem[{{Bailey}, {Batygin} \& {Brown}(2016){Bailey}, {Batygin}, \&
  {Brown}}]{2016AJ....152..126B}
{Bailey} E., {Batygin} K., {Brown} M.~E., 2016, AJ, 152, 126

\bibitem[{{Batygin} \& {Brown}(2016{\natexlab{a}})}]{BaBroAJ2016}
{Batygin} K., {Brown} M.~E., 2016{\natexlab{a}}, AJ, 151, 22

\bibitem[{{Batygin} \& {Brown}(2016{\natexlab{b}})}]{2016arXiv161004992B}
{Batygin} K., {Brown} M.~E., 2016{\natexlab{b}}, arXiv:1610.04992

\bibitem[{{Bertotti}, {Farinella} \& {Vokrouhlick\'{y}}(2003){Bertotti},
  {Farinella}, \& {Vokrouhlick\'{y}}}]{2003ASSL..293.....B}
{Bertotti} B., {Farinella} P., {Vokrouhlick\'{y}} D., 2003, {Physics of the
  Solar System - Dynamics and Evolution, Space Physics, and Spacetime
  Structure.} Kluwer, Dordrecht

\bibitem[{{Beust}(2016)}]{2016A&A...590L...2B}
{Beust} H., 2016, A\& A, 590, L2

\bibitem[{{Blanchet} \& {Novak}(2011)}]{2011MNRAS.412.2530B}
{Blanchet} L., {Novak} J., 2011, MNRAS, 412, 2530

\bibitem[{{Bromley} \& {Kenyon}(2016)}]{2016ApJ...826...64B}
{Bromley} B.~C., {Kenyon} S.~J., 2016, ApJ, 826, 64

\bibitem[{{Brown} \& {Batygin}(2016)}]{BroBaAJ2016}
{Brown} M.~E., {Batygin} K., 2016, ApJL, 824, L23

\bibitem[{{Buscaino} {et~al}\mbox{.}(2015){Buscaino}, {DeBra}, {Graham},
  {Gratta}, \& {Wiser}}]{2015PhRvD..92j4048B}
{Buscaino} B., {DeBra} D., {Graham} P.~W., {Gratta} G., {Wiser} T.~D., 2015,
  Phys. Rev. D, 92, 104048

\bibitem[{{Capistrano}, {Penagos} \& {Al{\'a}rcon}(2016){Capistrano},
  {Penagos}, \& {Al{\'a}rcon}}]{2016arXiv160605263C}
{Capistrano} A.~J.~S., {Penagos} J.~A.~M., {Al{\'a}rcon} M.~S., 2016, MNRAS,
  463, 1587

\bibitem[{{Capistrano}, {Roque} \& {Valada}(2014){Capistrano}, {Roque}, \&
  {Valada}}]{2014MNRAS.444.1639C}
{Capistrano} A.~J.~S., {Roque} W.~L., {Valada} R.~S., 2014, MNRAS, 444, 1639

\bibitem[{{Chen} {et~al}\mbox{.}(2016){Chen}, {Lin}, {Holman}, {Payne},
  {Fraser}, {Lacerda}, {Ip}, {Chen}, {Kudritzki}, {Jedicke}, {Wainscoat},
  {Tonry}, {Magnier}, {Waters}, {Kaiser}, {Wang}, \&
  {Lehner}}]{2016ApJ...827L..24C}
{Chen} Y.-T. {et~al.}, 2016, ApJL, 827, L24

\bibitem[{{Cheung} \& {Xu}(2013)}]{2013ApJ...774...65C}
{Cheung} Y.-K.~E., {Xu} F., 2013, ApJ, 774, 65

\bibitem[{{Chicone} \& {Mashhoon}(2016)}]{2016CQGra..33g5005C}
{Chicone} C., {Mashhoon} B., 2016, Classical Quant. Grav., 33, 075005

\bibitem[{{Cowan}, {Holder} \& {Kaib}(2016){Cowan}, {Holder}, \&
  {Kaib}}]{2016ApJ...822L...2C}
{Cowan} N.~B., {Holder} G., {Kaib} N.~A., 2016, ApJL, 822, L2

\bibitem[{{Dark Energy Survey Collaboration} {et~al}\mbox{.}(2016){Dark Energy
  Survey Collaboration}, {Abbott}, {Abdalla}, {Aleksi{\'c}}, {Allam}, {Amara},
  {Bacon}, {Balbinot}, {Banerji}, {Bechtol}, {Benoit-L{\'e}vy}, {Bernstein},
  {Bertin}, {Blazek}, {Bonnett}, {Bridle}, {Brooks}, {Brunner}, {Buckley-Geer},
  {Burke}, {Caminha}, {Capozzi}, {Carlsen}, {Carnero-Rosell}, {Carollo},
  {Carrasco-Kind}, {Carretero}, {Castander}, {Clerkin}, {Collett}, {Conselice},
  {Crocce}, {Cunha}, {D'Andrea}, {da Costa}, {Davis}, {Desai}, {Diehl},
  {Dietrich}, {Dodelson}, {Doel}, {Drlica-Wagner}, {Estrada}, {Etherington},
  {Evrard}, {Fabbri}, {Finley}, {Flaugher}, {Foley}, {Fosalba}, {Frieman},
  {Garc{\'{\i}}a-Bellido}, {Gaztanaga}, {Gerdes}, {Giannantonio}, {Goldstein},
  {Gruen}, {Gruendl}, {Guarnieri}, {Gutierrez}, {Hartley}, {Honscheid}, {Jain},
  {James}, {Jeltema}, {Jouvel}, {Kessler}, {King}, {Kirk}, {Kron}, {Kuehn},
  {Kuropatkin}, {Lahav}, {Li}, {Lima}, {Lin}, {Maia}, {Makler}, {Manera},
  {Maraston}, {Marshall}, {Martini}, {McMahon}, {Melchior}, {Merson}, {Miller},
  {Miquel}, {Mohr}, {Morice-Atkinson}, {Naidoo}, {Neilsen}, {Nichol}, {Nord},
  {Ogando}, {Ostrovski}, {Palmese}, {Papadopoulos}, {Peiris}, {Peoples},
  {Percival}, {Plazas}, {Reed}, {Refregier}, {Romer}, {Roodman}, {Ross},
  {Rozo}, {Rykoff}, {Sadeh}, {Sako}, {S{\'a}nchez}, {Sanchez}, {Santiago},
  {Scarpine}, {Schubnell}, {Sevilla-Noarbe}, {Sheldon}, {Smith}, {Smith},
  {Soares-Santos}, {Sobreira}, {Soumagnac}, {Suchyta}, {Sullivan}, {Swanson},
  {Tarle}, {Thaler}, {Thomas}, {Thomas}, {Tucker}, {Vieira}, {Vikram},
  {Walker}, {Wechsler}, {Weller}, {Wester}, {Whiteway}, {Wilcox}, {Yanny},
  {Zhang}, \& {Zuntz}}]{2016MNRAS.460.1270D}
{Dark Energy Survey Collaboration} {et~al.}, 2016, MNRAS, 460, 1270

\bibitem[{{de la Fuente Marcos} \& {de la Fuente
  Marcos}(2016{\natexlab{a}})}]{2016MNRAS.460L..64D}
{de la Fuente Marcos} C., {de la Fuente Marcos} R., 2016{\natexlab{a}}, MNRAS,
  460, L64

\bibitem[{{de la Fuente Marcos} \& {de la Fuente
  Marcos}(2016{\natexlab{b}})}]{2016MNRAS.459L..66D}
{de la Fuente Marcos} C., {de la Fuente Marcos} R., 2016{\natexlab{b}}, MNRAS,
  459, L66

\bibitem[{{de la Fuente Marcos} \& {de la Fuente
  Marcos}(2016{\natexlab{c}})}]{2016MNRAS.462.1972D}
{de la Fuente Marcos} C., {de la Fuente Marcos} R., 2016{\natexlab{c}}, MNRAS,
  462, 1972

\bibitem[{{de la Fuente Marcos}, {de la Fuente Marcos} \& {Aarseth}(2016){de la
  Fuente Marcos}, {de la Fuente Marcos}, \& {Aarseth}}]{2016MNRAS.460L.123D}
{de la Fuente Marcos} C., {de la Fuente Marcos} R., {Aarseth} S.~J., 2016,
  MNRAS, 460, L123

\bibitem[{{Deng} \& {Xie}(2014)}]{2014Ap&SS.350..103D}
{Deng} X.-M., {Xie} Y., 2014, Astrophys. Space Sci., 350, 103

\bibitem[{{Deng} \& {Xie}(2015{\natexlab{a}})}]{2015EPJC...75..539D}
{Deng} X.-M., {Xie} Y., 2015{\natexlab{a}}, Eur. Phys. J. C, 75, 539

\bibitem[{{Deng} \& {Xie}(2015{\natexlab{b}})}]{2015NewA...35...36D}
{Deng} X.-M., {Xie} Y., 2015{\natexlab{b}}, New Astronomy, 35, 36

\bibitem[{{Deng} \& {Xie}(2015{\natexlab{c}})}]{2015AnPhy.361...62D}
{Deng} X.-M., {Xie} Y., 2015{\natexlab{c}}, Annals of Physics, 361, 62

\bibitem[{{Deng} \& {Xie}(2015{\natexlab{d}})}]{2015IJTP...54.1739D}
{Deng} X.-M., {Xie} Y., 2015{\natexlab{d}}, International Journal of
  Theoretical Physics, 54, 1739

\bibitem[{{Deng} \& {Xie}(2016)}]{2016MPLA...3150021D}
{Deng} X.-M., {Xie} Y., 2016, Mod. Phys. Lett. A, 31, 1650021

\bibitem[{{Einstein}(1915)}]{1915SPAW...47..831E}
{Einstein} A., 1915, Sitzber. Preuss. Akad., 831

\bibitem[{{Feldman} \& {Anderson}(2015)}]{2015IJMPD..2450066F}
{Feldman} M.~R., {Anderson} J.~D., 2015, Int. J. Mod. Phys. D, 24, 50066

\bibitem[{{Fienga} {et~al}\mbox{.}(2015){Fienga}, {Laskar}, {Exertier},
  {Manche}, \& {Gastineau}}]{2015CeMDA.123..325F}
{Fienga} A., {Laskar} J., {Exertier} P., {Manche} H., {Gastineau} M., 2015,
  Celest. Mech. Dyn. Astr., 123, 325

\bibitem[{{Fienga} {et~al}\mbox{.}(2011){Fienga}, {Laskar}, {Kuchynka},
  {Manche}, {Desvignes}, {Gastineau}, {Cognard}, \&
  {Theureau}}]{2011CeMDA.111..363F}
{Fienga} A., {Laskar} J., {Kuchynka} P., {Manche} H., {Desvignes} G.,
  {Gastineau} M., {Cognard} I., {Theureau} G., 2011, Celestial Mechanics and
  Dynamical Astronomy, 111, 363

\bibitem[{{Fienga} {et~al}\mbox{.}(2009){Fienga}, {Laskar}, {Kuchynka},
  {Manche}, {Gastineau}, \& {Le Poncin-Lafitte}}]{2009sf2a.conf..105F}
{Fienga} A., {Laskar} J., {Kuchynka} P., {Manche} H., {Gastineau} M., {Le
  Poncin-Lafitte} C., 2009, in SF2A-2009: Proceedings of the Annual meeting of
  the French Society of Astronomy and Astrophysics, {Heydari-Malayeri} M.,
  {Reyl'E} C., {Samadi} R., eds., Soci\'{e}t\'{e} Fran\c{c}aise d' Astronomie
  et d' Astrophysique, pp. 105--109,
  \url{http://sf2a.eu/proceedings/2009/2009sf2a.conf..0105F.pdf}

\bibitem[{{Fienga} {et~al}\mbox{.}(2016{\natexlab{a}}){Fienga}, {Laskar},
  {Manche}, \& {Gastineau}}]{2016A&A...587L...8F}
{Fienga} A., {Laskar} J., {Manche} H., {Gastineau} M., 2016{\natexlab{a}}, A\&
  A, 587, L8

\bibitem[{{Fienga} {et~al}\mbox{.}(2016{\natexlab{b}}){Fienga}, {Laskar},
  {Manche}, \& {Gastineau}}]{2016arXiv160100947F}
{Fienga} A., {Laskar} J., {Manche} H., {Gastineau} M., 2016{\natexlab{b}},
  Proc. 14th Marcel Grossmann Meeting, Rome, July 12-18, 2015. arXiv:1601.00947

\bibitem[{{Fienga} {et~al}\mbox{.}(2012){Fienga}, {Laskar}, {Verma}, {Manche},
  \& {Gastineau}}]{2012sf2a.conf...25F}
{Fienga} A., {Laskar} J., {Verma} A., {Manche} H., {Gastineau} M., 2012, in
  SF2A-2012: Proceedings of the Annual meeting of the French Society of
  Astronomy and Astrophysics, {Boissier} S., {de Laverny} P., {Nardetto} N.,
  {Samadi} R., {Valls-Gabaud} D., {Wozniak} H., eds., Soci\'{e}t\'{e}
  Fran\c{c}aise d' Astronomie et d' Astrophysique, pp. 25--33,
  \url{http://sf2a.eu/proceedings/2012/2012sf2a.conf..0025F.pdf}

\bibitem[{{Folkner}(2009)}]{Folkner09}
{Folkner} W.~M., 2009, Bull. Am. Astron Soc., 41, 880

\bibitem[{{Fortney} {et~al}\mbox{.}(2016){Fortney}, {Marley}, {Laughlin},
  {Nettelmann}, {Morley}, {Lupu}, {Visscher}, {Jeremic}, {Khadder}, \&
  {Hargrave}}]{2016ApJ...824L..25F}
{Fortney} J.~J. {et~al.}, 2016, ApJL, 824, L25

\bibitem[{{Fountain} {et~al}\mbox{.}(2008){Fountain}, {Kusnierkiewicz},
  {Hersman}, {Herder}, {Coughlin}, {Gibson}, {Clancy}, {Deboy}, {Hill},
  {Kinnison}, {Mehoke}, {Ottman}, {Rogers}, {Stern}, {Stratton}, {Vernon}, \&
  {Williams}}]{2008SSRv..140...23F}
{Fountain} G.~H. {et~al.}, 2008, Space Sci. Rev., 140, 23

\bibitem[{{Francisco} {et~al}\mbox{.}(2012){Francisco}, {Bertolami}, {Gil}, \&
  {P{\'a}ramos}}]{2012PhLB..711..337F}
{Francisco} F., {Bertolami} O., {Gil} P.~J.~S., {P{\'a}ramos} J., 2012, Phys.
  Lett. B, 711, 337

\bibitem[{{Galiazzo} {et~al}\mbox{.}(2016){Galiazzo}, {de la Fuente Marcos},
  {de la Fuente Marcos}, {Carraro}, {Maris}, \&
  {Montalto}}]{2016Ap&SS.361..212G}
{Galiazzo} M., {de la Fuente Marcos} C., {de la Fuente Marcos} R., {Carraro}
  G., {Maris} M., {Montalto} M., 2016, Astrophys. Space Sci., 361, 212

\bibitem[{{Gingerich}(1958)}]{1958ASPL....8....9G}
{Gingerich} O., 1958, Leaflet No. 352 of the Astronomical Society of the
  Pacific, 8, 9

\bibitem[{{Ginzburg}, {Sari} \& {Loeb}(2016){Ginzburg}, {Sari}, \&
  {Loeb}}]{2016ApJ...822L..11G}
{Ginzburg} S., {Sari} R., {Loeb} A., 2016, ApJL, 822, L11

\bibitem[{{Gomes}, {Deienno} \& {Morbidelli}(2016){Gomes}, {Deienno}, \&
  {Morbidelli}}]{2016arXiv160705111G}
{Gomes} R., {Deienno} R., {Morbidelli} A., 2016, arXiv:1607.05111

\bibitem[{{Gratia} \& {Fabrycky}(2017)}]{2017MNRAS.464.1709G}
{Gratia} P., {Fabrycky} D., 2017, MNRAS, 464, 1709

\bibitem[{{Halilsoy}, {Gurtug} \& {Mazharimousavi}(2015){Halilsoy}, {Gurtug},
  \& {Mazharimousavi}}]{2015APh....68....1H}
{Halilsoy} M., {Gurtug} O., {Mazharimousavi} S.~H., 2015, Astroparticle
  Physics, 68, 1

\bibitem[{{Hees} {et~al}\mbox{.}(2015){Hees}, {Bailey}, {Le Poncin-Lafitte},
  {Bourgoin}, {Rivoldini}, {Lamine}, {Meynadier}, {Guerlin}, \&
  {Wolf}}]{2015PhRvD..92f4049H}
{Hees} A. {et~al.}, 2015, Phys. Rev. D, 92, 064049

\bibitem[{{Hees} {et~al}\mbox{.}(2014){Hees}, {Folkner}, {Jacobson}, \&
  {Park}}]{2014PhRvD..89j2002H}
{Hees} A., {Folkner} W.~M., {Jacobson} R.~A., {Park} R.~S., 2014, Phys. Rev. D,
  89, 102002

\bibitem[{{Hogg}, {Quinlan} \& {Tremaine}(1991){Hogg}, {Quinlan}, \&
  {Tremaine}}]{1991AJ....101.2274H}
{Hogg} D.~W., {Quinlan} G.~D., {Tremaine} S., 1991, AJ, 101, 2274

\bibitem[{{Holman} \& {Payne}(2016{\natexlab{a}})}]{2016AJ....152...80H}
{Holman} M.~J., {Payne} M.~J., 2016{\natexlab{a}}, AJ, 152, 80

\bibitem[{{Holman} \& {Payne}(2016{\natexlab{b}})}]{2016AJ....152...94H}
{Holman} M.~J., {Payne} M.~J., 2016{\natexlab{b}}, arXiv:1604.03180

\bibitem[{{Iorio}(2007)}]{2007FoPh...37..897I}
{Iorio} L., 2007, Found. Phys., 37, 897

\bibitem[{{Iorio}(2010)}]{2010OAJ.....3....1I}
{Iorio} L., 2010, The Open Astronomy Journal, 3, 1

\bibitem[{{Iorio}(2012)}]{2012CeMDA.112..117I}
{Iorio} L., 2012, Celest. Mech. Dyn. Astr., 112, 117

\bibitem[{{Iorio}(2013)}]{2013CeMDA.116..357I}
{Iorio} L., 2013, Celest. Mech. Dyn. Astr., 116, 357

\bibitem[{{Iorio} \& {Giudice}(2006)}]{2006NewA...11..600I}
{Iorio} L., {Giudice} G., 2006, New Astronomy, 11, 600

\bibitem[{{Kaib} \& {Sheppard}(2016)}]{2016arXiv160701777K}
{Kaib} N.~A., {Sheppard} S.~S., 2016, arXiv:1607.01777

\bibitem[{{Kalinowski}(2015)}]{2015ForPh..63..711K}
{Kalinowski} M.~W., 2015, Fortschritte der Physik, 63, 711

\bibitem[{{Kenyon} \& {Bromley}(2016)}]{2016ApJ...825...33K}
{Kenyon} S.~J., {Bromley} B.~C., 2016, ApJ, 825, 33

\bibitem[{{Lai}(2016)}]{2016arXiv160801421L}
{Lai} D., 2016, arXiv:1608.01421

\bibitem[{{Lawler} {et~al}\mbox{.}(2016){Lawler}, {Shankman}, {Kaib},
  {Bannister}, {Gladman}, \& {Kavelaars}}]{2016arXiv160506575L}
{Lawler} S.~M., {Shankman} C., {Kaib} N., {Bannister} M.~T., {Gladman} B.,
  {Kavelaars} J.~J., 2016, arXiv:1605.06575

\bibitem[{{Le Verrier}(1859)}]{LeVer1859}
{Le Verrier} U., 1859, Cr. Hebd. Acad. Sci., 49, 379

\bibitem[{{Lee} \& {Chiang}(2016)}]{2016ApJ...827..125L}
{Lee} E.~J., {Chiang} E., 2016, ApJ, 827, 125

\bibitem[{{Li} \& {Adams}(2016)}]{2016ApJ...823L...3L}
{Li} G., {Adams} F.~C., 2016, ApJL, 823, L3

\bibitem[{{Li} {et~al}\mbox{.}(2014){Li}, {Yuan}, {Lu}, \&
  {Xie}}]{2014RAA....14..139L}
{Li} Z.-W., {Yuan} S.-F., {Lu} C., {Xie} Y., 2014, Research in Astronomy and
  Astrophysics, 14, 139

\bibitem[{{Lin} \& {Wang}(2013)}]{2013PhRvD..87h4041L}
{Lin} K., {Wang} A., 2013, Phys. Rev. D, 87, 084041

\bibitem[{{Linder} \& {Mordasini}(2016)}]{2016A&A...589A.134L}
{Linder} E.~F., {Mordasini} C., 2016, A\& A, 589, A134

\bibitem[{{Liu} {et~al}\mbox{.}(2014){Liu}, {Zhong}, {Han}, {Wang}, {Yang}, \&
  {Xie}}]{2014RAA....14.1019L}
{Liu} M.-Y., {Zhong} Z.-H., {Han} Y.-C., {Wang} X.-Y., {Yang} Z.-S., {Xie} Y.,
  2014, Research in Astronomy and Astrophysics, 14, 1019

\bibitem[{{Malhotra}, {Volk} \& {Wang}(2016){Malhotra}, {Volk}, \&
  {Wang}}]{2016ApJ...824L..22M}
{Malhotra} R., {Volk} K., {Wang} X., 2016, ApJL, 824, L22

\bibitem[{{Mart{\'{\i}}nez-Barbosa}
  {et~al}\mbox{.}(2017){Mart{\'{\i}}nez-Barbosa}, {J{\'{\i}}lkov{\'a}},
  {Portegies Zwart}, \& {Brown}}]{2017MNRAS.464.2290M}
{Mart{\'{\i}}nez-Barbosa} C.~A., {J{\'{\i}}lkov{\'a}} L., {Portegies Zwart} S.,
  {Brown} A.~G.~A., 2017, MNRAS, 464, 2290

\bibitem[{{Michaely} \& {Loeb}(2016)}]{2016arXiv160908614M}
{Michaely} E., {Loeb} A., 2016, arXiv:1609.08614

\bibitem[{{Milgrom}(2009)}]{2009MNRAS.399..474M}
{Milgrom} M., 2009, MNRAS, 399, 474

\bibitem[{{Modenini} \& {Tortora}(2014)}]{2014PhRvD..90b2004M}
{Modenini} D., {Tortora} P., 2014, Phys. Rev. D, 90, 022004

\bibitem[{{Mustill}, {Raymond} \& {Davies}(2016){Mustill}, {Raymond}, \&
  {Davies}}]{2016MNRAS.460L.109M}
{Mustill} A.~J., {Raymond} S.~N., {Davies} M.~B., 2016, MNRAS, 460, L109

\bibitem[{{Nesvorn{\'y}}, {Vokrouhlick{\'y}} \& {Roig}(2016){Nesvorn{\'y}},
  {Vokrouhlick{\'y}}, \& {Roig}}]{2016ApJ...827L..35N}
{Nesvorn{\'y}} D., {Vokrouhlick{\'y}} D., {Roig} F., 2016, ApJL, 827, L35

\bibitem[{{Nobili} \& {Will}(1986)}]{1986Natur.320...39N}
{Nobili} A.~M., {Will} C.~M., 1986, Nature, 320, 39

\bibitem[{{Nordtvedt}(1987)}]{1987ApJ...320..871N}
{Nordtvedt} K., 1987, ApJ, 320, 871

\bibitem[{{Nyambuya}(2015)}]{2015MNRAS.451.3034N}
{Nyambuya} G.~G., 2015, MNRAS, 451, 3034

\bibitem[{{Pau{\v c}o} \& {Kla{\v c}ka}(2016)}]{2016A&A...589A..63P}
{Pau{\v c}o} R., {Kla{\v c}ka} J., 2016, A\& A, 589, A63

\bibitem[{{Petit} {et~al}\mbox{.}(2016){Petit}, {Kavelaars}, {Gladman},
  {Jones}, {Parker}, {Van Laerhoven}, {Pike}, {Nicholson}, {Bieryla}, {Ashby},
  \& {Lawler}}]{2016arXiv160802873P}
{Petit} J. {et~al.}, 2016, arXiv:1608.02873

\bibitem[{{Philippov} \& {Chobanu}(2016)}]{2016PASA...33...33P}
{Philippov} J.~P., {Chobanu} M.~I., 2016, Publ. Astron. Soc. Aus., 33, e033

\bibitem[{{Pireaux} \& {Rozelot}(2003)}]{2003Ap&SS.284.1159P}
{Pireaux} S., {Rozelot} J.-P., 2003, Astrophys. Space Sci., 284, 1159

\bibitem[{{Pitjeva}(2006)}]{PitPio}
{Pitjeva} E.~V., 2006, {Limitations on some physical parameters from position
  observations of planets}. {Poster presented at Session 3 of \textit{Joint
  Discussion 16. Nomenclature, Precession and New Models in Fundamental
  Astronomy. 26th meeting of the IAU, 22-23 August 2006, Prague, Czech
  Republic}, International Astronomical Union},
  \url{http://syrte.obspm.fr/iauJD16/Astro2006$\_$0537.html}

\bibitem[{{Pitjeva} \& {Pitjev}(2013)}]{2013MNRAS.432.3431P}
{Pitjeva} E.~V., {Pitjev} N.~P., 2013, MNRAS, 432, 3431

\bibitem[{{Raymond} {et~al}\mbox{.}(2016){Raymond}, {Izidoro}, {Bitsch}, \&
  {Jacobson}}]{2016MNRAS.458.2962R}
{Raymond} S.~N., {Izidoro} A., {Bitsch} B., {Jacobson} S.~A., 2016, MNRAS, 458,
  2962

\bibitem[{{Rievers} \& {L{\"a}mmerzahl}(2011)}]{2011AnP...523..439R}
{Rievers} B., {L{\"a}mmerzahl} C., 2011, Ann. Phys.-Berlin, 523, 439

\bibitem[{{Rincon}(2006)}]{PLUTO}
{Rincon} P., 2006, {BBC News},
  http://news.bbc.co.uk/2/hi/science/nature/4596246.stm

\bibitem[{{Rodrigues}, {Mauro} \& {de Almeida}(2016){Rodrigues}, {Mauro}, \&
  {de Almeida}}]{2016arXiv160903613R}
{Rodrigues} D.~C., {Mauro} S., {de Almeida} {\'A}.~O.~F., 2016, Phys. Rev. D

\bibitem[{{Ruggiero}(2015)}]{2015IJMPD..2450060R}
{Ruggiero} M.~L., 2015, Int. J. Mod. Phys. D, 24, 50060

\bibitem[{{Ruggiero} \& {Radicella}(2015)}]{2015PhRvD..91j4014R}
{Ruggiero} M.~L., {Radicella} N., 2015, Phys. Rev. D, 91, 104014

\bibitem[{{Saillenfest} {et~al}\mbox{.}(2016){Saillenfest}, {Fouchard},
  {Tommei}, \& {Valsecchi}}]{2016CeMDA.tmp...31S}
{Saillenfest} M., {Fouchard} M., {Tommei} G., {Valsecchi} G.~B., 2016, Celest.
  Mech. Dyn. Astr.

\bibitem[{{Shankman} {et~al}\mbox{.}(2016){Shankman}, {Kavelaars}, {Lawler},
  {Gladman}, \& {Bannister}}]{2016arXiv161004251S}
{Shankman} C., {Kavelaars} J., {Lawler} S.~M., {Gladman} B.~J., {Bannister}
  M.~T., 2016, arXiv:1610.04251

\bibitem[{{Shao} \& {Wex}(2013)}]{2013CQGra..30p5020S}
{Shao} L., {Wex} N., 2013, Classical Quant. Grav., 30, 165020

\bibitem[{{Sheppard} \& {Trujillo}(2016)}]{2016arXiv160808772S}
{Sheppard} S.~S., {Trujillo} C., 2016, arXiv:1608.08772

\bibitem[{{Sheppard}, {Trujillo} \& {Tholen}(2016){Sheppard}, {Trujillo}, \&
  {Tholen}}]{2016ApJ...825L..13S}
{Sheppard} S.~S., {Trujillo} C., {Tholen} D.~J., 2016, ApJL, 825, L13

\bibitem[{{Sivaram}, {Kenath} \& {Kiren}(2016){Sivaram}, {Kenath}, \&
  {Kiren}}]{2016Ap&SS.361..230S}
{Sivaram} C., {Kenath} A., {Kiren} O.~V., 2016, Astrophys. Space Sci., 361, 230

\bibitem[{{Standish}(2008)}]{2008AIPC..977..254S}
{Standish} E.~M., 2008, in American Institute of Physics Conference Series,
  Vol. 977, Recent Developments in Gravitation and Cosmology, {Macias} A.,
  {L{\"a}mmerzahl} C., {Camacho} A., eds., pp. 254--263

\bibitem[{{Standish}(2010)}]{2010IAUS..261..179S}
{Standish} E.~M., 2010, in Proceedings of the International Astronomical Union,
  Vol. 261, Relativity in Fundamental Astronomy, {Klioner} S.~A., {Seidelmann}
  P.~K., {Soffel} M.~H., eds., pp. 179--182

\bibitem[{{Stern}(2008)}]{2008SSRv..140....3S}
{Stern} S.~A., 2008, Space Sci. Rev., 140, 3

\bibitem[{{Tangen}(2007)}]{2007PhRvD..76d2005T}
{Tangen} K., 2007, Phys. Rev. D, 76, 042005

\bibitem[{{Toth}(2016)}]{2016A&A...592A..86T}
{Toth} I., 2016, A\& A, 592, A86

\bibitem[{{Trujillo} \& {Sheppard}(2014)}]{2014Natur.507..471T}
{Trujillo} C.~A., {Sheppard} S.~S., 2014, Nature, 507, 471

\bibitem[{{Turyshev} \& {Toth}(2010)}]{2010LRR....13....4T}
{Turyshev} S.~G., {Toth} V.~T., 2010, Living Rev. Relativ., 13

\bibitem[{{Turyshev} {et~al}\mbox{.}(2012){Turyshev}, {Toth}, {Kinsella},
  {Lee}, {Lok}, \& {Ellis}}]{2012PhRvL.108x1101T}
{Turyshev} S.~G., {Toth} V.~T., {Kinsella} G., {Lee} S.-C., {Lok} S.~M.,
  {Ellis} J., 2012, Phys. Rev. Lett., 108, 241101

\bibitem[{{Tyler} {et~al}\mbox{.}(2008){Tyler}, {Linscott}, {Bird}, {Hinson},
  {Strobel}, {P{\"a}tzold}, {Summers}, \&
  {Sivaramakrishnan}}]{2008SSRv..140..217T}
{Tyler} G.~L., {Linscott} I.~R., {Bird} M.~K., {Hinson} D.~P., {Strobel} D.~F.,
  {P{\"a}tzold} M., {Summers} M.~E., {Sivaramakrishnan} K., 2008, Space Sci.
  Rev., 140, 217

\bibitem[{{Veras}(2016)}]{2016MNRAS.463.2958V}
{Veras} D., 2016, MNRAS, 463, 2958

\bibitem[{{Weryk} {et~al}\mbox{.}(2016){Weryk}, {Lilly}, {Chastel}, {Denneau},
  {Jedicke}, {Magnier}, {Wainscoat}, {Chambers}, {Flewelling}, {Huber},
  {Waters}, \& {PS1 Builders}}]{2016arXiv160704895W}
{Weryk} R.~J. {et~al.}, 2016, arXiv:1607.04895

\bibitem[{{Wilhelm} \& {Dwivedi}(2014)}]{2014NewA...31...51W}
{Wilhelm} K., {Dwivedi} B.~N., 2014, New Astronomy, 31, 51

\bibitem[{{Xie} \& {Deng}(2013)}]{2013MNRAS.433.3584X}
{Xie} Y., {Deng} X.-M., 2013, MNRAS, 433, 3584

\bibitem[{{Xie} \& {Deng}(2014)}]{2014MNRAS.438.1832X}
{Xie} Y., {Deng} X.-M., 2014, MNRAS, 438, 1832

\bibitem[{{Zhao} \& {Xie}(2015)}]{2015PhRvD..92f4033Z}
{Zhao} S.-S., {Xie} Y., 2015, Phys. Rev. D, 92, 064033

\end{thebibliography}

\end{document}